\documentclass[twocolumn]{aastex631}

\usepackage{hyperref}
\usepackage[version=4]{mhchem} 
\usepackage{hyperref}

\begin{document}

\title{JWST resolves jet-driven H$_2$ and ionized outflows in radio galaxy 3C305}

\correspondingauthor{Biny Sebastian}
\email{bsebastian@stsci.edu}
\author[0000-0001-8428-6525]{Biny Sebastian}
\affiliation{Space Telescope Science Institute, 3700 San Martin Drive, Baltimore, MD 21218, USA}

\author[0000-0002-3471-981X]{Patrick M. Ogle}
\affiliation{Space Telescope Science Institute, 3700 San Martin Drive, Baltimore, MD 21218, USA}

\author[0000-0002-2421-1350]{P. Guillard}
\affiliation{CNRS, Institut d’Astrophysique de Paris, 98 bis boulevard Arago, 75014 Paris, France}
\author[0000-0002-3249-8224]{L. Lanz}
\affiliation{The College of New Jersey, 2000 Pennington Road, Ewing, NJ 08618, USA}

\author[0000-0002-9482-6844]{R. Morganti}
\affiliation{ASTRON, the Netherlands Institute for Radio Astronomy, Oude Hoogeveensedijk 4, 7991 PD Dwingeloo, The Netherlands.}
\affiliation{Kapteyn Astronomical Institute, University of Groningen, P.O. Box 800,
9700 AV Groningen, The Netherlands}

\author[0000-0002-6472-6711]{V. Reynaldi}
\affiliation{Instituto de Astrofísica de La Plata (CONICET–UNLP), Paseo del Bosque, 1900 La Plata, Argentina}
\affiliation{Facultad de Ciencias Astronómicas y Geofísicas, Universidad Nacional de La Plata, Argentina}

\author{I. E. López}
\affiliation{INAF - Osservatorio di Astrofisica e Scienza dello Spazio di Bologna, via Gobetti 93/3, 40129, Bologna, Italy }
\author{B. Emonts}
\affiliation{National Radio Astronomy Observatory (Associated Universities, Inc.), 520 Edgemont Road, Charlottesville, VA 22903-2475, USA}

\author{S. Garcia-Burillo}
\affiliation{Observatorio Astronómico Nacional (IGN), C/ Alfonso XII, 3, 28014 Madrid, Spain}

\author{C. Tadhunter}
\affiliation{Department of Physics and Astronomy, University of Sheffield, Hicks Building, Hounsfield Road, Sheffield S3 7RH, United Kingdom}

\author{C. P. O’Dea}
\affiliation{University of Manitoba, 66 Chancellors Circle, Winnipeg, MB R3T 2N2, Canada}

\affiliation{Center for Space Plasma \& Aeronomic Research, 
University of Alabama in Huntsville
Huntsville, AL 35899, USA}
\author{S. Baum}
\affiliation{University of Manitoba, 66 Chancellors Circle, Winnipeg, MB R3T 2N2, Canada}

\affiliation{Center for Space Plasma \& Aeronomic Research, 
University of Alabama in Huntsville
Huntsville, AL 35899, USA}
\author{F. R. Faifer}
\affiliation{Facultad de Ciencias Astron\'omicas y Geof\'isicas, Universidad Nacional de La Plata, Paseo del Bosque S/N, B1900FWA La Plata, Argentina}
\affiliation{Instituto de Astrof\'isica de La Plata (CONICET), Paseo del Bosque S/N, B1900FWA La Plata, Argentina}
\author[0000-0002-0690-8824]{A. Labiano}
\affiliation{Telespazio UK S.L. for ESA, ESAC, Camino Bajo del Castillo s/n, 28692 Villanueva de la Ca\~nada, Madrid, Spain}

\author{M. Lehnert}
\affiliation{Centre de Recherche Astrophysique de Lyon (CRAL), Observatoire de Lyon, 9 Avenue Charles André, 69561 Saint-Genis-Laval Cedex, France}

\author{A. Togi}
\affiliation{Department of Physics, Texas State University, 601 University Drive, San Marcos, TX 78666, USA
}
\author{K. Alatalo}
\affiliation{Space Telescope Science Institute, 3700 San Martin Drive, Baltimore, MD 21218, USA}

\author{P. Appleton}
\affiliation{California Institute of Technology, 1200 E. California Blvd., Pasadena, CA 91125, USA}

\author{R. Bhutkar}
\affiliation{University of Manitoba, 66 Chancellors Circle, Winnipeg, MB R3T 2N2, Canada}
\author{E. Egami}
\affiliation{Department of Astronomy/Steward Observatory, University of Arizona, 933 N. Cherry Avenue, Tucson, AZ 85721-0065, USA}




\begin{abstract}

We present JWST MIRI MRS, NIRSpec, NIRCam, and MIRI imaging observations of 3C\,305, a radio galaxy with a compact jet that is confined within the galaxy.  We utilize the H$_2$ 0-0 S(1) - S(7) lines, several mid-IR fine structure lines, and PAH emission in the MIRI MRS spectrum to conduct a multi-phase study of the radio jet's impact on the interstellar medium. Multiple tracers, including H$_2$/PAH 11.3 $\mu$m and [Fe\,II] 5.34$\mu$m provide evidence for shocks at the jet termination locations. Two Gaussian components are required to reproduce the kinematics of warm H$_2$ adequately, with one representing the bulk low-velocity component, while the other corresponds to an outflow. 
The ionized gas attains higher outflow velocities than the H$_2$ outflow. The stark increase in velocities at the jet hotspots points to jet-driven outflows. We fit the H$_2$ excitation diagram with a power-law temperature distribution and find that the hotspots exhibit flatter slopes, indicating a larger warm/hot gas mass fraction at these locations. 
 Our MAPPINGS line-ratio analysis indicates that most of the mid-IR ionized gas can be fit by a `shock-plus-precursor' model. We find that strong radiative losses dominated by line cooling, along with moderate kinetic power of the molecular + ionized gas outflows, can account for all of the estimated jet power, indicating high jet coupling efficiency in 3C\,305. 
 Our results, in tandem with other studies on multiphase gas, show that jets efficiently shock-heat and accelerate the gas that comes in contact with them, driving massive, kiloparsec-scale, multiphase outflows.

\end{abstract}


\section{Introduction} \label{sec:intro}

 Active galactic nucleus (AGN) feedback has been identified as a potential solution to several long-standing discrepancies between observations of galaxy evolution and theoretical predictions \citep{Harrison2024}. For example, the over-prediction of galaxies at the high mass ends of the luminosity functions, the observed correlation of the black hole mass and the velocity dispersion of the stellar bulge, the metal enrichment of the inter cluster medium, and the lower gas cooling rates around galaxy clusters compared to theoretical predictions were some of the major problems successfully resolved by invoking AGN feedback \citep{Ferrarese2000,Gebhardt2000,silk1998,dimatteo2005,Springel2005,Croton2006,Schaye2015,Weinberger2018,Fabian2012}.
 
 Observational studies have indeed supported this idea and have made significant progress in understanding AGN feedback over the last few decades. Both positive and negative feedback modes are known to operate in galaxies. Positive feedback operates via triggering of star formation by compression of dense gas by outflows/jets, and is primarily a localized phenomenon  \citep{Silk2013}, whereas negative feedback is the more widely accepted form of feedback with a more global impact \citep{Cresci2015}.
 
 However, some questions still remain unanswered. For example, how is AGN feedback operating? Is it mainly via outflows, where the fuel required for star formation is ejected out of the galaxy, or via turbulent heating, where the AGN injects energy back into the ISM and IGM through heating via shocks and turbulence? In practice, outflows can also generate shocks and turbulence and contribute to the heating. Furthermore, it is challenging to disentangle the factors that influence galaxy feedback; for example, the effects of star formation, radio jets, and AGN radiation-driven outflows.

Multiphase gas outflows have routinely been detected in AGN and are also associated with radio galaxies \citep[e.g.,][]{Nesvadba2006,Nesvadba2017,Harrison2018,Veilleux2020,Morganti2023}. Feedback on the molecular gas phase is particularly interesting due to its direct link to star formation. It has been observed that molecular gas-phase outflows often dominate the outflow gas budget and can even match or exceed star formation rates in several galaxies \citep{Alatalo2015,Cicone2014,Feruglio2010}. Warm molecular hydrogen (H$_2$), traced by mid-IR rotational lines, is a good tracer of the warm molecular phase with temperatures above 200~K. Due to the lack of a permanent dipole moment, it is hard to detect cold molecular gas. In contrast, warm molecular hydrogen emission can be probed using MIR rotational and NIR ro-vibrational lines. Although cold molecular gas is the most critical factor for predicting star formation, the presence of warm H$_2$ suggests that some form of heating or feedback is at play, exciting the cold interstellar medium through mechanisms like cosmic ray heating, UV or X-ray photon heating, or shock interactions from jets \citep{Lopex2025}.

The mechanical power of jets estimated from X-ray cavity studies exceeds their radio luminosity by at least two orders of magnitude \citep{Birzan2008,cavagnolo2010}. 
This is in sharp contrast to accretion-mode feedback, where only a few percent of the accretion luminosity is coupled to the host galaxy. Hence, the jet's mechanical power can exceed the accretion-related feedback even when jet-related radio emission is weak. Recent observations and simulations at low redshift have indeed been pointing to this major role played by radio jets in quasars at low redshift \citep{Mukherjee2018,jarvis2019,Molyneux2019}. 

\citet{Ogle2010}, observing 3CR radio galaxies with \textit{Spitzer} IRS spectroscopy, found that about 30\% of these systems show excess warm molecular hydrogen emission. Owing to the bright radio emission and the relatively weak AGN accretion disk-related emission or star formation, low-redshift 3CR radio sources provide a clean sample for unambiguously investigating the role of radio jet feedback. The study indicated that the greatest predictor of this warm molecular gas excess is the presence of extended radio jets that remain confined within the galaxy's ISM, suggesting significant interaction between the jet and the surrounding medium. \cite{Odea2021} compile ample evidence to show that jets confined within the galaxy indeed are particularly effective in transferring energy into the galaxy. 

Following up on some of the systems in \cite{Ogle2010} that hosted H\,I outflows, \cite{Guillard2012a} found that warm H$_2$ was indeed excited by the jets, but it was rarely found to be outflowing. However, the non-detection of molecular outflows may have been because \textit{Spitzer} did not have the spatial and spectral resolution required to distinguish the outflowing material from the rotating disc.

The IFUs aboard JWST have enabled spatially resolved studies of the impact of the radio jet feedback at both high and low redshifts \citep{Pereira-Santaella2022,Dasyra2024,Roy2024,Ogle2025,Riffel2025}. The local systems have the advantage of enabling spatially resolved studies of jet feedback. Moreover combined MIRI and NIRSpec spectral coverage provides a sensitive, higher-resolution view of the impact of jets on the molecular and ionized phases.






\subsection{3C\,305}
3C\,305 is a nearby radio galaxy ($z\sim$0.042) whose host galaxy is a disturbed early-type galaxy exhibiting signatures of a recent gas-rich merger \citep{Heckman1985,Jackson2003}. 3C\,305 has a peculiar $z$-shaped radio morphology \citep{Heckman1982}. The radio luminosity at 178~MHz of 5.5$\times$10$^{24}$~W~Hz$^{-1}$~sr$^{-1}$ lies on the boundary of the Fanaroff-Riley divide \citep[FR;][]{Fanaroff1974}. However, higher resolution images reveal bright jets and hotspots indicative of a classical FR\,II radio morphology \citep{Jackson2003,Hardcastle2012}. The host galaxy also presents evidence for tidal tails and merger-induced disturbances \citep{Emonts2016}. 

The compact jet, with a projected hotspot-to-hotspot size of 3.5~kpc in 3C\,305 shows evidence for prominent jet-ISM interactions as revealed by multi-wavelength studies of the various phases of gas. In particular, several shock indicators, excess warm H$_2$, and disturbed multi-phase gas kinematics make 3C\,305 particularly interesting. This galaxy hosts an extended emission line region (EELR) aligned with the radio jet axis. The strong [O\,{\small{III}}] emission in 3C\,305 was found to have a shock-ionized origin \citep{Reynaldi2013}. \cite{Jackson2003} reported [Fe\,II] emission coincident with the hotspot regions, pointing towards the presence of shock-heated gas. Extended X-ray emission coincident with the optical emission line gas was also found to be shock heated rather than photoionized by the central AGN \citep{Massaro2009,Hardcastle2012}. \cite{Hardcastle2012} find that this X-ray emitting gas dominates the energy budget of the multi-phase ISM. We note that AGN photoionization cannot be ruled out entirely based on X-ray spectroscopy alone, as the primary evidence for shock origin is morphological.

Extreme kinematics implying jet-driven outflows are found in both radio H{\sc i} absorption lines \citep{Morganti2005} and optical emission lines \citep{Reynaldi2013}. Excess warm molecular H$_2$  ($\sim$10$^8$ M$_{\odot}$ at 200–3000~K) was found in 3C\,305 \citep{Guillard2012a}. While it is hypothesized that the shock heating by the jet is responsible for the excitation of the molecular gas \citep{Ogle2010}, the \textit{Spitzer} spectra did not reveal evidence for extreme kinematics or outflows similar to those found in the cold neutral gas \citep{Guillard2012a}. \cite{Hardcastle2012} note that such an absence could be attributed to sensitivity limitations or dissociation of the gas. 


Using our new JWST NIRSpec and MIRI medium-resolution spectrometer (MRS) observations, we have mapped the spatial distribution and kinematics of the multi-phase gas at spatial resolutions, $\sim$60--130~pc for NIRSpec IFU and $\sim$170--600~pc for MIRI MRS. 
In this paper, we carry out a detailed analysis of the impact of the jets on the warm molecular and ionized gas as well as polycyclic aromatic hydrocarbons (PAHs). We address the following outstanding science questions concerning jet feedback in radio galaxies (i) What fraction of the ISM is heated versus ejected in outflows? and (ii) What is the net efficiency of the energy transfer from the jets to various gas phases? and (iii) What is the energy budget transferred onto various gas phases? 

This paper is organized as follows. In section~\ref{sec:observations}, we describe the JWST NIRCam, NIRSpec, and the MIRI MRS data along with archival VLA data and the data analysis procedures used. We then detail results from our kinematic analysis, shock diagnostics, the excitation mechanism analysis, and a comparison of the H$_2$ and ionized gas in section~\ref{sec:results}. In section~\ref{sec:discussion}, we attempt to answer the questions raised in the previous paragraph and then finally conclude in section~\ref{sec:conclusion}.

In this paper, we adopt a $\Lambda$CDM cosmology, with
$H_0 = 67.4~\mathrm{km~s^{-1}~Mpc^{-1}}$, $\Omega_m = 0.315$,
and $\Omega_\Lambda = 0.685$. We use a luminosity distance
of 190~Mpc and an angular scale of
0.85~kpc~arcsec$^{-1}$ at $z = 0.04161$ for 3C\,305.

\section{Observations and Data Analysis}
\label{sec:observations}
The radio galaxy, 3C\,305 was observed as a part of the cycle-2 General Observer (GO) program, (Program ID: \href{https://www.stsci.edu/jwst/science-execution/program-information?id=4237}{4237}; PI: P. Ogle) using NIRCam, NIRSpec IFU \citep{nirspec2023,Nirspec2022},  MIRI MRS \citep{miri2015,miri2023} and MIRI imaging \citep{Dicken2024} instruments on board JWST \citep{JWST2023}. The details of the observations from each of the instruments are provided below. 
\subsection{NIRCam Imaging}
\begin{figure*}
    \centering
    \includegraphics[height=0.325\linewidth]{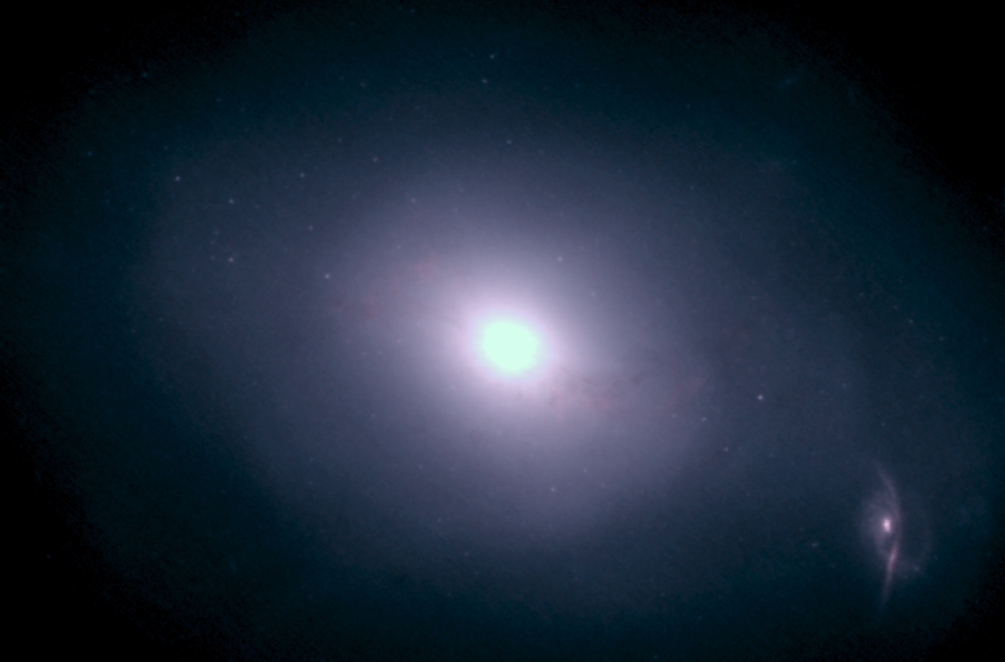}
    \includegraphics[height=0.325\linewidth]{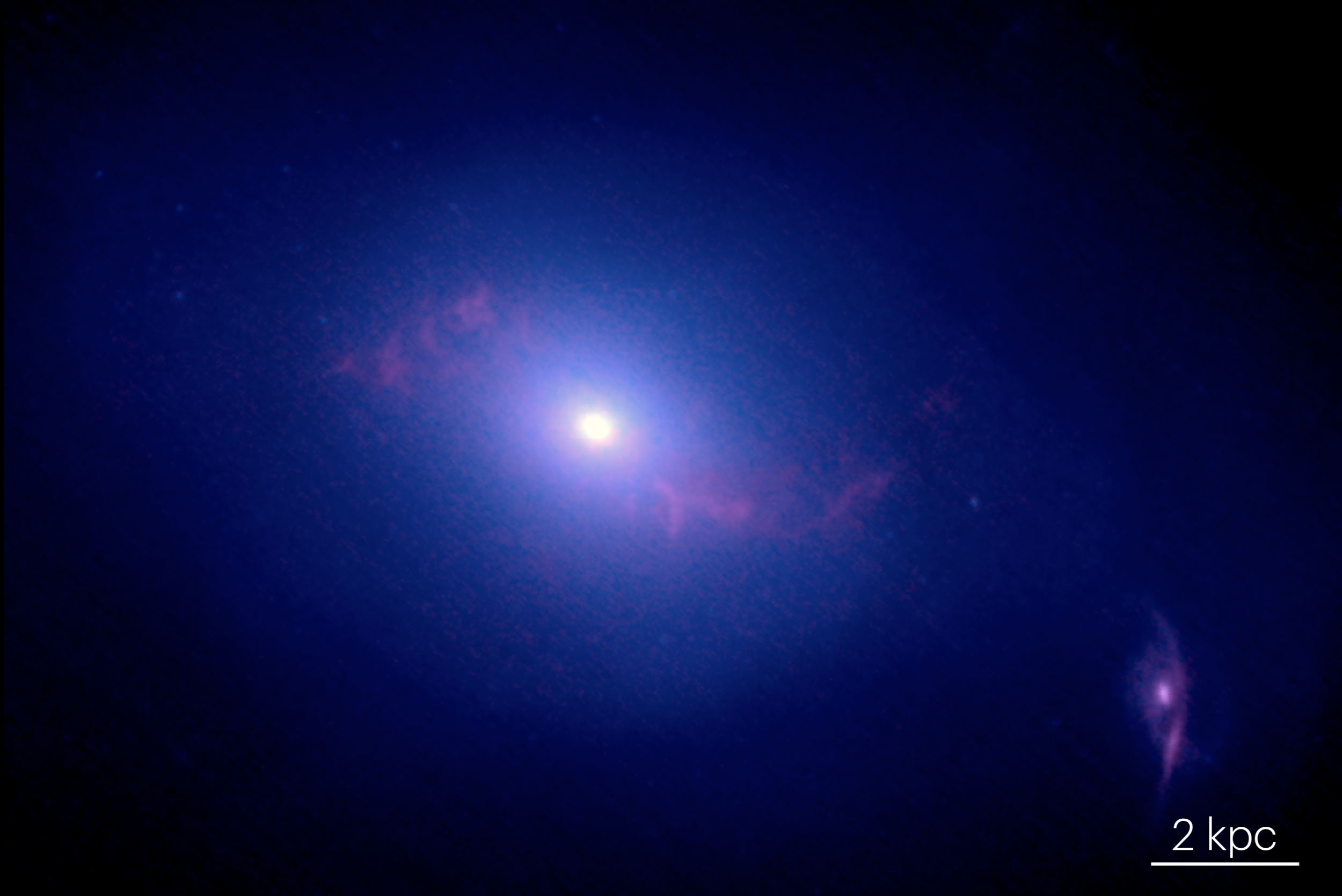}
    \includegraphics[height=0.325\linewidth]{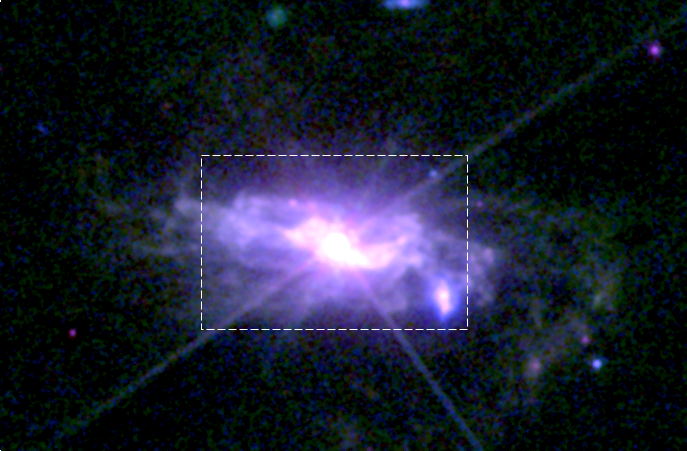}
    \includegraphics[height=0.325\linewidth]{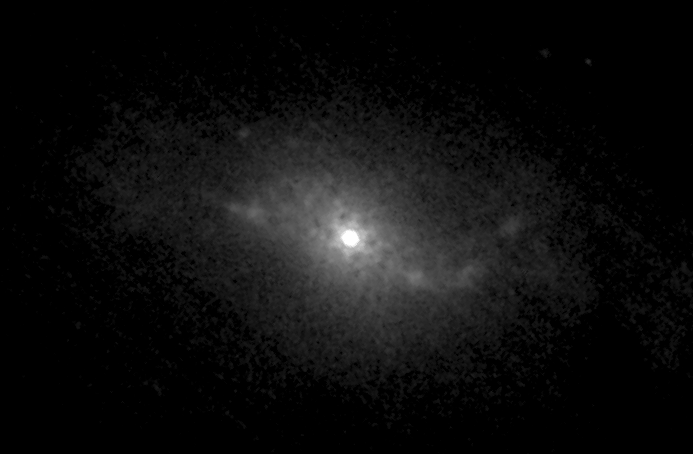}

    \caption{
\textit{Top Left:} JWST/NIRCam color composite image of 3C\,305 combining F150W (blue+green) and F182M (red), showing the stellar continuum and the dust lanes. The object to the SW appears to be an unrelated background galaxy pair.
\textit{Top Right:} Continuum-subtracted Pa$\alpha$ map derived as NIRCam F182M -- F150W (red), together with F150W continuum(blue). Extended Pa$\alpha$ emission traces ionized gas in filamentary structures along the jet axis. 
\textit{Bottom Left:} MIRI F770W, F1280W, and F1000W (r, g, b), highlighting PAH 7.7 $\mu m$, PAH 11.3 $\mu m$, and dust continuum emission with a visible contribution from H$_2$ and ionized gas lines. The dashed box outlines the extent of the other three panels in this figure.
\textit{Bottom Right:} Continuum-subtracted PAH 3.3 $\mu$m map derived from NIRCam F335W -- 0.78 F277W. The spatial distribution of PAH emission is similar to the dust lanes. AGN hot-dust continuum emission is apparent in the galaxy core. All the images are oriented with north up and east to the left.
}
    \label{fig:imaging}
\end{figure*}

NIRCam imaging was carried out using four filters, namely, F335M, F182M, F277W, and F150W on 2024-Jun-18. A total exposure time of 515.364 seconds was spent on each of the filters. The F182M and F335M  filters targeted the Pa$\alpha$ emission line at 1.875$\mu$m and the PAH 3.3$\mu$m dust emission feature, respectively. Both F277W and F150W act as the continuum reference filters to trace the stellar emission. NIRCam images were aligned and reprojected to a common world coordinate system using JWST/HST Alignment Tool \citep[JHAT; ][]{Rest2023} and astropy reproject \citep{Robitaille2020}.  Continuum-subtracted NIRCam images of Pa$\alpha$ and PAH 3.3 $\mu$m are shown together with continuum images of the galaxy and its dust lanes in Figure~\ref{fig:imaging}.
\begin{figure*}
    \centering
    \includegraphics[width=\linewidth]{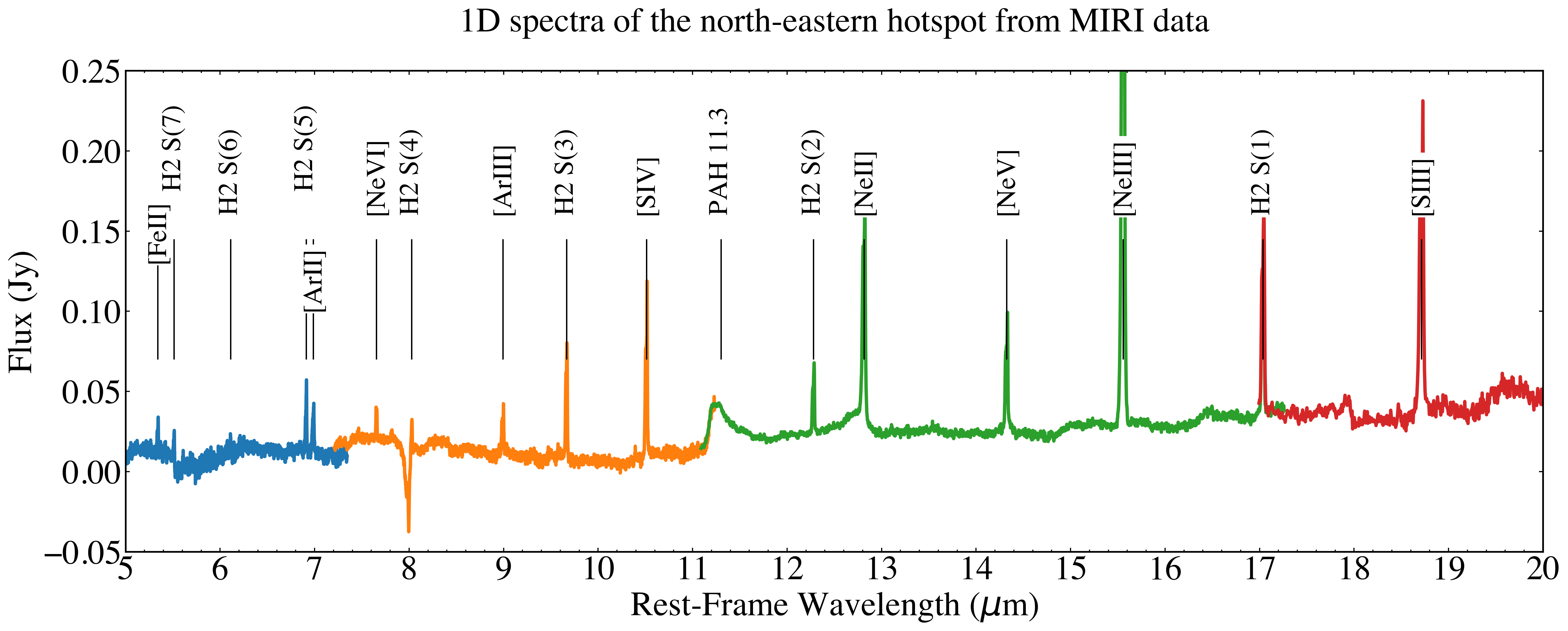}
    \caption{MIRI spectrum of a circular region extracted with a size of $\sim$ 0.9$^{\prime\prime}$at the north-east hotspot of 3C\,305. Strong emission lines and the PAH 11.3 $\mu$m feature are marked. The different colors denote the four MIRI MRS spectral channels. The region from which the spectra was extracted is marked as NE1 in the top left panel of Figure~\ref{fig:ionizedgas_comparison}.}
    \label{fig:1D_MIRI_spectra}
\end{figure*}


\subsection{MIRI Imaging}
 MIRI imaging was carried out on 2024-Apr-02 using filters F560W, F770W, F1000W, and F1280W. Together with the imaging taken in MRS mode, these filters cover the redshifted PAH 7.7 $\mu$m and 11.3 $\mu$m features (in F770W and F1280W) and adjacent continuum bands. A 4-point dither pattern and a FASTR1 readout pattern were utilized. A total exposure time of 111~s was devoted to each configuration. An off-galaxy background observation was also performed. A three-color rendition of the MIRI F770W, F1000W, and F1280W images is presented in Figure~\ref{fig:imaging}, featuring PAH and dust continuum emission from the galaxy disk, including spiral arms, with a minor contribution from H$_2$ and ionized gas emission lines.

\subsection{MIRI MRS}
The mid-infrared IFU observations were procured using the MIRI Medium-Resolution Spectrometer (MRS) on 2024-Apr-01. A 2$\times$2 mosaic was utilized to cover the inner jet interaction region of central 6.5 $\times$ 8.0~$^{\prime\prime}$ (5.5 $\times$ 6.8~kpc). Observations were taken in a four-point dither pattern optimized for an extended source to improve sampling. A single-point dithered off-galaxy observation was also obtained to perform reliable MIR background subtraction. Data were obtained in all the channels and sub-bands of MIRI-MRS. The readout mode used was FASTR1. An exposure time of 333~s was spent on each of these configurations.  Simultaneous imaging of 3C 305 was taken in the F560W, F770W, and F1130W filters during the background observations.
 
The MIRI MRS data were reprocessed using the Python module {\tt{JWST}} pipeline version 1.15.1 and CRDS context {\tt{jwst\_1293.pmap}}.
Figure~\ref{fig:1D_MIRI_spectra} shows the composite 1D spectra extracted from a region located at the NE hotspot of MIRI MRS, with prominent lines marked.
For all comparisons, including ratio maps and kinematic analyses, we reproject the data to the largest spaxel size and smooth to the coarsest PSF to ensure matched sampling and resolution across all bands.

\subsection{NIRSpec IFU Spectroscopy}
NIRSpec IFU observations were conducted on 2024-Apr-02 using F170LP/G235H and F290LP/G395H filter/grating combinations with a spectral coverage of 1.66--5.27~$\mu$m and wavelength-dependent spectral resolution, R of 1700-3500, i.e., a velocity resolution, $\Delta$v  $\sim$85-160~km~s$^{-1}$ (FWHM). The wavelength coverage by each grating is nearly seamless, except for a small gap caused by the physical separation between the detectors. These observations were taken using a 2$\times$2 mosaic covering the inner 6.2$''$ $\times$ 6.2$''$ (5.3$\times$5.3~kpc$^2$). The total science exposure time, including all configurations, amounts to 7002.656 seconds. The data were obtained using a small dither pattern with four dithers to improve spatial sampling of the PSF and minimize detector artifacts. NRSIRS2RAPID readout mode was employed to reduce the 1/f correlated noise and prevent saturation. 

\begin{figure*}
    \centering
    \includegraphics[width=\linewidth]{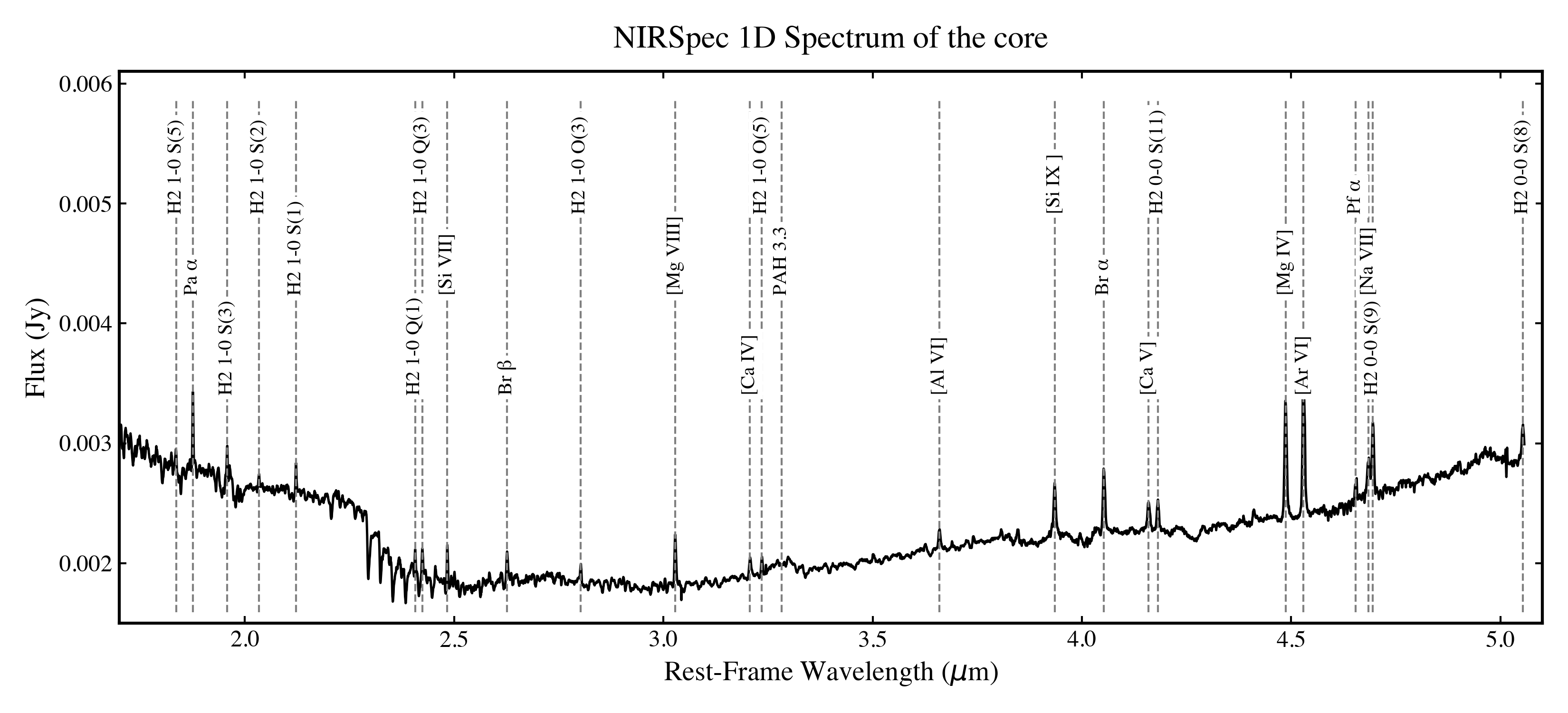}
    \caption{NIRSpec spectrum of the core. Strong emission lines and the PAH 3.3$\mu$m feature are marked. Prominent H$_2$ rovibrational and pure rotational lines mark emission from shock-heated warm molecular gas. High-ionization potential (coronal) lines such as [Mg VIII] and [Si IX] are likely photoionized by the AGN. CO absorption bands in the stellar continuum are apparent at 2.3-2.5$\mu$m.}
    \label{fig:1D_NIRspec_spectra}
\end{figure*}

We obtained the NIRSpec IFU data from the MAST portal, which was processed using the JWST calibration pipeline version 1.16.1 and the calibration reference files set using 1303 pmap. Because the source is relatively bright, our attempt to further reduce 1/f noise with the NSCLEAN algorithm \citep{Rauscher2024} did not significantly improve the data quality, so we utilize the standard pipeline-processed data.
We use the {\tt{JDAVIZ}} \citep{jdaviz2024} python module for cube visualization and creation of moment maps. The NIRSpec spectrum of the galaxy core is shown in Figure~\ref{fig:1D_NIRspec_spectra}, showing a variety of ionized and molecular gas emission lines, including coronal lines and H$_2$ rovibrational lines, and the PAH 3.3 $\mu$m feature.

\subsection{VLA data}
We used Karl G. Jansky Very Large Array (VLA) observations of 3C\,305 at X-band in the A-array configuration \citep[project ID: AH982;][]{Hardcastle2012}. Two intermediate frequencies (IFs) with a bandwidth of 100~MHz centered at 8.435 GHz and 8.735 GHz were used. 3C\,286, and J1436+636 were used as the primary flux density and phase calibrator, respectively. Data reduction was carried out using the standard pipeline VLARUN in \texttt{AIPS}, which applies automated flagging, calibration, and imaging. After calibration, the target was split and imaged using the task `IMAGR' with robust weighting with a robust parameter of 0.  The final image has a resolution of 0.065$''$ $\times$ 0.055$''$ at a position angle of -12.9$^{\circ}$ and rms noise of 33 $\mu$Jy.

\section{Results}
\begin{figure*}
    \centering
    \includegraphics[width=\linewidth]{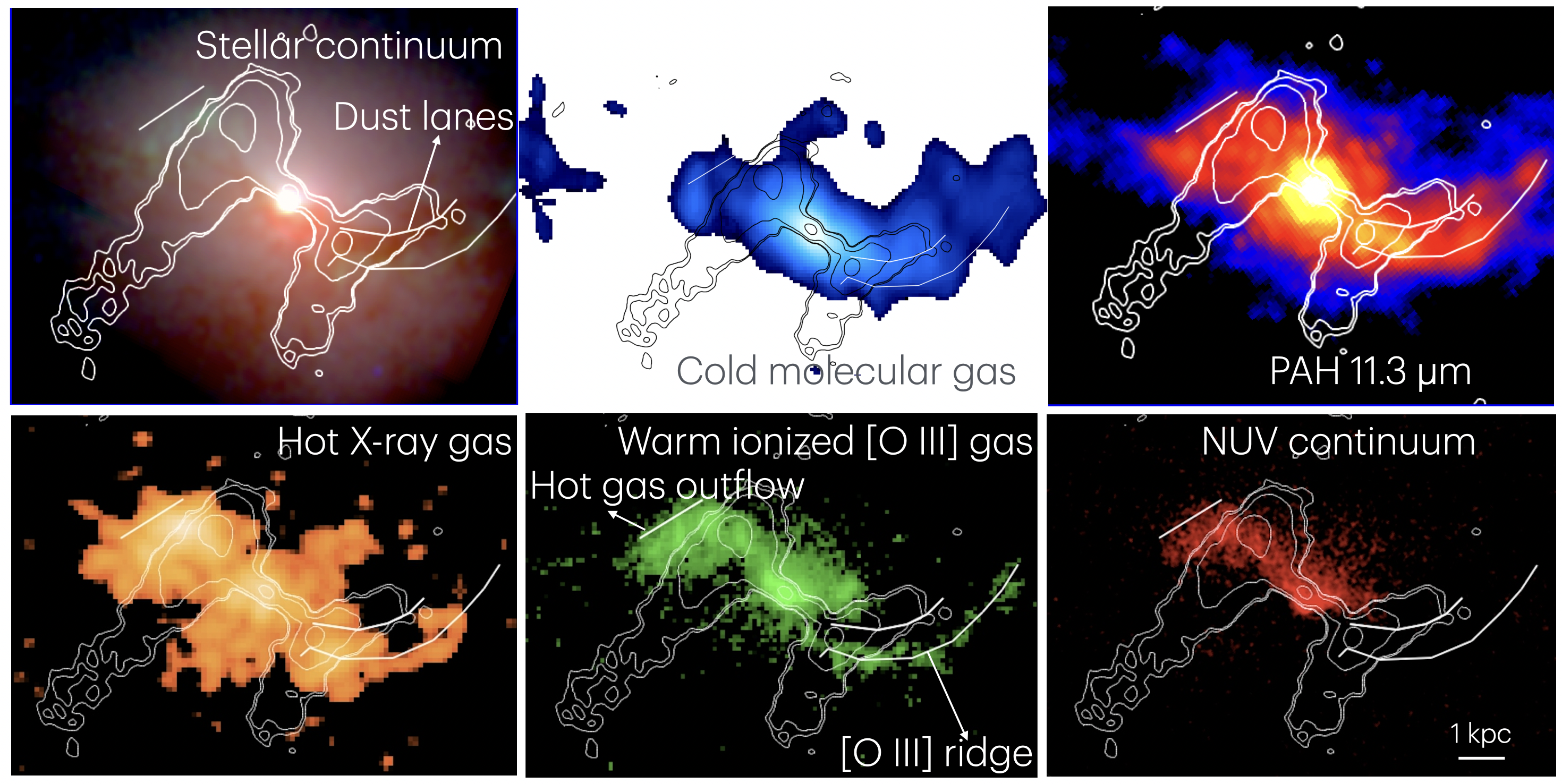}
    \caption{Context: Multi-phase gas distribution compiled from archival and current JWST data. Top row: (left) rgb color composite of HST NIR (F160W), optical V (F555W), and optical R (F702W) bands showing prominent dust lanes, (middle) NOEMA CO (1-0) gas tracing cold molecular gas shown as a moment-0 map with a synthesized beam of 1.5$''$ × 0.9$''$ at PA $\approx$ 12$^\circ$  \citep{Morganti2023}, (right) JWST/MIRI F1130W image tracing the PAH 11.3 µm feature (similar to PAH 3.3 $\mu$m in Figure~\ref{fig:imaging}). Bottom row: (left) hot X-ray emitting plasma emission (0.5–7 keV) from \textit{Chandra}/ACIS,\citep{Massaro2009,Hardcastle2012}, (middle) HST/WFPC2 narrowband image (F521N) centered at [O\,{\small{III}}] $\lambda$5007 emission tracing warm ionized gas, and (right) near-UV (NUV) continuum emission using HST/STIS NUV-MAMA. X-band radio contours are overlaid in all panels to highlight the jet morphology. Contour levels are $[0.8,\,1.2,\,5.7,\,50]\times10^{-4}~\mathrm{Jy}$.
 The marked dust lane may obscure some of the X-ray, [O\,{\small{III}}], and NUV emission in the southwest.}
    \label{fig:collage}
\end{figure*}

The main goal of this work is to characterize the impact of the radio jets on the various gas phases. Figure~\ref{fig:1D_MIRI_spectra} shows emission lines in the MIRI-MRS spectra that enable spatially resolved characterization of feedback on multi-phase gas. The rotational H$_2$ S(1)-S(5) lines, tracing warm molecular gas (200-3000~K), can constrain molecular gas excitation by jet-driven shock-heating. (H$_2$ S(6)-H$_2$ S(8) were omitted due to poor SNR). Ionized gas emission lines with a range of ionization states (e.g., [Fe II], [Ne II], [Ne III], [Ne V], [Ar II], [Ar III], [S III], [S IV], [O IV]) can help to distinguish among ionization mechanisms, such as shock-ionization by jets or winds versus AGN photoionization. The absorption feature seen at 8$\mu$m results from residual instrumental systematics. 


\label{sec:results}
\subsection{Context: Multi-phase gas in 3C\,305}
Figure~\ref{fig:collage}
presents the distribution of the gas in various phases in 3C\,305, and its correlation with the jet. The X-band radio contours at 8.5~GHz are overlaid on all images. Each panel marks the dust lanes seen in HST wideband images and the path of the warm ionized gas as seen in the [O\,III]~$\lambda$5007 narrowband image.  The RGB color composite of the HST wideband images depicts the prominent dust lanes (also seen in the JWST NIRCam image; Figure~\ref{fig:imaging}). The cold molecular gas phase follows the dust lanes closely, whereas the neutral PAH features occupy regions marked by both dust lanes and the warm ionized gas. While the 3.3 and 7.7 $\mu$m PAH features may be used to trace star formation, they and the 11.3 $\mu$m PAH feature may also trace extended neutral and molecular gas. There are extended emission regions seen in the [O\,III] image, including an [O\,III] ridge towards the southwest, and a diffuse emitting region with an apparent shared starting point at the NE hotspot inclined at $\sim$15$^{\circ}$ with the northern lobe. Very similar features are seen in the Pa-$\alpha$ image, although more diffuse emission is detected to the south, probably because of lesser extinction (Figure~\ref{fig:imaging}).  

The ionized gas emission appears to avoid regions overlapping with the dust lanes and is more concentrated near jet-termination regions or in regions perpendicular to jet propagation.  One possibility is that the dust lanes heavily obscure any [O III] or NUV emission in those regions. Alternatively, if this corresponds to a shocked region, it would be consistent with multiphase stratification where ionized gas is on the outskirts of the molecular gas \citep[e.g., ][]{Fischer2025,Ogle2025}. The fact that the X-ray southern ridge peaks slightly to the south relative to the [O\,III] emission favors the latter interpretation.

It is also of interest that the X-ray emission extends in a direction perpendicular to the jet. While most other gas phases are roughly aligned along the jet/disk axis, weak transverse emission is also seen in the PAHs. \cite{Hardcastle2012}
find that the X-ray-emitting gas is collisionally excited. Hence, the excess X-ray emission in the transverse direction may trace the jet-heated cocoon of hot gas predicted by simulations \citep{Mukherjee2018b}. 
While the near UV image shows extended emission towards the northeast, it shows no extended emission towards the southwest, possibly because of extinction from the dust lanes. The extinction pattern suggests that the SW hotspot is inclined further away from us than the disk of the galaxy, assuming that the warm/hot gas emission is jet-related. Similarly, the NE hotspot must be inclined more towards us than the galactic disk. 
Towards the north, NUV emission seems to coincide with the diffuse NE warm gas and the photo-ionization bicone. The NUV emission could point to either the presence of young stars or shock ionization. 

\subsection{Warm H$_2$ and PAH Emission: Morphology and Excitation Ratios}
In this section, we discuss the distribution of the warm H$_2$ and PAH molecules.  
We present the moment-0 (total flux) map of H$_2$ ~0-0~S(2) with the radio flux density contours at X-band overlaid on top in Figure~\ref{fig:morphology}. We made use of the {\tt{JDAVIZ}} \citep{jdaviz2024} module to generate the moment maps. To avoid redundancy, we only present the H$_2$~S(2) moment-0 map, which strikes a good balance between resolution and sensitivity among various rotational transitions. The other H$_2$ rotational transitions show similar distribution and kinematics. 
The moment-0 map shows that the warm H$_2$ gas spatially coincides with both the warm ionized gas seen in the [O\,{\small{III}}] narrow band image and the dust lane (Figure~\ref{fig:collage}).

\subsubsection{H$_2$ S(3)-to-S(1) ratio}
Figure~\ref{fig:morphology} also shows the line ratio map of H$_2$ 0-0 S(3)/S(1). Since the lower levels are often thermalized, high S(3)/S(1) ratios suggest an overabundance of higher-energy H$_2$ molecules relative to lower-energy states, implying higher fractions of hot gas in these regions. Such enhancements are observed near both hotspots, even though the NE hotspot shows a higher ratio compared to the SW hotspot. The global S(3)/S(1) in 3C\,305 has a median value of 0.44, which is consistent with what is usually found in nearby star-forming galaxies (median $\sim$0.6; 16--84\% range $\sim$0.4--1.3;
\citealt{Roussel2007}). On the other hand, the NE hotspot shows median values of $\sim$0.8 and peak values reaching
$\sim$1.8, whereas the SW hotspot shows somewhat modest enhancements (median $\sim$0.5; maximum $\sim$0.9). Shock model grids predict S(3)/S(1) ratios of the order of unity or higher for moderate-to-fast shocks \citep{Kristensen2023,Lopex2025}.
Slightly elevated ratios are also seen in the direction perpendicular to the jet and in the core. 

Elevated temperatures can result from jet-driven shocks and turbulence that heat the H$_2$ gas and, consequently, excite it locally. 3C\,305, classified as a molecular hydrogen emission galaxy \citep[MOHEG; ][]{Guillard2012a,Ogle2010} exhibits unusually bright mid IR H$_2$ lines. \cite{Ogle2010} find that the H$_2$ emission in MOHEGs far exceeds the expected amount from UV photo-dissociation regions and X-ray heating.
The elevated S(3)/S(1) ratio at the location of the hotspots further supports this scenario, where shocks that are possibly jet-driven are responsible for the H$_2$ excitation in these radio galaxies.

\subsubsection{PAH distribution}

PAH emission at 11.3~$\mu$m, mainly tracing dusty and neutral molecular gas, is also observed across the galaxy disk. On the other hand, the ionized PAH emission at 7.7~$\mu$m is suppressed in comparison. 

The H$_2$ S(3) to PAH ratio is another good diagnostic for identifying shocked regions. 
The H$_2$ S(3) emission line is excited either by jet-driven shocks or by UV/X-ray photoionization \citep{Ogle2010}. On the other hand, PAH emission traces dusty molecular gas and can often be destroyed by shocks or intense radiation \citep{Voit1992}. 

Figure~\ref{fig:morphology} shows the spatial distribution of the ratio between H$_2$ S(3) to PAH~11.3$\mu$m in relation to the radio jet. It is interesting to note that the jet-termination region shows higher H$_2$ S(3)-to-PAH ratios, whereas the inner regions do not. 
The higher values of H$_2$ S(3)-to-PAH ratio coincide with higher values of H$_2$ S(3)-to-H$_2$ S(1) emission. Hence, from our maps, the regions with elevated H$_2$ S(3)-to-PAH imply higher levels of H$_2$ S(3) and do not necessarily imply PAH molecules are being destroyed.

\subsection{Warm H$_2$ kinematics}
\label{sec:kinematics}
The line profile of the warm H$_2$ component is complex and does not resemble a simple rotating-disk model. The spaxel-by-spaxel line profiles often show double-humped features, and the moment-2 maps show high velocity dispersion values (median$\sim$150~km~s$^{-1}$, see Figure~\ref{fig:gaussianfits-spaxel}), pointing to the presence of additional components other than the molecular gas disk rotation.

\subsubsection{Disk rotation model}

To identify deviations from the bulk disk rotation of the galaxy, we generate a simple rotating disk model assuming an arctangent parameterization. We use the arctan parameterization because it is commonly adopted in the literature for simple rotation models that reproduce the inner linear rise and eventual flattening at outer radii. We note that the goal is not to reconstruct a perfect disk model, but rather to construct a representative model against which excess velocity components can be identified.

To constrain the characteristic rotation amplitude, we compare the warm H$_2$ kinematics with published CO and optical emission-line data. We overlay the CO(1-0) position-velocity (PV) diagram (see more details in Section~\ref{sec:PV}) derived from the CO cube of \cite{Morganti2023}, represented by green dashed contours, over the H$_2$ S(1) PV diagram in Figure~\ref{fig:pv_comparison}. 
However, we find that the CO moment-1 field kinematics also appear to be influenced by non-circular motions and possible contamination from outflows, as suggested by asymmetries in the PV diagram. Additionally, we compared our adopted rotation curve model to the [O\,III] velocity curve, deprojected along the major axis, presented in \cite{Reynaldi2013}. At larger radii, around 10--15$^{\prime\prime}$, the [O\,III] velocities were in the range of $\pm 150$--$200~\mathrm{km~s^{-1}}$.

Based on these comparisons, we choose an arctan model with a conservative upper limit on the asymptotic velocity, $v_{\mathrm{flat}} = \pm240~\mathrm{km~s^{-1}}$ and major axis along PA=80$^{\circ}$, shown in Figures~\ref{fig:pv_comparison} and \ref{fig:H2S1_residuals_diskmodel}, to represent the disk rotation. We choose this value to prevent the misclassification of rotating material as outflowing. We also note that the model only reaches its asymptotic value at radii larger than the JWST field of view.

\subsubsection{Position-velocity diagram}
\label{sec:PV}
We generated position–velocity (PV) diagrams to assess the complex kinematics of the warm molecular H$_2$ in 3C\,305. The PV diagrams were extracted from a continuum-subtracted spectral slab sampled along two axes with a finite width ($0.6^{\prime\prime}$): the major axis of the host galaxy at ${\rm PA} \approx 80.5^\circ$ and the radio jet axis at ${\rm PA} \approx 42^\circ$ (see Figure~\ref{fig:pv_comparison}). The extraction apertures are overlaid on the H$_2$ S(2) moment-0 map in Figure~\ref{fig:morphology}.

In addition to the warm H$_2$, we also extracted PV diagrams for the [Ne III] emission to compare the kinematics of the ionized gas with those of the warm H$_2$. Along the major axis, the warm H$_2$ gas roughly follows disk rotation, with several indications of outflowing features at various offsets, particularly in the NE direction. In contrast, the PV diagram extracted along the radio jet axis shows enhanced velocities at the SW hotspot and slightly beyond the NE hotspot. The ionized gas PV diagram exhibits more extreme kinematics than the warm H$_2$ gas. These velocity enhancements cannot be explained by disk rotation alone and indicate the presence of multiple kinematic components.
To disentangle these components, we model the H$_2$ line profiles using multiple Gaussian components, as described below.


\subsubsection{Multi-component analysis}
To study the distribution of the various components, we carry out a double-Gaussian fitting of all the rotational transitions of the H$_2$ emission line profiles on a spaxel-by-spaxel basis in {\tt{PYTHON}} using {\tt{ASTROPY}} and {\tt{SPECUTILS}} modules. Based on the redshift, we identify the center of the emission line and an appropriate spectral region for emission line fitting. In addition, on either side of the selected spectral region, we select two regions for continuum subtraction. We fit a linear local continuum and then subtract it from the spectrum before carrying out Gaussian fitting. We then carried out a single, double, and triple Gaussian fit to the spectrum. We set initial parameters and boundary constraints on the models to ensure physically meaningful fits.
The output parameters of the fitting routine are saved as FITS images.

\begin{figure}
    \centering
    \includegraphics[width=\columnwidth]{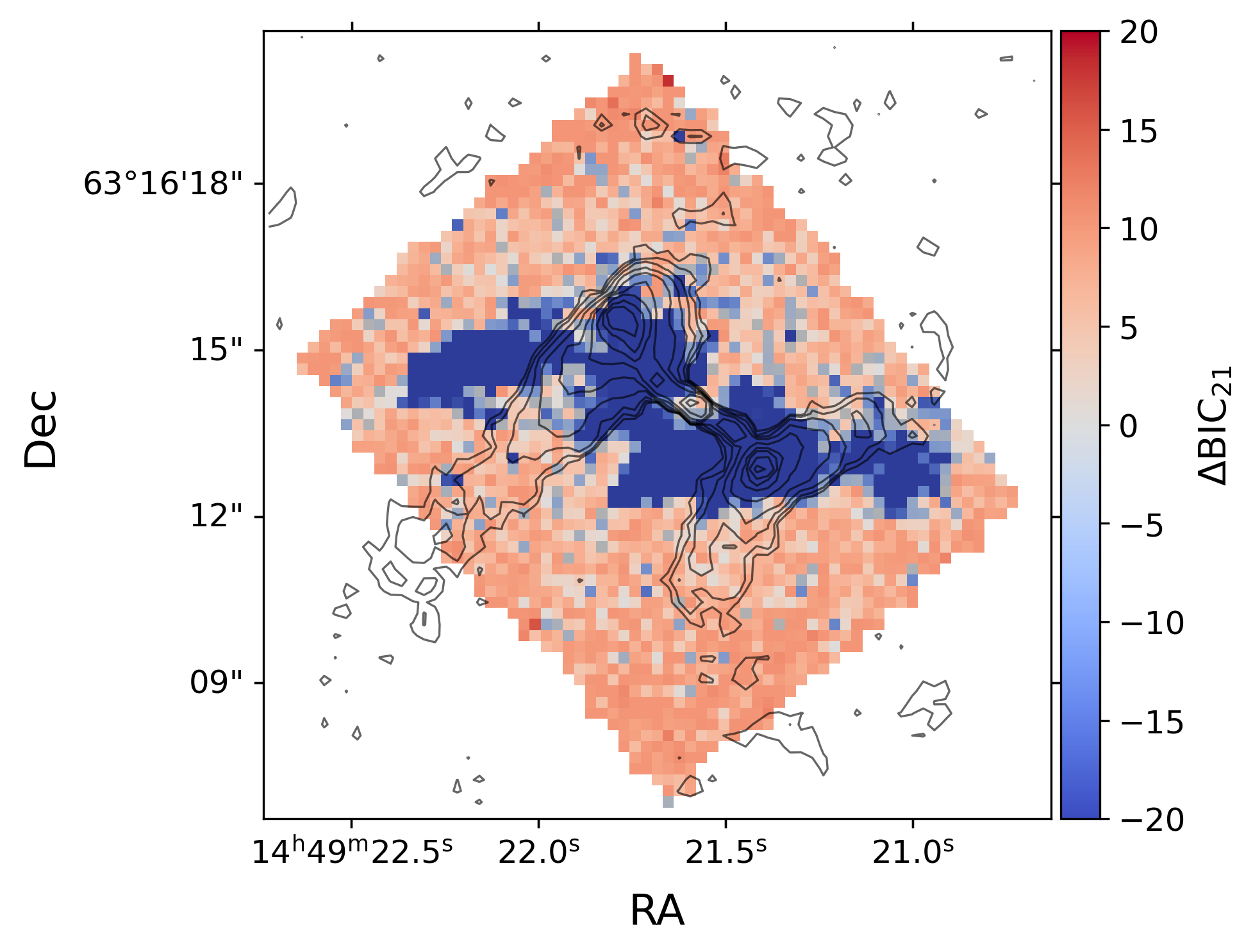}
    \caption{
    Map of $\Delta\mathrm{BIC}_{21} \equiv \mathrm{BIC}_2 - \mathrm{BIC}_1$ for the H$_2$ S(1) emission line. Negative values indicate a statistical preference for the two-component model. Radio continuum contours are overlaid. Shaded regions mark spaxels where the two-component model is adopted ($\Delta\mathrm{BIC}_{21} < 0$).
    }
    \label{fig:bic_map}
\end{figure}

To determine whether a two-Gaussian fit is required to reproduce the H$_2$ kinematics in 3C\,305, we estimate the Bayesian Information Criterion (BIC) on a spaxel-by-spaxel basis. The BIC is defined as
\begin{equation}
\mathrm{BIC} = \chi^2 + k \ln(n),
\end{equation}

where $\chi^2$ is the chi-square statistic, $k$ is the number of free parameters, and $n$ is the number of spectral channels used in the fit \citep{Chu2022}. This formula assumes Gaussian error bars and is equivalent to the original likelihood expression with the exception of a constant which cancels out during comparison between two models.
$\Delta\mathrm{BIC}_{21}$ is defined as $ \equiv \mathrm{BIC}_2 - \mathrm{BIC}_1$, where negative values imply that the two Gaussian component model provides the better fit. In Figure~\ref{fig:bic_map}, we present the $\Delta\mathrm{BIC}_{21}$ map for 3C305. We find that for most spaxels, the double Gaussian fitting yields the best statistics. 
In this paper, we limit the double-Gaussian fitting and our outflow analysis for spaxels with $\Delta\mathrm{BIC}_{21}<0$ and have high enough SNR.

In Figure~\ref{fig:gaussianfits-spaxel}, we present the spectra and the Gaussian fits to four representative spaxels, which roughly coincide with the radio jet in 3C\,305. Both the broad and narrow components in the spectra of both spaxels towards the SW are centered around the same central velocity. However, towards the NE, the narrow component appears to be more offset from the rest-frame velocity of the galaxy in comparison to the broad component by $\sim$70~km~s$^{-1}$ on average. 
Figure~\ref {fig:double_gaussian_mom0} presents a comparison of the amplitudes (peak line surface brightness), velocities, and the velocity dispersion, respectively, of the narrow and broad components for the H$_2$ S(1) emission line.  From a comparison of the amplitude (peak line surface brightness) images of the narrow and the broad component, the narrow component appears to be prominent across the length of the source in the direction of the radio jet except at the SW jet termination region, where the broad component is noticeably more dominant. 

There are several details that make the H$_2$ kinematics peculiar, especially in comparison with the radio jet/lobe structures. First, the narrow component velocities in a region just beyond the NE hotspot appear enhanced (see Figure~\ref{fig:double_gaussian_mom0}). The broad component, on the other hand, seems to follow a smooth velocity gradient. 
A similar enhancement is visible beyond the SW hotspot, albeit to a smaller degree, and it is gradual rather than abrupt, unlike the northern feature. 

The higher velocities in the narrow component relative to the broad component were surprising, as the broad component is usually the outflowing component. Throughout this paper, we refer to the components as narrow and broad based on their velocity dispersions. Hence, while the narrow component can have larger velocity offsets from the rotating disk, implying larger systematic bulk velocities, the broad component might have larger widths due to other factors such as turbulence. 

To further explore the bulk velocities and the nature of the narrow and broad components, we subtracted them from a simple rotating-disk model. Figure~\ref{fig:H2S1_residuals_diskmodel}
illustrates a simple arctan disk model and the residuals of the narrow and the broad components. 
The middle and the right panels show the residual velocities after subtracting the disk model from the observed narrow and broad components, respectively. We use the same color scale limits to enable direct comparison. After disk subtraction, the narrow component residual image shows an extended region just beyond the radio hotspot with coherent, high-velocity values that reach $|v_{\rm resid}|\sim$ a few $\times 10^2\ \mathrm{km~s^{-1}}$. The morphology, coherence, and spatial overlap with several shock tracers all suggest that the velocity enhancement represents true bulk velocity, indicating bulk outflow. On the other hand, the broad component shows much lower residuals and is more consistent with overall disk rotation. Hence, we interpret the narrow component as tracing the outflowing component, in locations where two Gaussian components are statistically required to explain the kinematics.

The maximum velocities of the narrow-component reach up to 400~km~s$^{-1}$.
Such high velocities ($\sim$400~km~s$^{-1}$) were also reported in optical emission line studies \citep{Reynaldi2013}.
The enhanced velocity regions beyond the hotspots preferentially lie alongside the [Fe\,II] emission (see Figure~\ref{fig:FEII}), which is a shock tracer (see Section~\ref{sec:feII}). 
Several gas phases (gaseous systems) coexist in these enhanced velocity regions: from the hot X-ray-emitting phase to warm H$_2$; the only exception being the cold gas, as accounted for by CO (Figure~\ref{fig:collage}). 

Our warm H$_2$ PV diagram resembles that of the CO(1--0) presented by \citet{Morganti2023}, pointing to a major contribution from a rotation-dominated disk. 

From optical emission-line studies \citep{Reynaldi2013}, the disk rotation plateaus at 150-200~km~s$^{-1}$ at larger radii beyond 5$^{\prime\prime}$. While the optical spectra also show localized velocity enhancements near the hotspots, velocities at larger radii are likely representative of disk rotation. Optical emission line studies by \cite{Heckman1982} also conclude that the kinematics are complex and are best explained by a hybrid model with rotation plus outflows. They assume that the solid body rotation ends at 2-3~kpc and the rotation velocity at this location is 260(sin $i$)$^{-1}$~km~s$^{-1}$. The high H$_2$ velocities correlated with the radio hotspots indicate the presence of extensive molecular outflows that are most likely jet-driven.

The velocity dispersion maps reveal other interesting aspects. The velocity dispersion appears broader perpendicular to the jet, as seen in the broad component image. The transverse enhancement of the velocity dispersion is mainly restricted to the inner section of the radio jet, and is not clearly associated with the radio hotspots. Simulations show that when a jet expands at a low inclination angle to the disk of the galaxy, it often blows up a jet-driven bubble with significantly higher velocity dispersion \citep{Mukherjee2018}. Several previous studies have reported similar velocity-dispersion enhancements perpendicular to the jet \citep{Couto2013,Schnorr2014,Schnorr2016,Lena2015,Riffel2023,Feruglio2010,Ali2025}. \cite{Riffel2026} find that such enhancements in the velocity dispersion of molecular and ionized gas perpendicular to the jet are commonplace, based on a sample study of radio-loud galaxies using JWST MIRI data.

The SW hotspot shows noticeably lower velocity dispersion values compared to the neighboring pixels. There is no outflow at the location of the SW hotspot, though there are considerable velocity enhancements to the west of that hotspot. 



\begin{figure*}
    \centering
    \includegraphics[width=\linewidth]{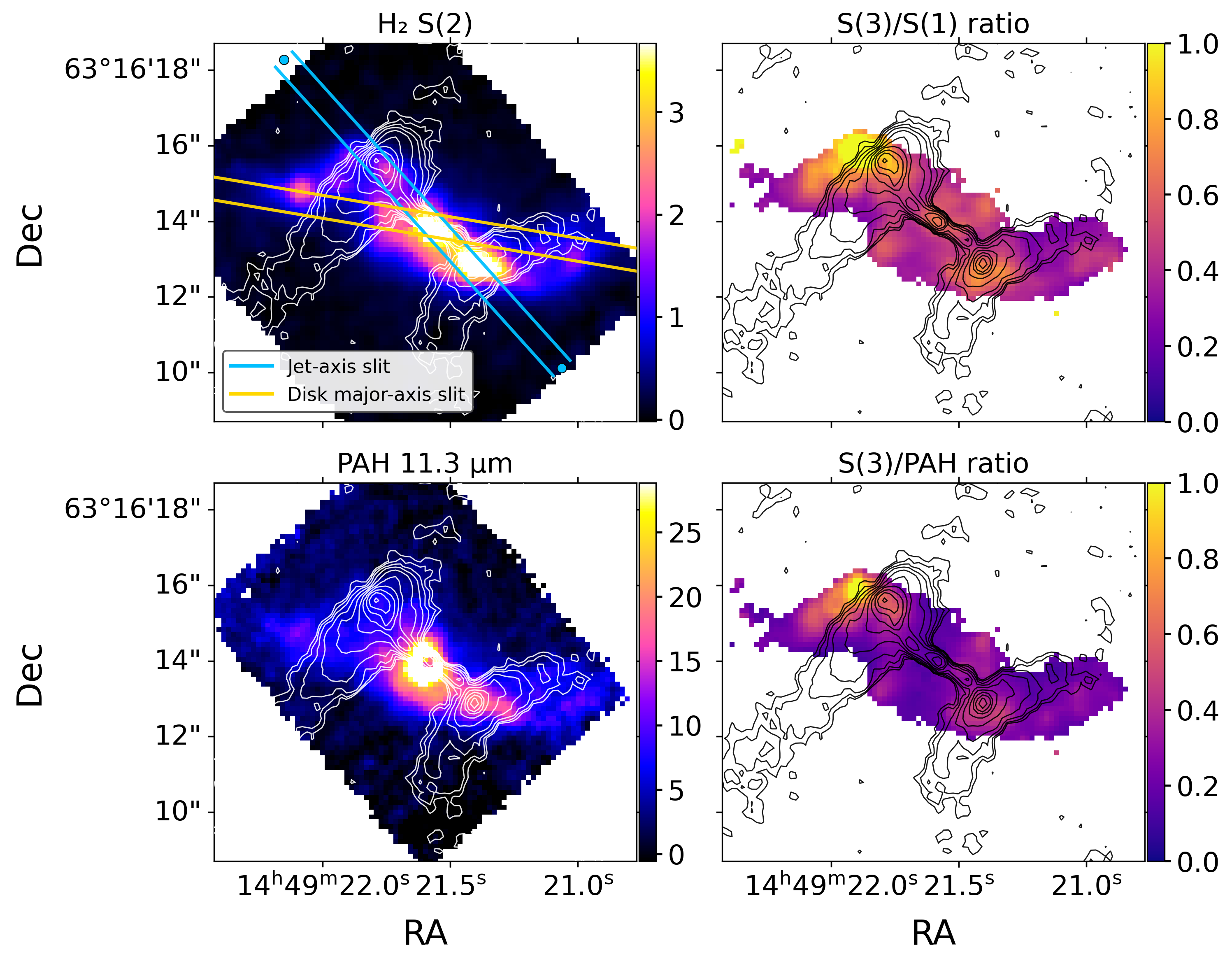}
    \caption{Top left: H$_2$ S(2) moment 0 map showing spatial distribution of warm molecular gas (color bar in MJy~$\mu$m$^{-1}$~sr$^{-1}$). Top right: H$_2$ S(3)/H$_2$ S(1) is a tracer of H$_2$ temperature. The region beyond the NE  radio hotspot exhibits the highest temperatures, whereas the AGN core and NE and SW radio hotspots also show relatively high temperatures, consistent with shock heating by the radio jet. There is tentative evidence for elevated temperature in the direction transverse to the radio jet. Bottom left: The PAH 11.3 $\mu$m map (color bar in MJy~$\mu$m$^{-1}$~sr$^{-1}$)
 has a similar overall spatial distribution to H$_2$, showing it is associated with the dusty molecular disk. Emission peaks near the center, where the ambient UV field from the bulge is most intense. The apparent central divot is an instrumental artifact. Bottom right: The H$_2$ S(3)/PAH 11.3  shock indicator, similar to the H$_2$ S(3)/S(1) ratio, peaks just outside of the radio hotspots.}
    \label{fig:morphology}
\end{figure*}

\begin{figure*}
    \centering
    \includegraphics[width=0.48\linewidth]{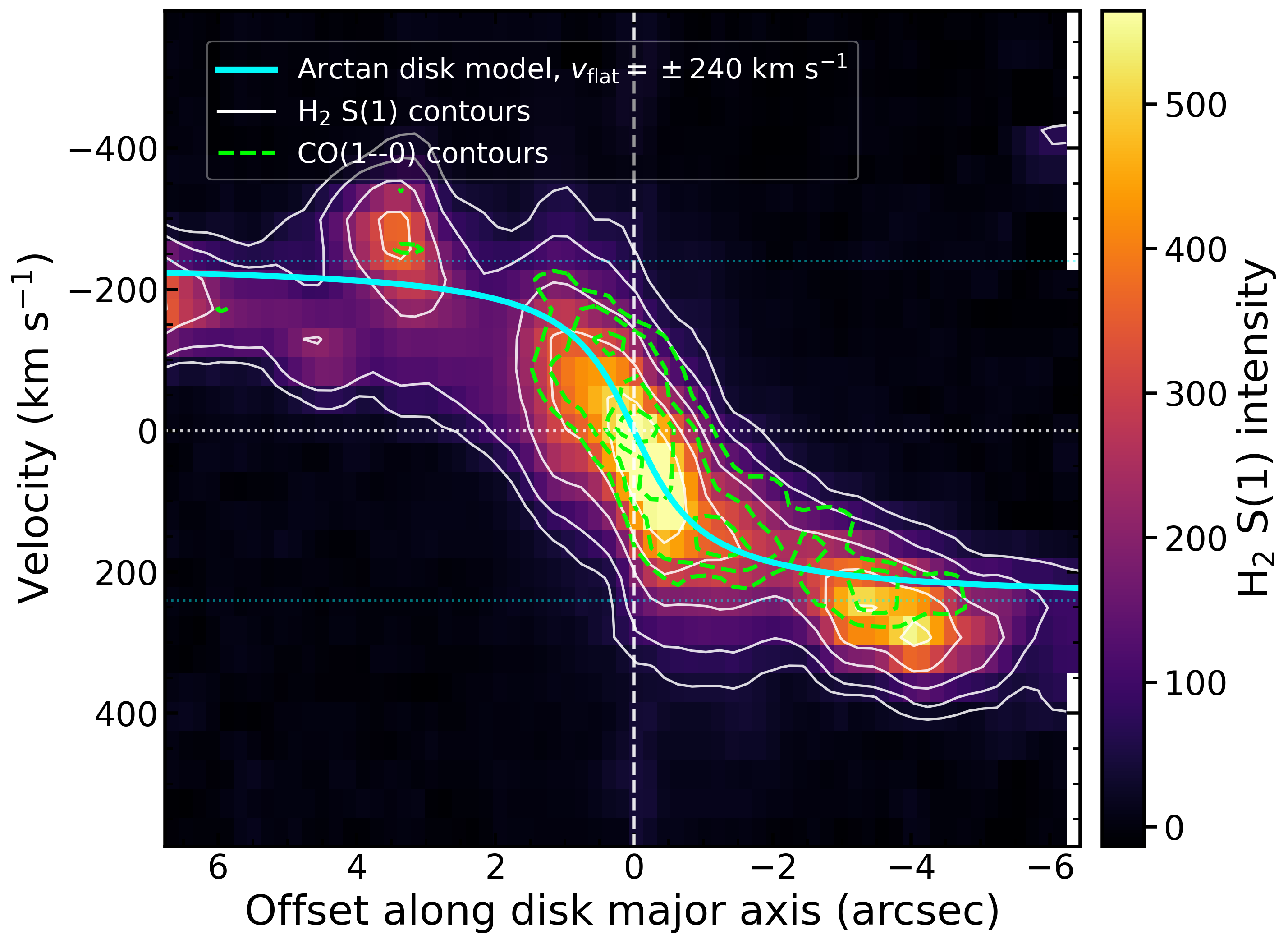}
    \includegraphics[width=0.48\linewidth]{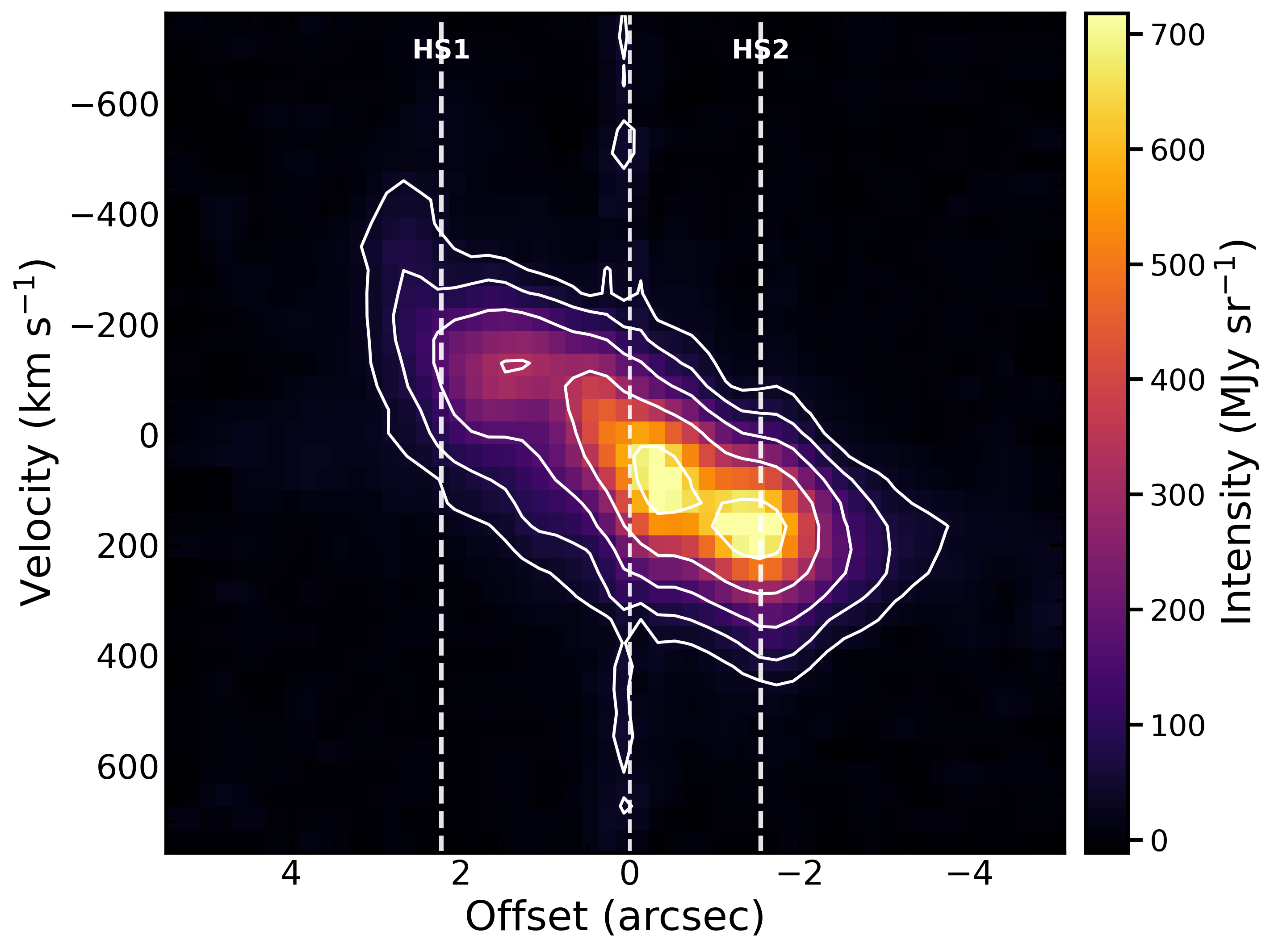}
    \includegraphics[width=0.48\linewidth]{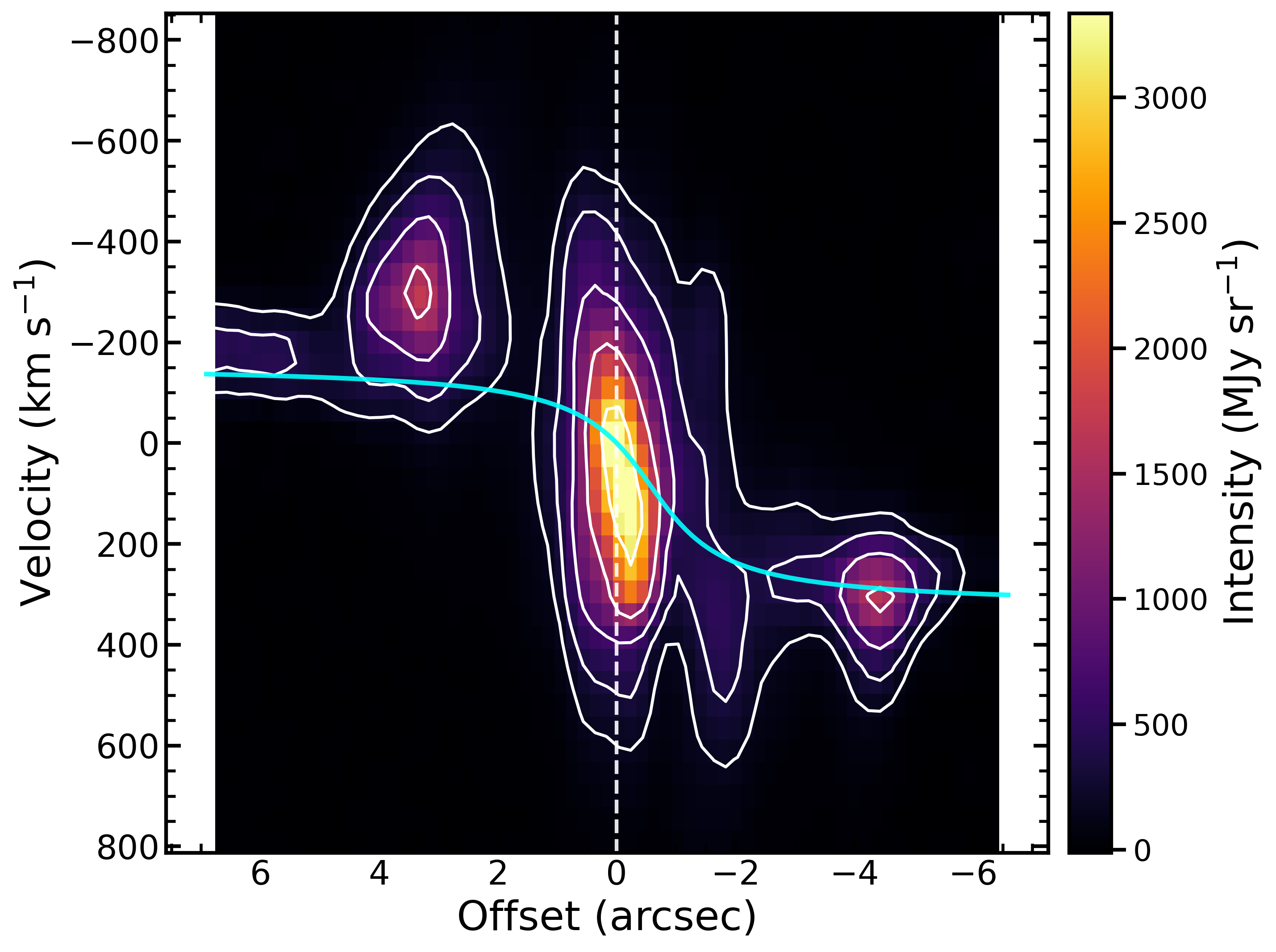}
    \includegraphics[width=0.48\linewidth]{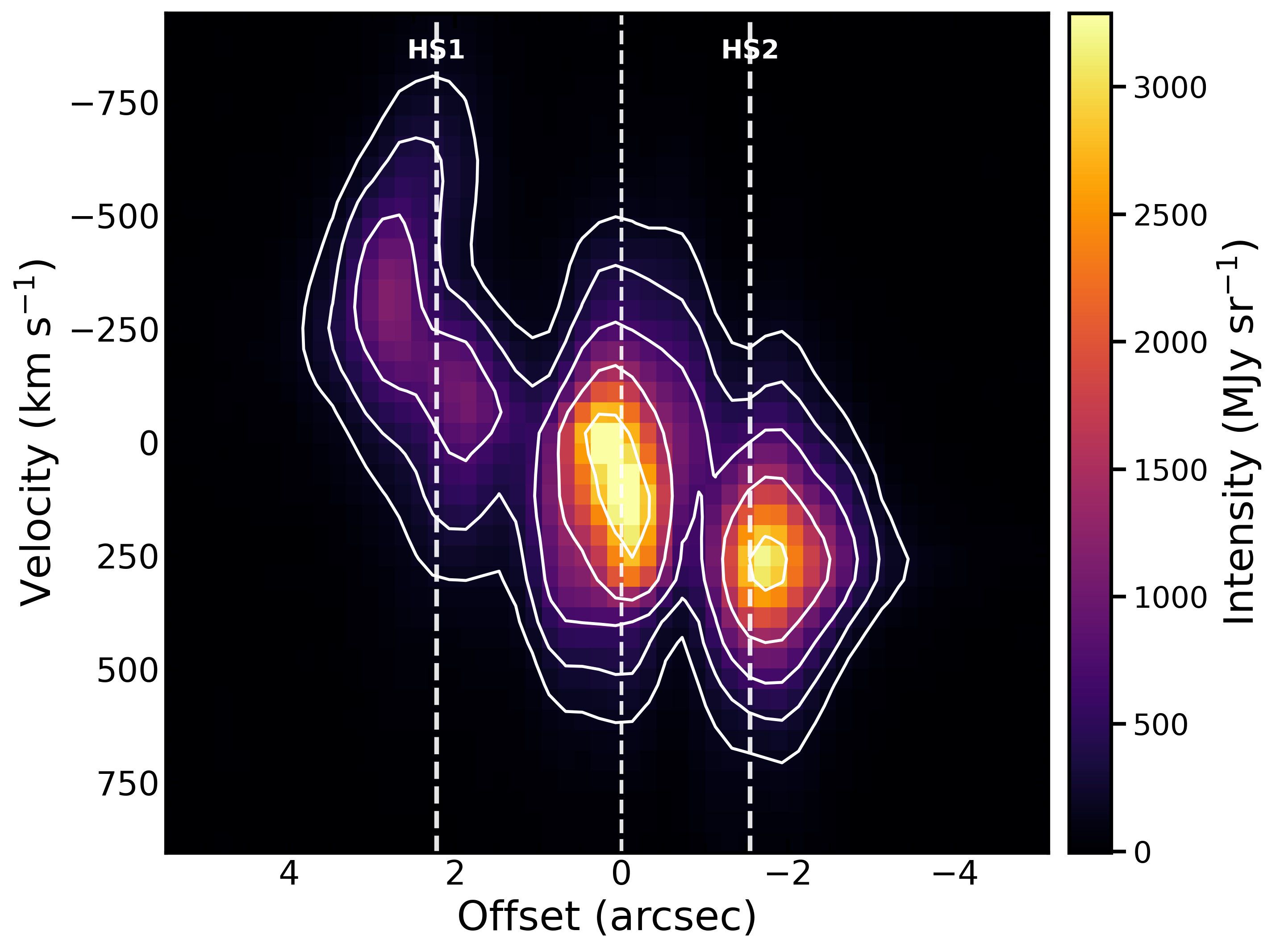}
    \caption{
    Position--velocity diagrams for molecular and ionized gas in 3C\,305.
    Top row: H$_2$ 0--0 S(1).
    Bottom row: [Ne\,{\sc iii}].
    Left column: Extraction line aligned with the galaxy major axis. A disk rotation curve, assuming an arctan model, is overlaid on the PV diagram.
    Right column: Extraction line aligned with the radio jet axis. The locations of the hotspots are marked as HS1 and HS2. The CO(1-0) PV diagram derived from the CO cube \citep{Morganti2023} is shown as the green dashed contours in the top-left panel.
    }
    \label{fig:pv_comparison}
\end{figure*}

\begin{figure*}
    \centering
    \includegraphics[width=\linewidth]{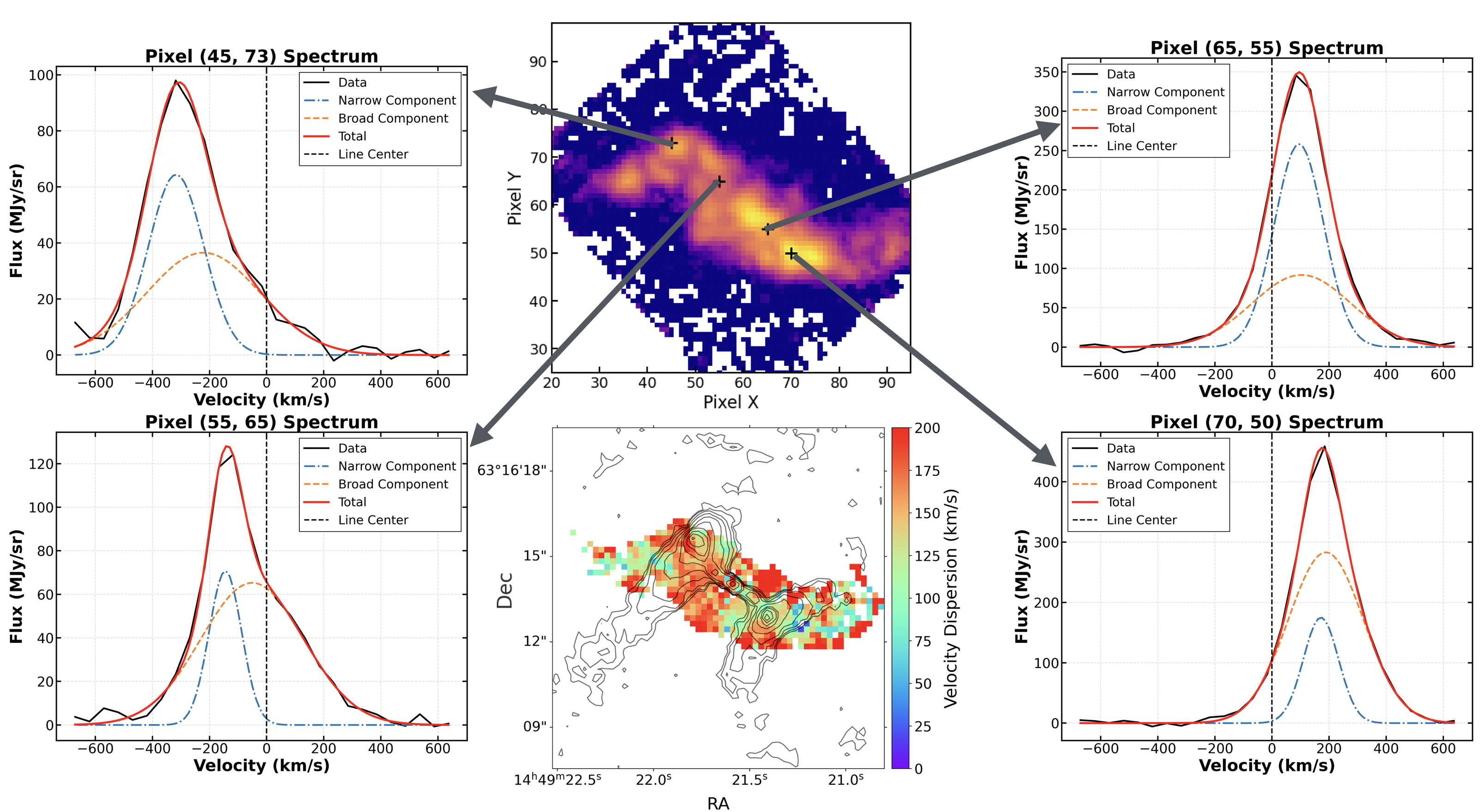}
    \caption{The figure shows the double-Gaussian fitting of H$_2$ S(2) emission lines at the marked spaxel locations. The arrows starting from the central moment-0 image show the location of the spaxels from where the corresponding spectra were extracted. Most spaxels require at least two Gaussian components, a broad and a narrow component to satisfactorily reproduce the spectral shape. More interestingly, the narrow emission lines in the north-eastern region seem to be systematically outflowing, suggesting potential jet-driven feedback. The bottom middle panel shows the moment-2 map of the  H$_2$ S(2) emission line. It displays larger-than-expected velocity dispersion values, especially in the central region, with a median value of $\sim 150\ \mathrm{km\ s^{-1}}$. This motivated the use of multi-component Gaussian analysis.}
    \label{fig:gaussianfits-spaxel}
\end{figure*}


\begin{figure*}[h]
    \centering
    \includegraphics[width=\linewidth]{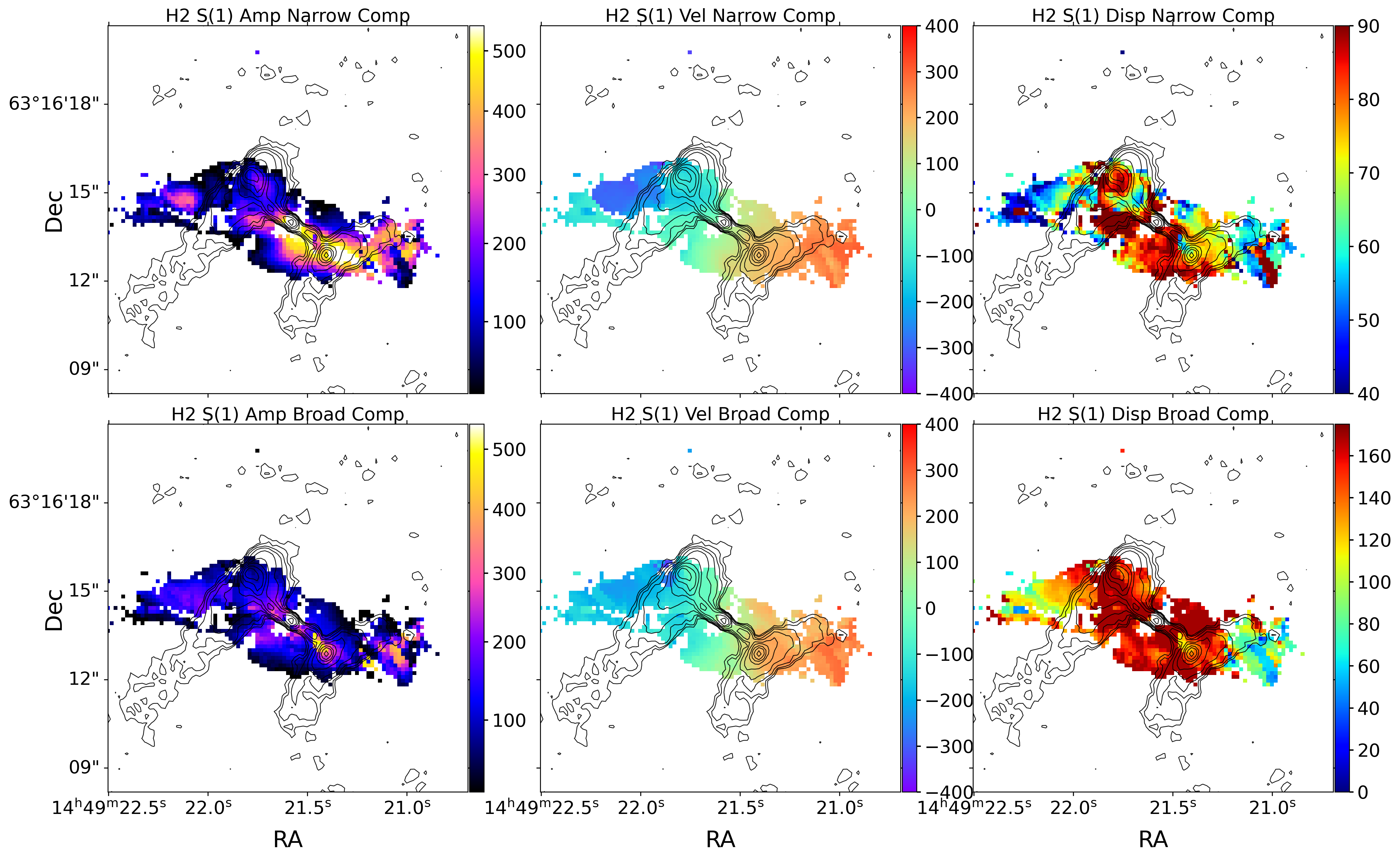}
    \caption{Best fit parameter maps from the two-component Gaussian decomposition of the H$_2$ S(1) emission line. Left panels: Peak line surface brightness maps for narrow (top) and broad (bottom) components, with color bars in units of MJy~sr$^{-1}$. Middle panels: line-of-sight velocity maps for narrow (top) and broad (bottom) components, with color bars in units of km~s$^{-1}$. Right panels: velocity dispersion maps for narrow (top) and broad (bottom) components, with color bars in units of km~s$^{-1}$. We note that the velocity dispersion maps displayed here are not corrected for instrumental resolution. The JWST MIRI MRS instrumental resolution ranges from 35 to 100 km~s$^ {-1}$, depending on the wavelength \citep{Argyriou2023}.}  Overlaid contours show the radio continuum emission at X-band.
    \label{fig:double_gaussian_mom0}
\end{figure*}
\begin{figure*}
    \centering
    \includegraphics[width=\linewidth]{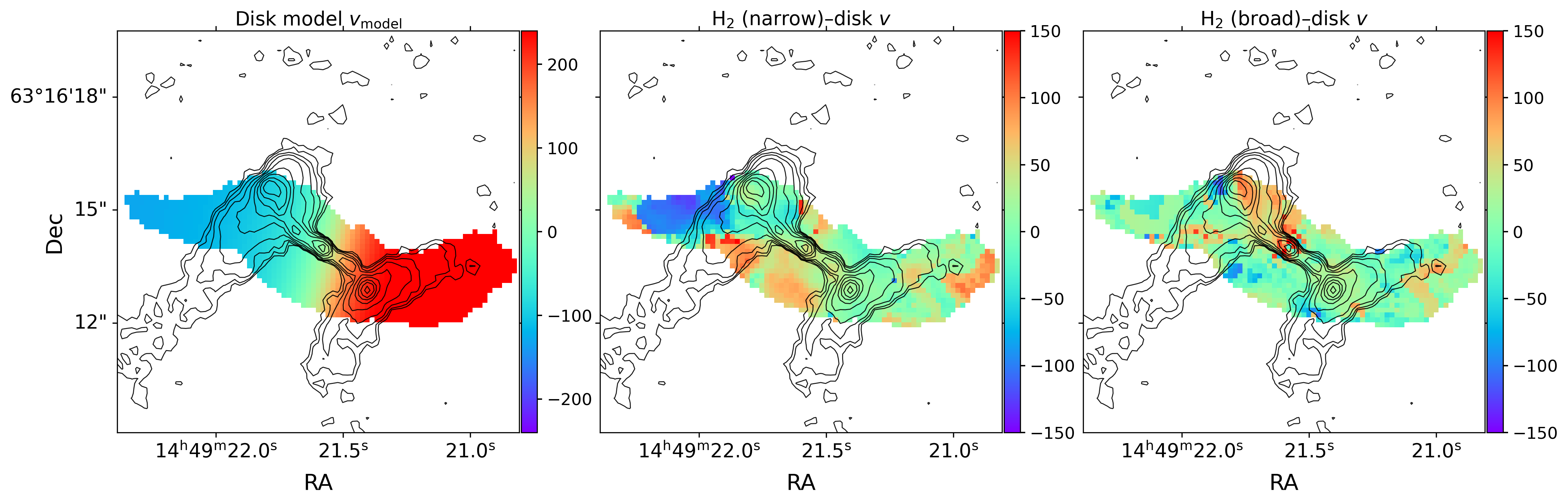}
    \caption{
    Left: Best-fitting rotating disk velocity field ($v_{\rm model}$) derived assuming the flat part of the rotation curve has a velocity of $\pm240~\mathrm{km\,s^{-1}}$ (also see Figure~\ref{fig:pv_comparison}), and a disk PA of 80$^{\circ}$, 
    Middle: residual velocity field obtained by subtracting the disk model from the narrow H$_2$ velocity. Right: residual velocity field obtained by subtracting the disk model from the broad H$_2$ velocity.
    The color scale limits are $\pm240~\mathrm{km\,s^{-1}}$ and 
    $\pm150~\mathrm{km\,s^{-1}}$ for the disk model  and both residual maps respectively.
    Radio continuum contours are overlaid in all panels.
    While both narrow and broad components show significant residuals, the narrow component surprisingly shows an extended region of high bulk velocities near the NE hotspot. 
    }
    \label{fig:H2S1_residuals_diskmodel}
\end{figure*}





\subsection{[Fe\,II] 5.34~$\mu$m emission}
\label{sec:feII}
One of the primary modes for jets to provide feedback into the ISM is via heating through shocks or by inducing turbulence.
[Fe\,II] emission is a well-established tracer of shock ionization in regions where jets interact with the surrounding medium. Shocks release iron from dust grains, and these atoms generate strong [Fe\,II] emission when excited by the ambient radiation field \citep{Zovaro2019} or via collisional excitation. 

\begin{figure*}
    \centering
    \includegraphics[width=\linewidth]{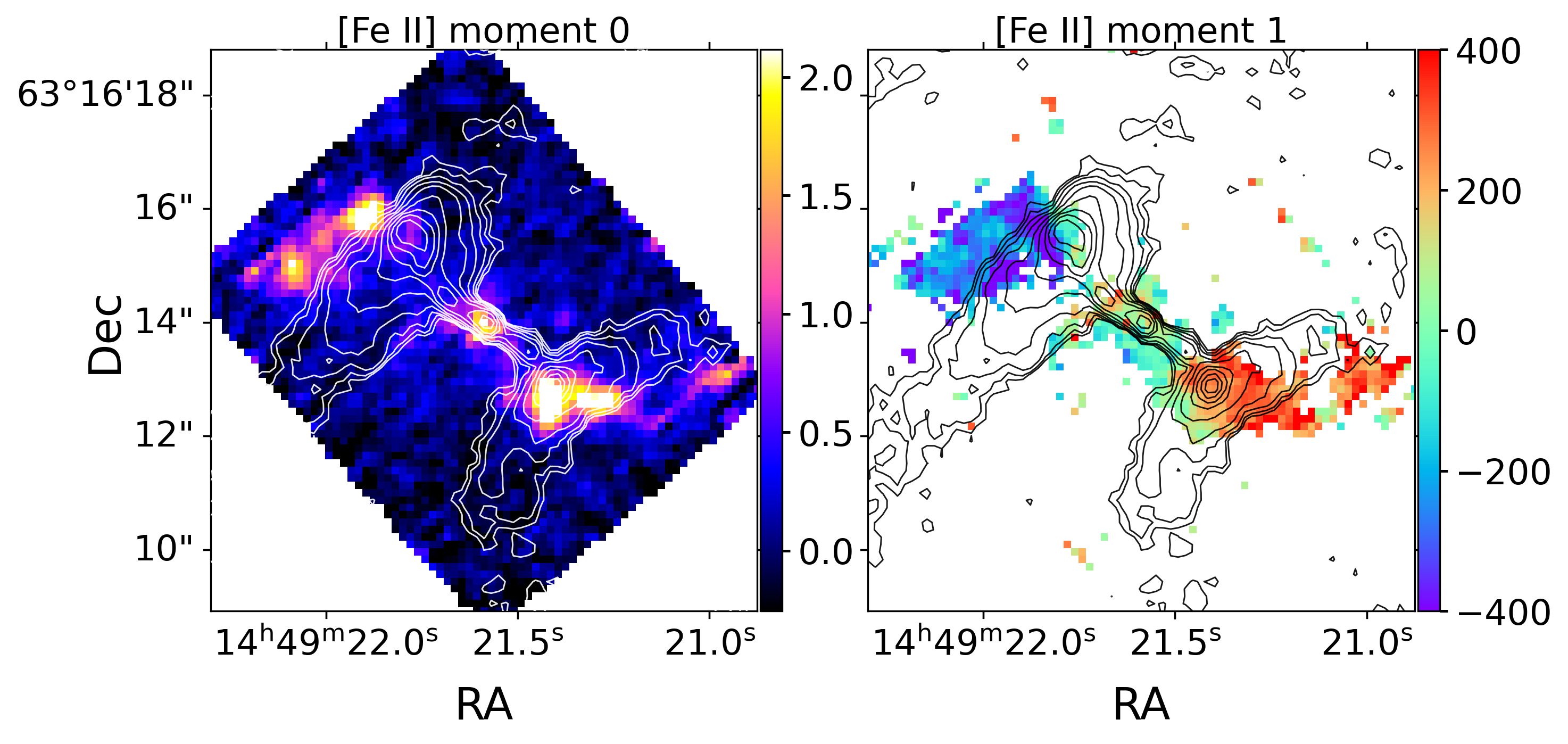}
    \caption{[Fe\,II] moment maps. Left: Integrated intensity (moment-0) map with color bar units in MJy~$\mu$m$^{-1}$~sr$^{-1}$. Right: Velocity (moment-1) map with color bar units in km~s$^{-1}$. X-band radio continuum contours are overlaid. Shocks release iron from dust grains. The gas-phase atoms emit strong [Fe\,II] lines when excited. [Fe\,II] emission peaks appear to be highly correlated with the jet terminal locations.}
    \label{fig:FEII}
\end{figure*}

Figure~\ref{fig:FEII} shows moment maps of [Fe\,II] at 5.34 $\mu$m, its spatial distribution, how it correlates with the radio jet, and an additional bright component at the core. We find [Fe\,II] emission spatially correlated with locations beyond the hotspot, especially in regions where outflows in the narrow component were identified in Section~\ref{sec:kinematics}. 
The spatial distribution of [Fe\,II] differs significantly from other ionized emission lines (see Figure~\ref{fig:collage} and \ref{fig:ionizedgas_comparison}). While ionized gas is present throughout the full jet extent, [Fe\,II] is primarily concentrated at the jet termination location or where the jet's impact onto the ISM is thought to be strongest.  
The spatial correspondence of [Fe~II] with jet hotspots thus reinforces the idea of jet-driven shocks. Previous near-infrared studies have shown similar enhancement of [Fe~II] \,1.257~$\mu$m at the northeastern hotspot \citep{Jackson2003}. However, it remains unclear why it peaks beyond the NE hotspot rather than at the hotspot itself. We note that this offset is observed across all hot/warm gas phases, although it is less pronounced in the cold molecular gas (see Figure~\ref{fig:collage}). Notably, this region also exhibits high outflow velocities and a high H$_2$ S(3)/S(1) ratio. Such an offset is not seen at the southern hotspot.

The right panel of Figure~\ref{fig:FEII} shows the Moment 1 map, illustrating the velocity distribution of [Fe\,II]. From the figure, it is clear that [Fe\,II] follows the sense of disk rotation in 3C\,305 \citep{Heckman1982,Morganti2023}, as well as the outflows observed in warm H$_2$ gas. The velocities exceed 400~km~s$^{-1}$, further suggesting that this gas is also outflowing. 


%

\subsection{H$_2$ excitation modeling and molecular mass estimates}

\label{sec:excitation_model}
\subsubsection{Methodology}
We use the \cite{Togi2016} prescription of an underlying power-law distribution of the temperature to fit the excitation diagram. 
According to their model, the warm H$_2$ gas spans a range of temperatures rather than a discrete distribution of temperatures, described as, 
\begin{equation}
\frac{dN}{dT} \propto  T^{-n}.
\label{eq:powerlaw}
\end{equation}
where N is the H$_2$ column density, and n is the power law index. 
While fitting the model to the data, we keep the upper cutoff temperature T$_u$, a constant at 2000~K, and allow the lower cutoff temperature (T$_l$) and n to vary.

For each spaxel, we first derived the line fluxes for all the transitions from H$_2$ S(1) to H$_2$ S(5). We do not consider H$_2$ S(6) to H$_2$ S(8) due to poor SNR. The H$_2$ line fluxes for the excitation analysis are measured from the data cubes (i.e., integrated line fluxes). Excitation analysis on the individual velocity components is not possible due to the poor SNR of the decomposed flux maps, especially the higher-excitation H$_2$ lines.
We then constructed the excitation diagram by computing the upper-level column density \( N_u \). We made use of the Python implementation of the \cite{Togi2016} model, as described in \cite{Jones2024} (also see the GitHub repo\footnote{
\href{https://github.com/astrolojo/H2Powerlaw}{https://github.com/astrolojo/H2Powerlaw}}). 
We then normalized \( N_u \) by the statistical weight \( g_u \) and plotted it against the upper-level energy \( E_u / k_B \), where \( k_B \) is the Boltzmann constant. 
The code fits the upper-level column densities for all transitions assuming a power-law temperature distribution model (see Equation~\ref{eq:powerlaw}), as discussed above, to determine the values of the slope \( n \) and the lower cut-off temperature \( T_{\ell} \). The value of \( \frac{N_u}{g_u} \) is normalized to that of the S(1) line while minimizing the difference between the observed and model column densities during fitting.

\begin{figure*}
    \centering

    \includegraphics[width=0.49\linewidth,trim= 0 0 0 400]{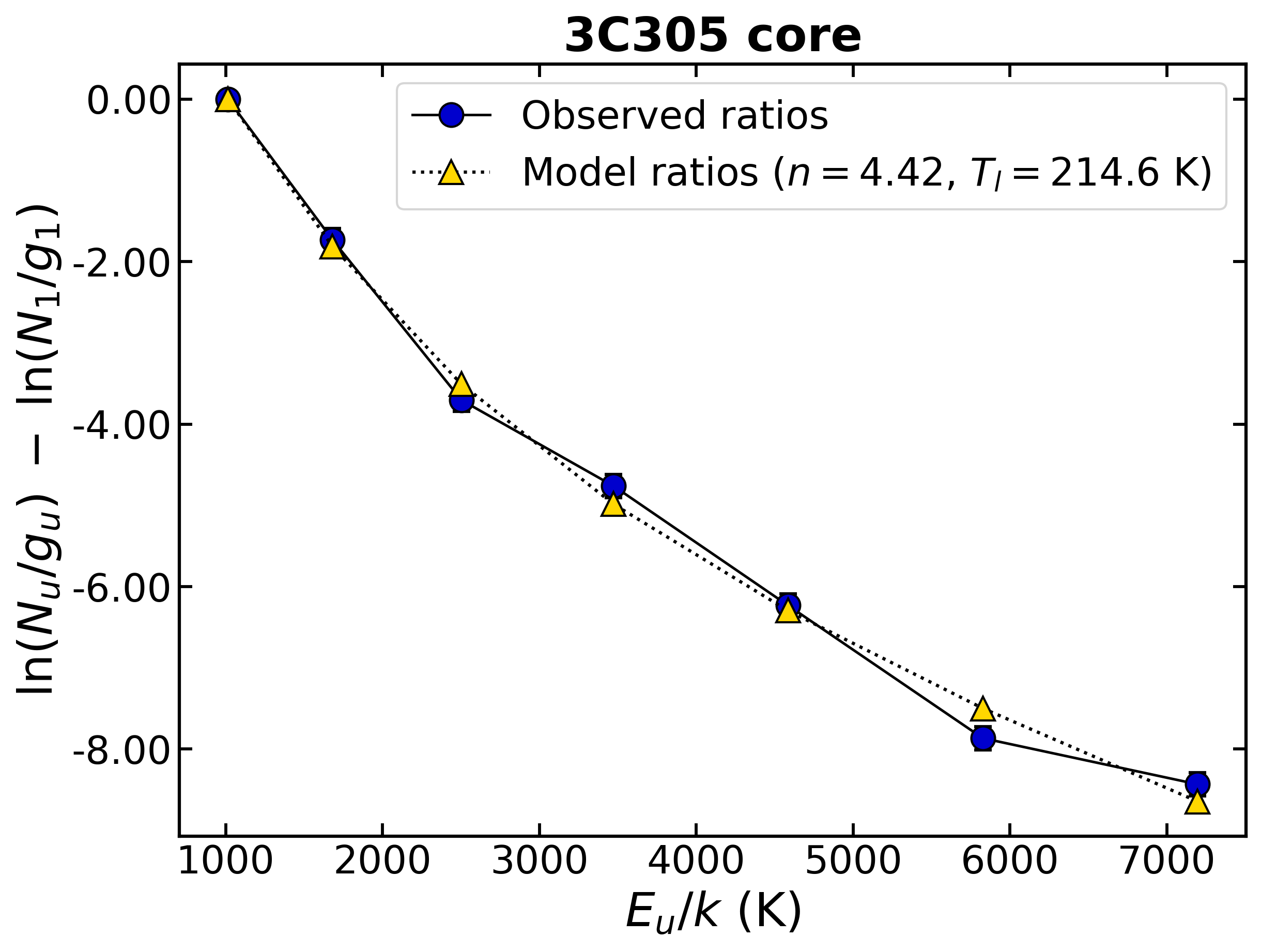}
    \includegraphics[width=0.48\linewidth]{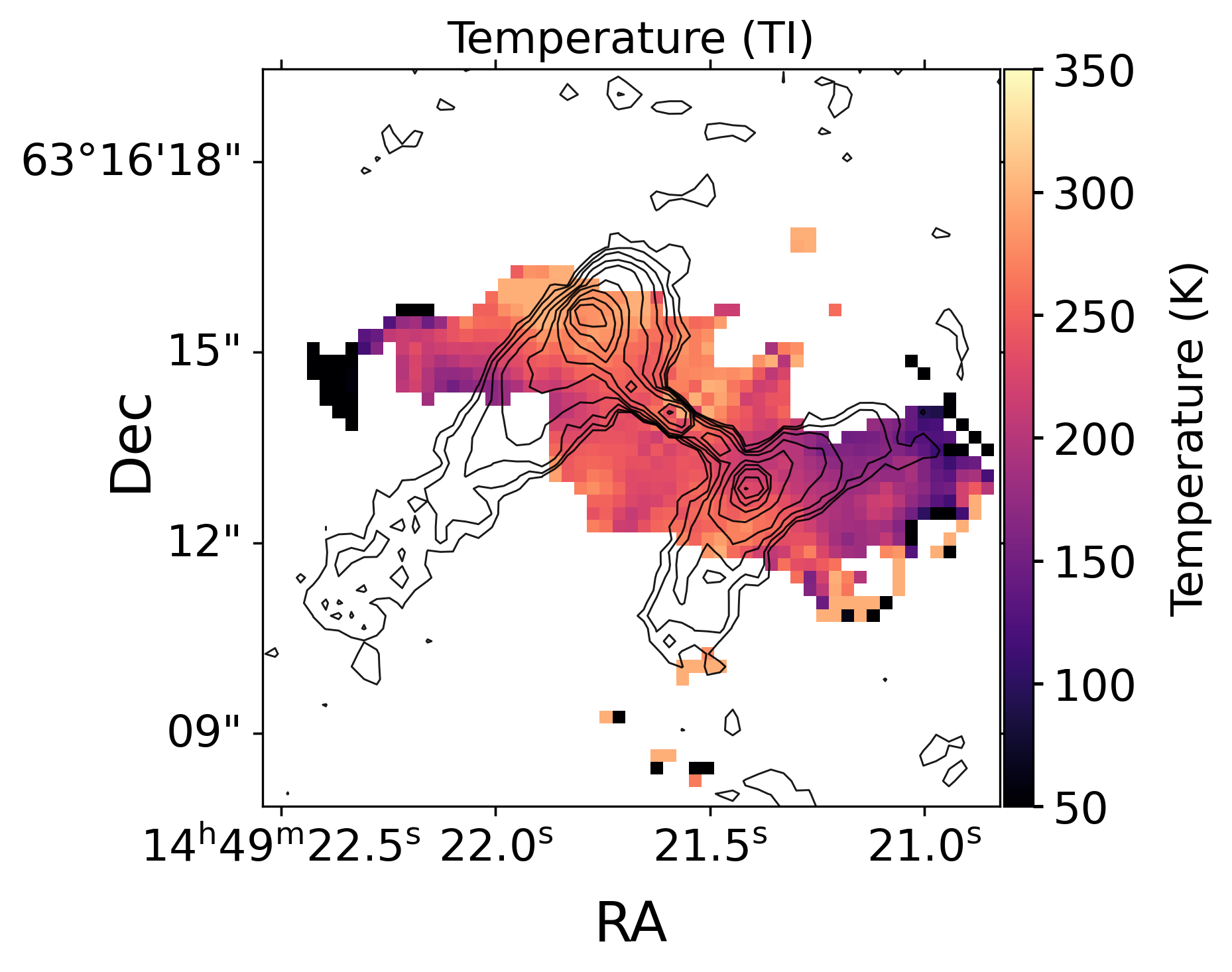}


\includegraphics[width=0.9\linewidth]{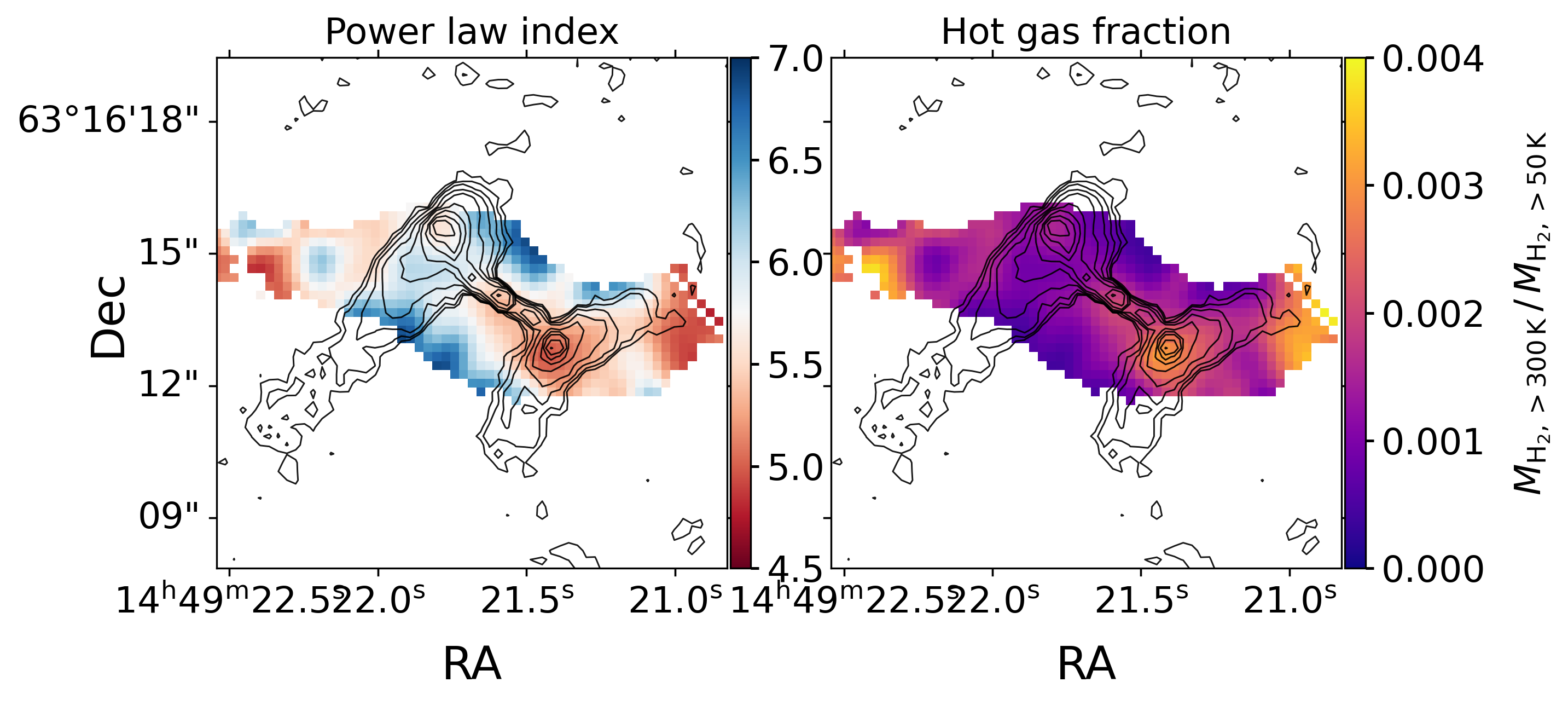}

    \caption{
        {H$_2$ excitation and temperature structure.}
        Top-left: H$_2$ excitation diagram of the core.
        Top-right: Map of the lower-bound temperature $T_\ell$.
        Bottom-left: Power-law slope $n$ of the H$_2$ temperature distribution.
        Bottom-right: Hot gas mass fraction map.
    }
    \label{fig:h2_4panel}
\end{figure*}





\subsubsection{Physical parameter maps: Slope, Temperature, and H$_2$ Mass}

We present maps of the power-law slope \( n \) and the lower bound temperature \( T_{\ell} \) (Figure~\ref{fig:h2_4panel}) along with the corresponding distributions of warm ($T >$ 50~K) and hot ($T >$ 300~K) H$_2$ mass. 
From the power-law slope \( n \) map, we find that it is skewed toward lower power-law indices along the radio jet, indicating a higher fraction of hot H$_2$ relative to cooler warm H$_2$ gas. 

There is significant spatial variation of $n$ across various regions, including the AGN core and the hotspots. It is evident from the power-law index maps that both hotspots show flatter power-law indices, indicating a higher fraction of warmer H$_2$ gas. The power-law index of the southwest hotspot is $n \approx 3.6 \pm 0.4$, indicating a very flat distribution, whereas the northeastern hotspot has $n \approx 4.4 \pm 0.7$, which is marginally flatter than the average temperature distribution. It is important to note that even though the power-law index at the northeast hotspot is steeper compared to the southwest hotspot, the lower temperature cutoff is higher at the northeast. The nucleus exhibits an index of $n =4.2$, which is intermediate between the two hotspots.
The correlation of the warmest H$_2$ temperatures with the jet hotspots indicates that jet-driven shocks are the primary excitation mechanism for warm hydrogen molecules. 

In addition to the temperature, we also estimate a warm gas mass of $2.6 \times 10^7$ M$_{\odot}$ (for $T \gtrsim 200$K), whereas CO observations reveal a gas mass of 2$\times$10$^9$ M$\odot$ \citep{Ocana2010} yielding a warm-to-cold gas fraction of $\approx$1.3$\%$. We note that systematic uncertainties, including differing cosmology or luminosity distance, CO-to-H$2$ conversion factor, and heavy-element mass corrections, can alter the warm-to-cold gas fraction.





\subsection{Comparison to Ionized Gas}
\label{sec:ionizedgas_results}
\begin{figure*}
    \includegraphics[width=\textwidth]{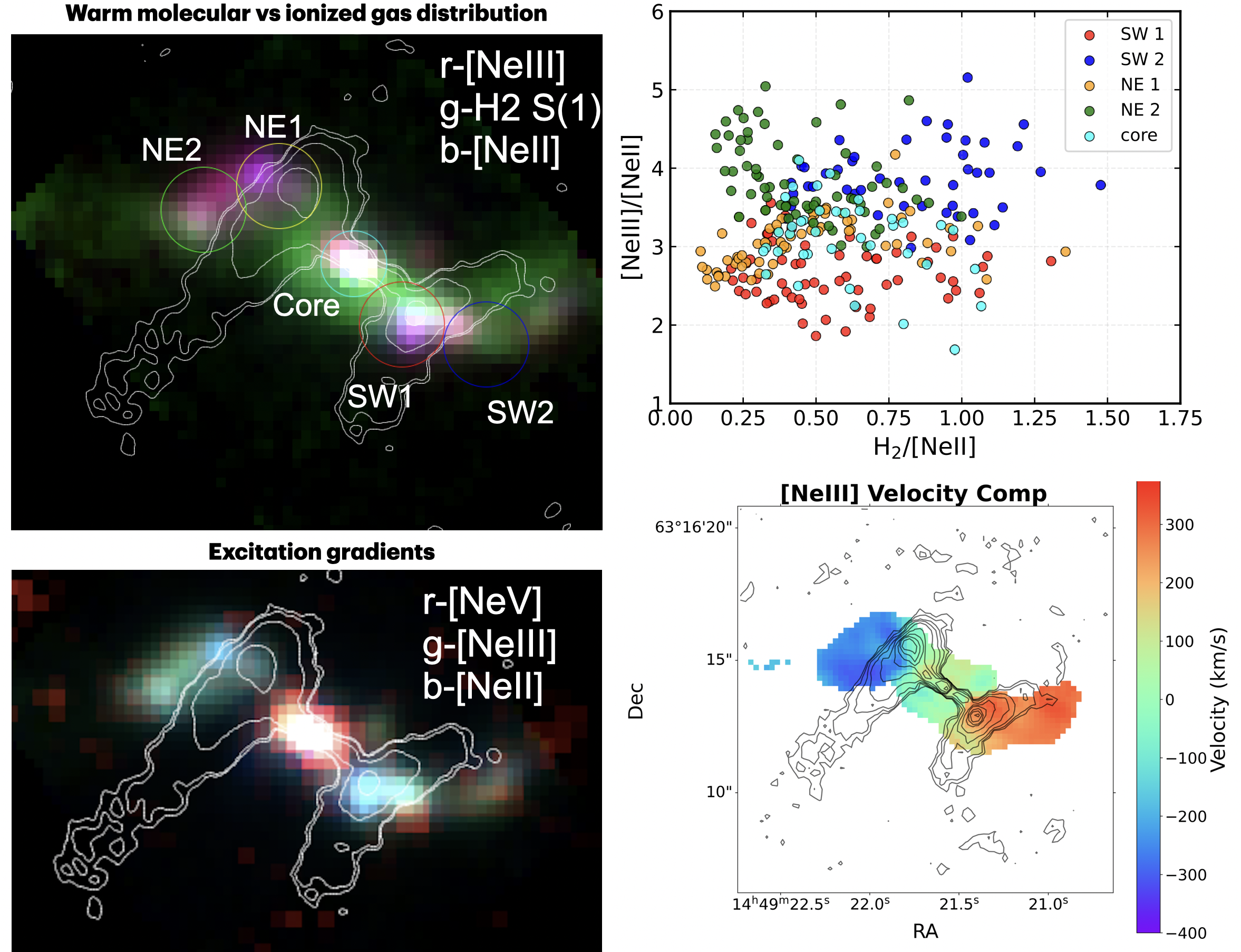}
\includegraphics[height=0.38\linewidth]{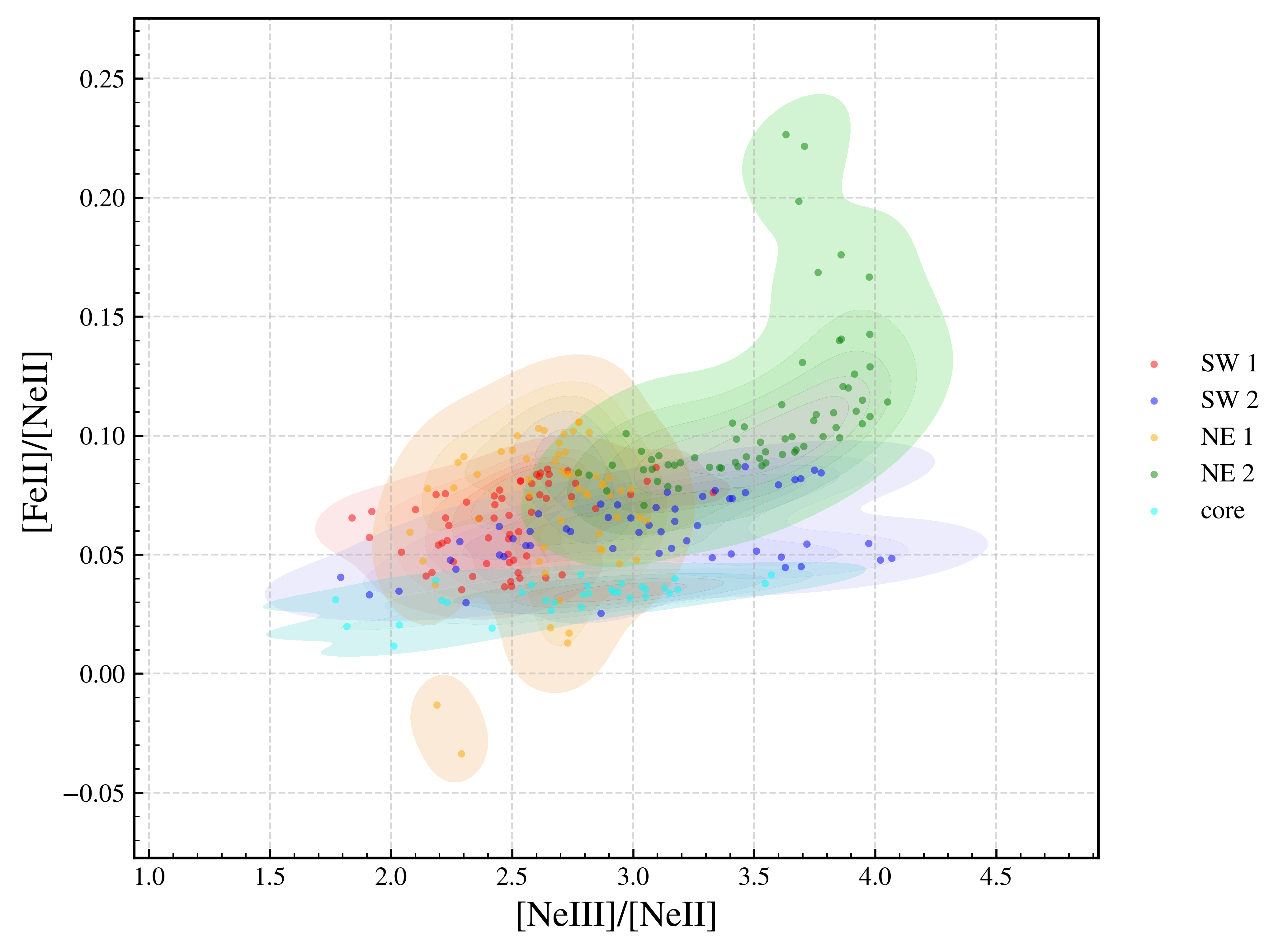}
\includegraphics[height=0.36\linewidth]{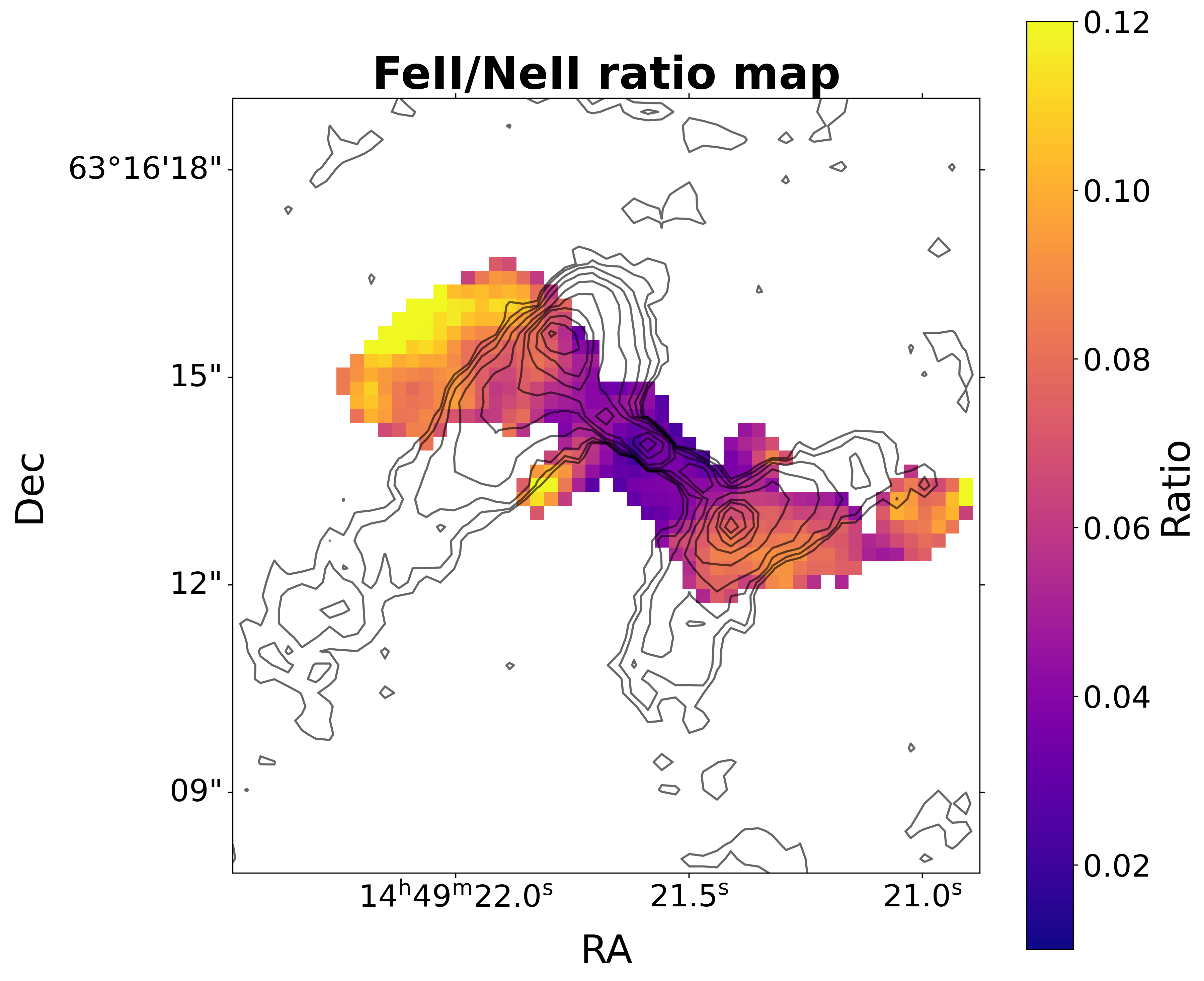}
            \caption{Top left: rgb color composite of [Ne\,III], H$_2$ S(1) and [Ne\,II]. We circle five regions of interest. Top right: Scatter plot showing the [Ne\,III]/[Ne\,II] ratio vs H$_2$ S(1)/[Ne\,II] ratio. Middle left: rgb color composite of [Ne\,V], [Ne\,III] and [Ne\,II] with radio contours overlaid. Middle right: [Ne\,III] moment 1 map with radio contours overlaid. Bottom left: Spaxel-by-spaxel diagnostic line ratio-plot of [Fe\,II]/[Ne\,II] vs [Ne\,III]/[Ne\,II] (see top left panel). The contours correspond to the kernel density distribution. Nucleus occupies the lower quadrant with low values in both [Fe\,II]/[Ne\,II] and [Ne\,III]/[Ne\,II] ratios consistent with being photoionized by the AGN. On the other hand, both NE1 and SW1 exhibit high values of [Fe\,II]/[Ne\,II] and suppressed [Ne\,III]/[Ne\,II]  ratios consistent with grain destruction in dense gas. The regions past the hotspot (NE2 and SW2) show high values for both ratios, consistent with fast shocks and a lower-density medium.
            Bottom right: [Fe\,II]/[Ne\,II] ratio map shows clear enhancement along the jet and in the vicinity of the hotspot.}
    \label{fig:ionizedgas_comparison}
\end{figure*}

In addition to the warm molecular hydrogen rotational lines, we also detect a plethora of ionized gas lines in the MIRI MRS data (see Figure~\ref{fig:1D_MIRI_spectra}), such as those from Ne, Ar, S, and O. The Ne emission lines are particularly strong. We compare the distribution and kinematics of the brightest Ne emission lines in the MIRI spectra in Figure~\ref{fig:ionizedgas_comparison}. 
The top left panel displays a direct comparison of the warm molecular gas H$_2$ S(1) emission line with the [Ne III]15.5$\mu$m and [Ne II]12.8$\mu$m lines. We have annotated five regions marked in five different colors, and the top right panel depicts the line ratio plot in all the pixels in these regions. The NE1 region is purple in color, and it corresponds to regions where both the [Ne\,\textsc{iii}] (red channel) and [Ne\,\textsc{ii}] (blue channel) emission are strong, while H$_2$ S(1) emission (green channel) is weak. However, at these locations past the hotspot, warmer transitions (e.g., S(3)) are bright even though the cooler H$_2$ S(1) appears to be low, possibly because of the localized heating and shock excitation near jet termination/interaction sites. 
This is also evident in the line ratio plot, where most of the yellow pixels occupy the low H$_2$/[Ne\,II] regime. More interestingly, it is evident from the line ratio plot that both the hotspots, NE1 and SW1, show a lower [Ne\,III]/[Ne\,II] ratio compared to the rest of the regions. This is visible in both the RGB images (top left and bottom left panels) as well. We note that H$_2$/[Ne\,II] is largely uncorrelated with the [Ne\,III]/[Ne\,II]. This might be because of the presence of already existing cold molecular hydrogen along the disk, which is not cospatial with the ionized gas. The H$_2$ gas might be governed by turbulence or shock impinging on the underlying cold molecular gas in the disk, whereas the ionized gas traces its origin to outflowing gas. We also note that H$_2$/[Ne\,II] can be affected by variations in molecular fraction and gas metallicity and may be sensitive to additional excitation mechanisms such as weak X-ray heating, unlike the [Ne\,III]/[Ne\,II] ratio.  

The middle left panel of Figure~\ref{fig:ionizedgas_comparison} presents an RGB color composite of the [Ne\,\textsc{v}], [Ne\,\textsc{iii}], and [Ne\,\textsc{ii}] images, with radio contours overlaid. 
This visualization highlights excitation gradients and can help distinguish between AGN photoionization and star formation.
Significant [Ne\,\textsc{v}] emission is detected only at the core,  pointing to AGN photoionization, whereas both hotspots display [Ne\,II] excess in comparison to both [Ne\,\textsc{v}] and [Ne\,\textsc{iii}]. Also [Ne\,\textsc{iii}] seems to be the most extended feature. Such a pattern is a result of both ionization potential and the effect of gas density on shock excitation. In the context of shock models, higher ionization lines such as [Ne\,\textsc{v}] and [Ne\,\textsc{iii}] are produced by fast low-density shocks or by direct AGN radiation, whereas 
[Ne\,\textsc{ii}] dominates in denser, slower-shock regions \citep{Dopita1996,Allen2008}.
In the case of pressure equilibrium \citep{Hardcastle2012}, denser regions possess lower temperature and slower shocks, which in turn favor [Ne\,II] emission. Hence, the NE1 and SW1 regions indicate jet or cocoon interaction with the dense molecular material, resulting in bright [Ne\,II] emission, while [Ne\,III]- dominant regions likely represent lower-density, higher-velocity regions.
Moreover, the jet ISM interaction regions are notably asymmetric, pointing towards a difference in ambient density. The southwestern jet also appears shorter, indicating higher deceleration by a denser medium. 

The velocity maps (middle right panel of Figure~\ref{fig:ionizedgas_comparison}) show higher velocities for ionized gas components compared to H$_2$ lines.
Such stratification of higher ionization lines achieving higher velocities has also previously been observed in other systems \citep[e.g., IC\,5063][]{Dasyra2024}. A clear jump in velocities is visible in the [Ne\,III] velocity maps at locations close to both hotspots. The velocity jump seen in the moment 1 map hence favors the model where the majority of the gas near the hotspot is outflowing. There is a striking resemblance between the integrated velocities from the moment 1 map of [Ne\,III] and the velocity field of the outflowing component of the molecular gas (see Figure~\ref {fig:double_gaussian_mom0} ), whereas the bulk of H$_2$ follows galactic rotation. All this evidence suggests that the differences in the distribution of gases in various ionization phases are probably a result of shocks, ambient density gradients, and the location of the pre-existing molecular gas structures.

\begin{figure*}
    \centering
    \includegraphics[width=0.95\linewidth]{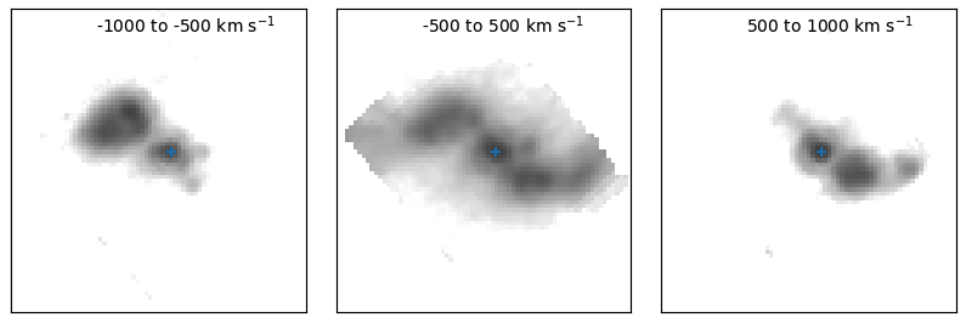}

    \includegraphics[width=0.48\linewidth]{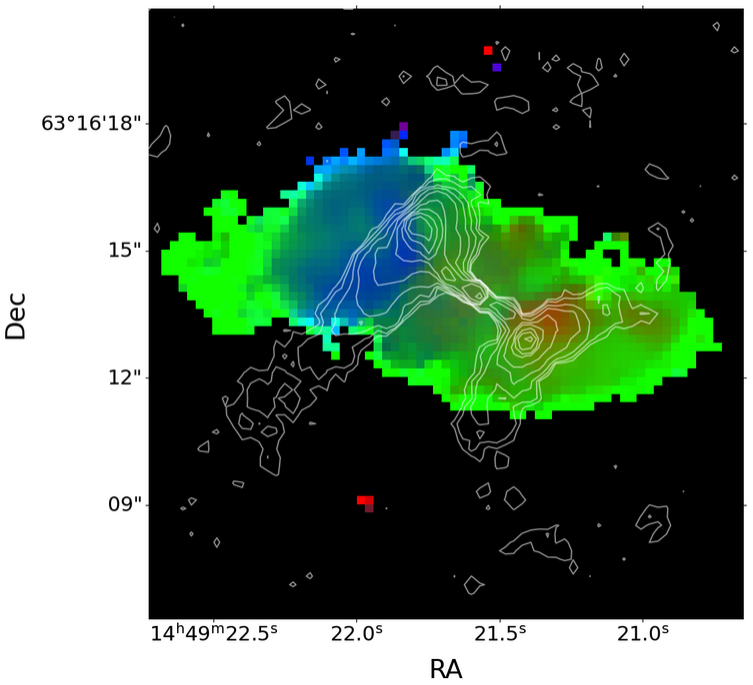}
    \includegraphics[width=0.40\linewidth]{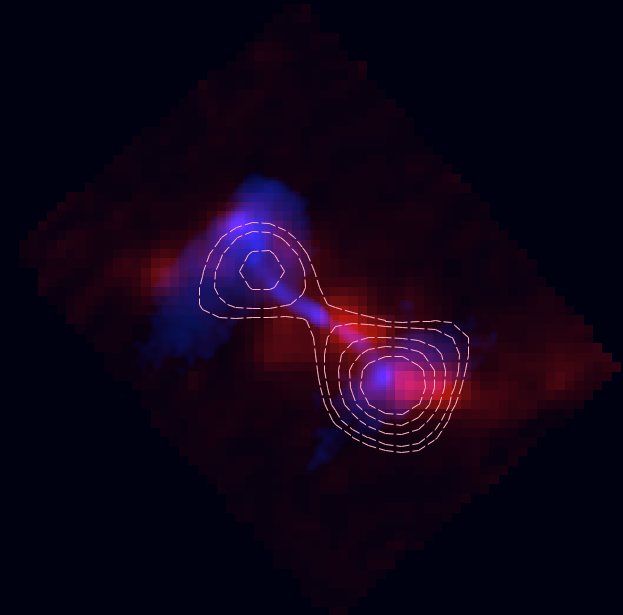}
    
    \caption{Top: Integrated [Ne III] channel maps, highlighting high-velocity outflows. Bottom Left: 3-color map for these 3 velocity ranges (where blue corresponds to $-1000$ to $-500~\mathrm{km~s^{-1}}$, green to $-500$ to $500~\mathrm{km~s^{-1}}$, and red to $500$ to $1000~\mathrm{km~s^{-1}}$). The dog-leg morphology indicates that the fastest outflows and radio lobes escape the galaxy disk together along the path of least resistance and greatest pressure gradient. Bottom Right: comparison of H$_2$ S(3) emission (red) to Merlin 1.5 GHz radio map (blue) and H I absorption optical depth (dashed contours). Systemic HI absorption is strongest at the SW hotspot, which is more deeply embedded in the molecular disk, exciting more warm H$_2$ emission, and driving stronger H$_2$ outflows. Weaker, faster HI outflow is seen against the NE radio lobe \citep{Morganti2005}.}
    \label{fig:fastoutflow}
\end{figure*}

While the [Ne III] moment map demonstrates a flux-weighted outflow velocity of $\sim$ 400 km s$^{-1}$, higher velocity ionized outflows can be seen in the channel maps (Figure~\ref{fig:fastoutflow}). In particular, ionized outflows of up to $\sim$ 1000 km s$^{-1}$ are found near the radio hotspots. Because there is little to no extinction at 15.55 $\mu$m, the observed distribution of [Ne III] should faithfully reflect the actual distribution of gas on both sides of the galaxy disk. Blue-shifted fast outflows are found preferentially on the front side of the disk, near the NE hotspot.  Redshifted outflows are found preferentially on the far side of the disk, near the SW hotspot. This front-back asymmetry of the disk outflows is consistent with the sense of the jet inclination to the galaxy disk inferred from HI, where shallow, fast, blue-shifted absorption is found in front of the NE radio lobe and deep systemic-velocity absorption in front of the SW hotspot \citep[Figure 11;][]{Morganti2005}. Fast outflows occur preferentially on the side of the disk occupied by the hotspot because outflows are generally driven away from the plane of the disk and not through it. Moreover, the fastest ionized outflows are more prevalent in the NE, where the jet length is greater, and the hotspot is therefore likely elevated higher above the disk plane, surrounded by less dense gas. Conversely, fast ionized outflows are less prevalent, and molecular outflows are stronger in the SW, where the hotspot is closer to the disk plane and more deeply embedded in the disk.

The bottom right panel of Figure~\ref{fig:ionizedgas_comparison} shows the spatially resolved line-ratio maps of [FeII]/[NeII]. Both [Fe\,II] and [Ne\,II] can be excited by shocks, although [Fe\,II] is more sensitive to dust grain destruction and sputtering. The higher values of [Fe\,II]/[Ne\,II] at hotspots hence imply efficient dust destruction by strong shocks, in contrast to the nucleus, where AGN photoionization tends to cause the [Ne\,II] emission. 
The bottom left panel presents the line ratio plot of [Fe\,II]/[Ne\,II] vs [Ne\,III]/[Ne\,II] for five distinct regions. Although the hotspot regions (NE1 \& SW1) show high values for  [Fe\,II]/[Ne\,II], suggesting the presence of strong fast shocks, while the relatively low values of [Ne\,III]/[Ne\,II] argue in favor of dense post shock environment where rapid recombination results in lower abundance of doubly ionized Ne. Our MAPPINGS model overpredicts the [Fe\,II]/[Ne\,II] ratio by about an order of magnitude.  This discrepancy indicates that while the [Fe\,II] emission is elevated, a large fraction of the Fe II remains depleted onto the dust grains, whereas fast radiative shock models assume efficient dust destruction and, as a result, maximal gas-phase Fe\,II \citep{Allen2008}.
Hence, it is possible that the jet is driving fast shocks into a clumpy dense ISM, where Fe grains are effectively released into the ISM even while maintaining the low ionization state of Ne ions.

\section{Discussion}
\label{sec:discussion}

Our spatially resolved JWST observations of warm H$_2$ confirm the strong jet-ISM interaction in 3C\,305. The H$_2$ emission line profiles require a double Gaussian model to be represented accurately, including an outflowing component and the rotating disk of the galaxy. The narrow component is identified as the outflowing component, as it is often observed to be farther offset from the systemic velocity (see Section~\ref{sec:kinematics} for a detailed rationale). The maximum velocities of the narrow component reach up to $\sim$400~km~s$^{-1}$, whereas typical AGN systems without much impact from the jet show velocities up to 150~km~s$^{-1}$ \citep{Esparza2025}. However, we note that there are exceptions even among low-power jet systems and the observed kinematic signatures strongly depend on the disk density and the jet inclination \citep[see for e.g., ][]{Lopex2025}.

Similar multi-component features were also previously reported in other radio galaxies (for example, 3C\,326 \cite{Leftley2024}, 3C\,293 \cite{deMellos2025}, and IC\,5063 \cite{Dasyra2024}), suggesting that jet-driven warm H$_2$ outflows might be commonplace. The outflowing component is coincident with the radio lobes and is not symmetric about the nucleus.
The striking correspondence of the outflow seen in warm H$_2$ with the already reported neutral and ionized gas outflows \citep{Morganti2005,Heckman1982}, demonstrates the jet's impact on the multiphase ISM. 

The high excitation and spatial distribution of H$_2$ gas also indicate shock heating by the jet. Moment zero maps of the rotational lines of warm H$_2$ show that they are aligned along the radio jet axis, with bright clumps near the jet termination sites. Jet-driven shocks and in-situ heating of the molecular hydrogen can explain such an enhancement in brightness. Spitzer studies of the mid-infrared spectrum of 3C\,305 had already shown that it is an H$_2$-luminous galaxy, similar to other shock-dominated systems like 3C\,326, where the PAH and other star formation indicators are weak \citep{Ogle2010}. We find that  1.3\% of the cold molecular gas has been heated to temperatures above 300~K. All of these pieces of evidence suggest that jet feedback is efficiently operating in 3C\,305 via both outflows as well as shocks and heating of the ISM. 

Observations of 3C\,305 fit very well with the theoretical expectations of a relatively low-power jet plunging through a dense interstellar medium. As the jet propagates through the interstellar medium, it drives a bow shock and a laterally expanding hot cocoon of gas \citep{Mukherjee2016}. Simulations of IC\,5063 show that as the jet passes through low-density channels in the disk, it ablates, shocks, and accelerates dense clouds up to velocities greater than 400~km\,s$^{-1}$ \citep{Mukherjee2018b}. The jet in 3C\,305 is not expected to carve out a narrow tunnel, but instead fill a wide cavity or cocoon of hot plasma. Indeed, we see evidence for laterally expanding hot gas in the soft X-rays \citep{Hardcastle2012}. The second component of H$_2$ shows broad velocity widths in directions perpendicular to the jet, which is probably a result of cocoon-driven turbulence (see bottom panel of Figure~\ref{fig:double_gaussian_mom0}). The impact of the jet in 3C\,305 is hence multi-faceted. 


\subsection{Shock ionization modeling}
\label{sec:shock_ionization}
In section~\ref{sec:ionizedgas_results}, we found that the ionized gas is profoundly impacted by the jet, driving ionized outflows as fast as 1000 km~s$^{-1}$. In this section, we analyze the role of the AGN jet in shock-ionizing the ISM based on the emission line ratios of various fine-structure lines in the MIR range.

\cite{Reynaldi2013} successfully reproduced the observed optical ionized gas emission lines using shock models. The models required the presence of autoionizing shocks (shock + precursor) in low-density ($ n_e \sim 1.0$ cm$^{-3}$) gas to reproduce the line ratios and intensities. Furthermore, a low metal abundance similar to the LMC was required, which seems unexpected for the massive 3C 305 host galaxy. However, given that 3C305 seems to be undergoing a fairly major merger, the lower abundances could reflect those of the merging galaxy. Interestingly, low metal abundance was also found from spectral fits to the X-ray data, but this might alternatively indicate a multi-temperature plasma \citep{Hardcastle2012}. However, we find that we can model the [O II]/[O III] and He~II/H$\beta$ line ratios measured by \cite{Reynaldi2013} in shocks with solar abundance at higher density. 


We use the library of shock models precomputed in the 3MdBs database \citep{Morisset2015} to compare the emission-line ratios obtained from JWST data in the mid-infrared and the near-infrared range. These models are computed using the {\tt MAPPINGS V} code \citep{Sutherland2017} and include both ``shock-only'' and ``shock+precursor'' emission models \citep{Allen2008}. We select precomputed model values available on a discrete grid, covering shock velocities from $200$ to $400~\mathrm{km\,s^{-1}}$, and magnetic field strengths from $0.01$ to $1000~\mu\mathrm{G}$. We assumed a fixed pre-shock density of $1000~\mathrm{cm^{-3}}$ and solar metallicity abundance. Although slightly lower pre-shock densities can reproduce the mid-IR line ratios, higher densities yield a more accurate simultaneous reproduction of many observed line ratios.

\begin{figure*}
    \includegraphics[width=0.5\linewidth,trim=0 100 0 0]{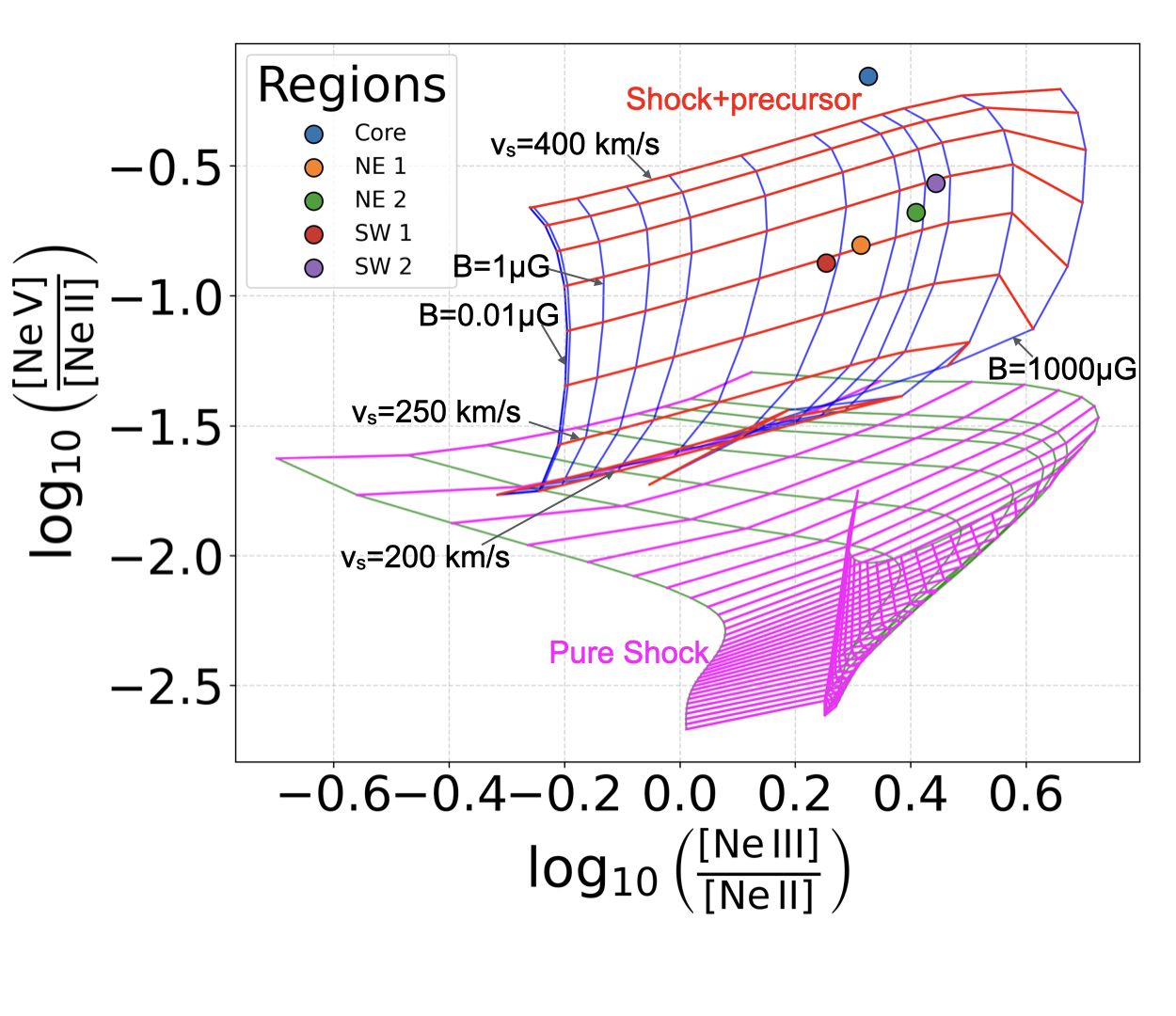}
    \includegraphics[width=0.5\linewidth,trim=0 100 0 0]{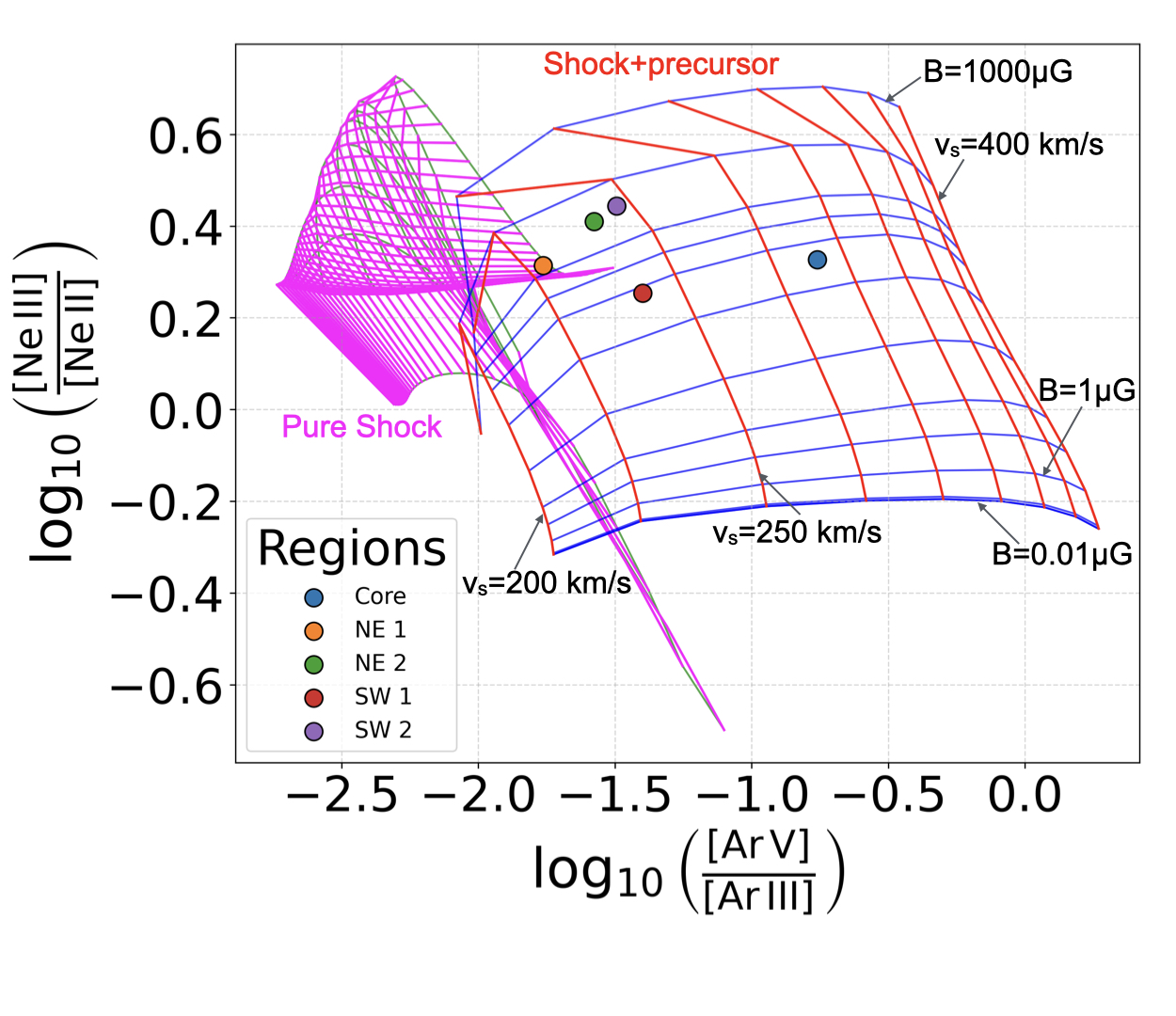}
    \includegraphics[width=0.5\linewidth,trim=0 100 0 0]{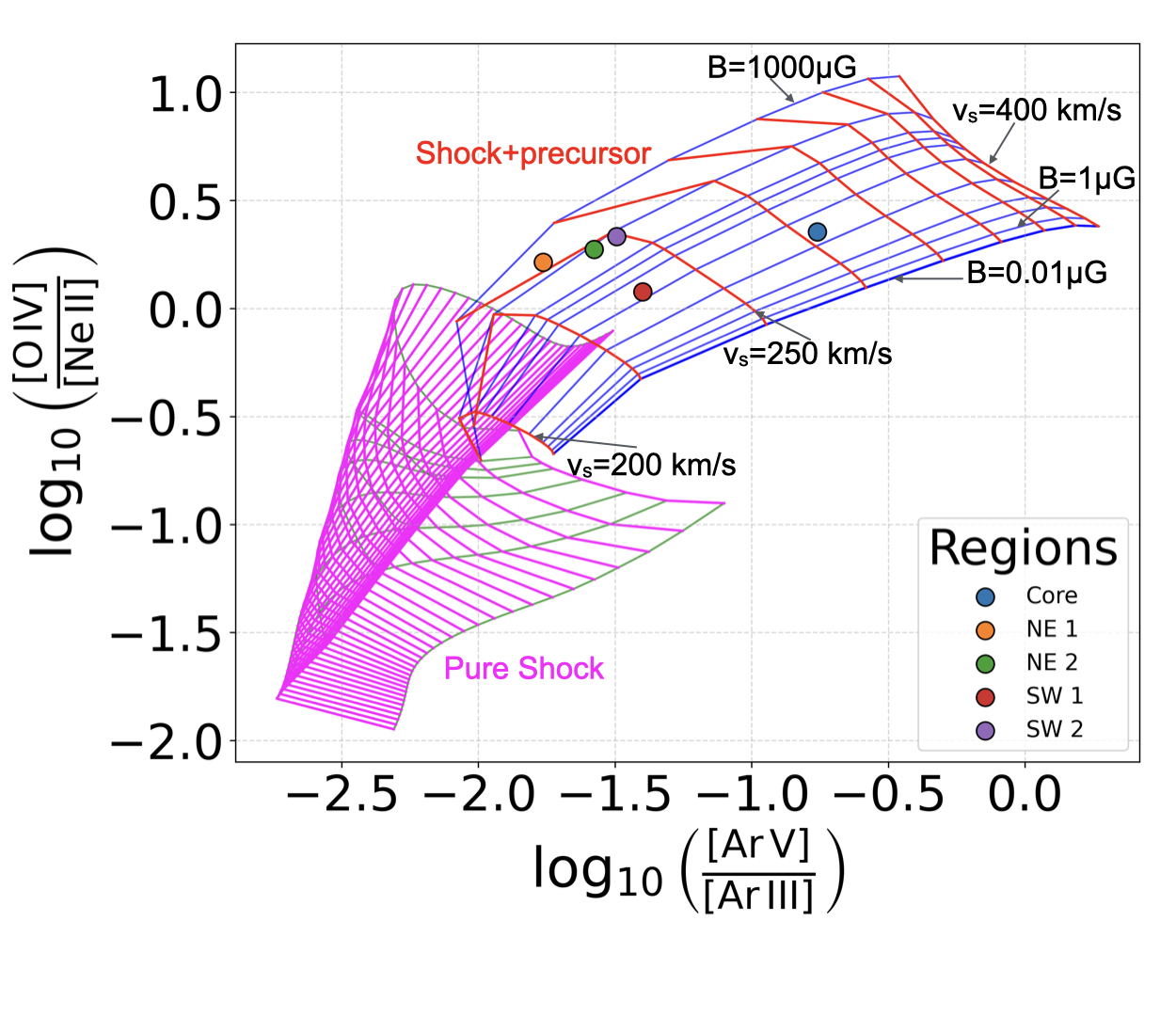}
    \includegraphics[width=0.5\linewidth,trim=0 100 0 0]{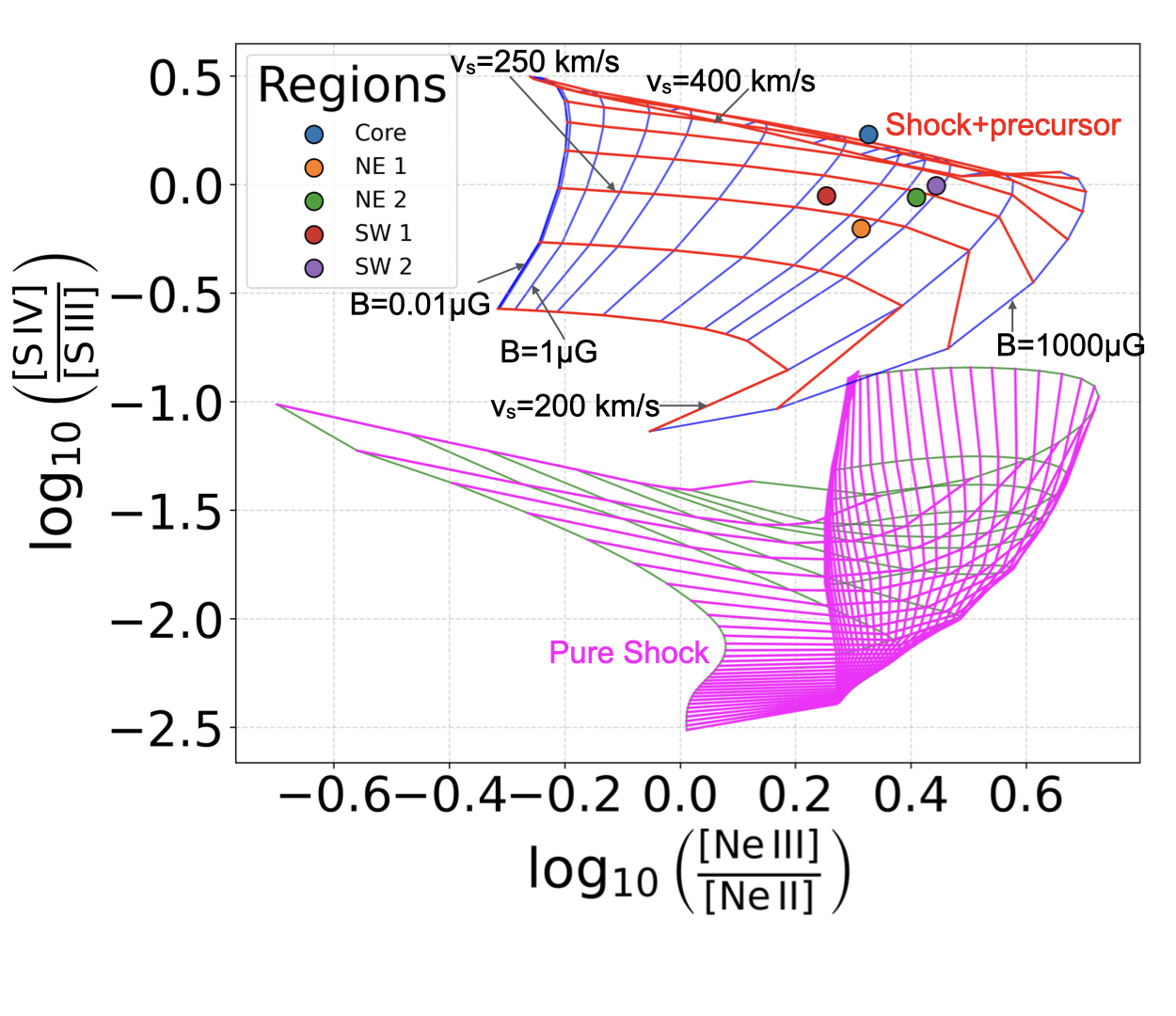}
    \caption{Diagnostic line ratio plots compared to fast shock models for collisionally ionized gas. Models with a photoionized precursor (red) match the data, while those without (magenta) do not. a) $\log_{10}([\mathrm{Ne\,V}]/[\mathrm{Ne\,II}])$ vs. $\log_{10}([\mathrm{Ne\,III}]/[\mathrm{Ne\,II}])$, demonstrating agreement with model predictions. b) $\log_{10}([\mathrm{Ne\,III}]/[\mathrm{Ne\,II}])$ vs. $\log_{10}([\mathrm{Ar\,V}]/[\mathrm{Ar\,III}])$  c) $\log_{10}([\mathrm{O\,IV}]/[\mathrm{Ne\,II}])$ vs. $\log_{10}([\mathrm{Ar\,V}]/[\mathrm{Ar\,III}])$, and d) $\log_{10}([\mathrm{S\,IV}]/[\mathrm{S\,III}])$ vs. $\log_{10}([\mathrm{Ne\,III}]/[\mathrm{Ne\,II}])$.  Models are shown for a grid with $v = 200, 225, 250, 275, 300, 325, 350, 375, 400$ km s$^{-1}$ and $B = 0.01, 0.1, 1.0, 5.0, 10.0, 16.0, 32.0, 63.0, 100.0, 126.0, 160.0, 316.0, 1000.0$ $\mu$G.}
    \label{fig:NeV_NeIII_diagnostic}
\end{figure*}

We present several line diagnostic plots, including bright line ratios such as [SIV]/[NeII], [NeIII]/[NeII], [ArV]/[ArIII],  [NeV]/[NeII], and [OIV]/[NeII] in Figure~\ref{fig:NeV_NeIII_diagnostic}. 
In 3C\,305, we observe [Ne\,III]15.55 $\mu$m / [Ne\,II] 12.81 $\mu$m line flux ratios that are notably higher (0.2--0.5; also see \cite{Guillard2012a}) than what is typically found in pure starburst galaxies \citep[$0.05$–$0.2$;][]{Salas2009ApJS..184..230B,Herrero2024}. Intense compact star-forming regions can sometimes lead to much larger [Ne\,III]/[Ne\,II] ratios \citep[up to $\sim10$; e.g., ][also see Figure~1 in \citealt{Reynaldi2020}]{Hao2009}. However, the [O\,IV]/[Ne\,II] versus [Ne\,III]/[Ne\,II] diagnostic rules out star formation (see Fig~\ref{fig:NeV_NeIII_diagnostic}).

Specifically, all of our diagnostic plots indicate that the `shock+precursor' model is a viable mechanism for explaining the line ratios at the assumed pre-shock density and abundances. In the `shock+precursor' model, the blue lines depict the variation in the magnetic field values, whereas the red lines show the variation in shock velocities. The observed line ratios are mostly consistent with shock velocities of  250-300~km~s$^{-1}$, in agreement with our kinematic analysis. Magnetic fields of $\sim 100$ $\mu$G are needed to match the line ratios for the assumed pre-shock density of 1000 cm$^{-3}$.

The core stands out from both hotspot regions in line diagnostic ratio plots that include high-ionization potential lines (e.g., [Ar V] (E$_\mathrm{ion} = 59.81\,\mathrm{eV}$) and [Ne V] (E$_\mathrm{ion} = 97.1\,\mathrm{eV}$)). This may indicate that the core is more strongly influenced by AGN photoionization than the hotspots or jets, where shocks are more relevant.

Away from the core, all diagnostic plots indicate the need for a radiative precursor to reproduce the observed line ratios, while `pure shock' models remain incompatible with the data unless extreme magnetic field values are chosen. According to the `shock+precursor' model, the UV photons generated in the shock travel ahead of the shock front and lead to an ionization front that pre-ionizes the gas before it enters the shock, giving rise to the observed enhancement in ionization. 

A common expression for the post-shock temperature in a strong (non-relativistic) shock is:

\begin{equation}
T_{s} \;\approx\; \frac{3}{16}\,\frac{\mu \, m_{p}}{k_{B}}\, v_{s}^{2},
\label{eq:shock_temp}
\end{equation}
where
$T_{s}$ is the post-shock temperature,
$\mu$ is the mean mass per particle (assumed to be 0.6; in units of the proton mass $m_{p}$),
$k_{B}$ is Boltzmann's constant, and
$v_{s}$ is the shock velocity (in cm\,s$^{-1}$).

For a shock velocity range of 200-400~km~s$^{-1}$, we predict a post-shock temperature in the range of 0.5$\times$10$^6$K (0.05 keV) to 2.2$\times$10$^6$K (0.19 keV).
This temperature range is considerably lower than that inferred by  \cite{Hardcastle2012} from spectral modeling of the $kT \simeq 0.78$--$1~{\rm keV}$ X-ray emitting hot plasma surrounding the radio jet, corresponding to shock velocities of $\sim800$--$950~{\rm km~s^{-1}}$.
This difference is likely due to the multi-phase nature of the shocked material, with varied densities. When an overpressured hot cocoon interacts with the denser disk material, a shock forms with lower velocities, leading to the cooler post-shock temperature we infer from mid-IR emission lines. In contrast, the X-ray-emitting gas traces the lower-density medium where the cocoon/jet pressure drives faster shocks (also see Section~\ref{sec:ionizedgas_results}).  However, we do find evidence of faster outflow velocities, exceeding $1000~{\rm km~s^{-1}}$, more comparable to the sound speed in the X-ray emitting plasma. Even higher, subrelativistic velocities may be achieved by the hottest ($10^8-10^{10}$ K ) plasma in the jet cocoon, but such gas is expected to be completely stripped of electrons, such that it is impossible to measure its velocity from electronic transitions. It is also possible that for the faster shocks in lower-density gas, the gas may not cool before clouds are destroyed by hydrodynamic instabilities \citep[e.g., ][]{Klein1994}.


Figure~\ref{fig:collage} shows the presence of extended UV and X-ray emission coinciding with the jet, the hotspot, and the extended emission line regions beyond the NE hotspot. Since the origin of the X-ray emission is attributed to shock heating \citep{Hardcastle2012}, the UV emission may be as well. 

\subsection{Power budget}

\begin{figure*}
    \centering
    \includegraphics[width=0.49\linewidth]{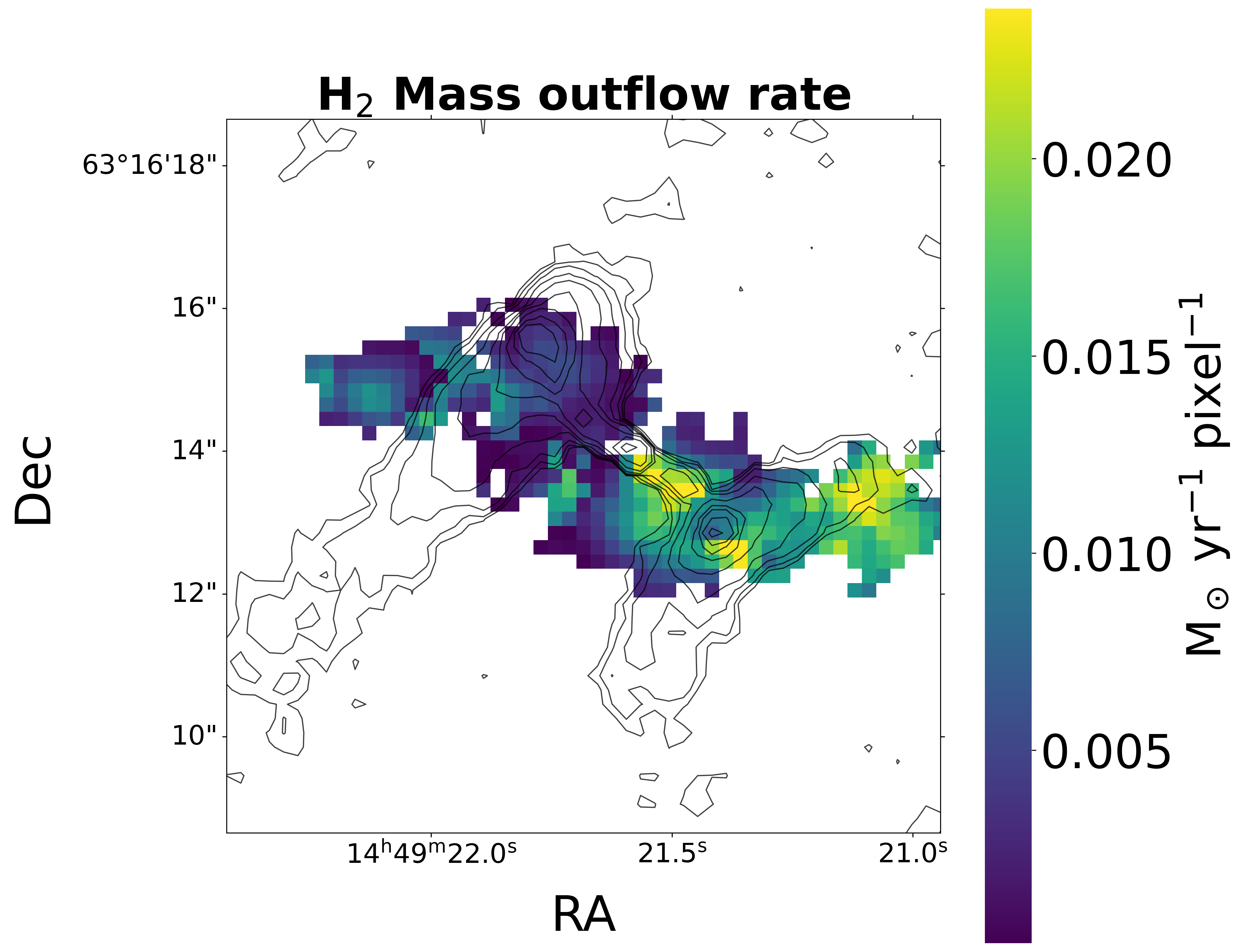}
    \includegraphics[width=0.49\linewidth]{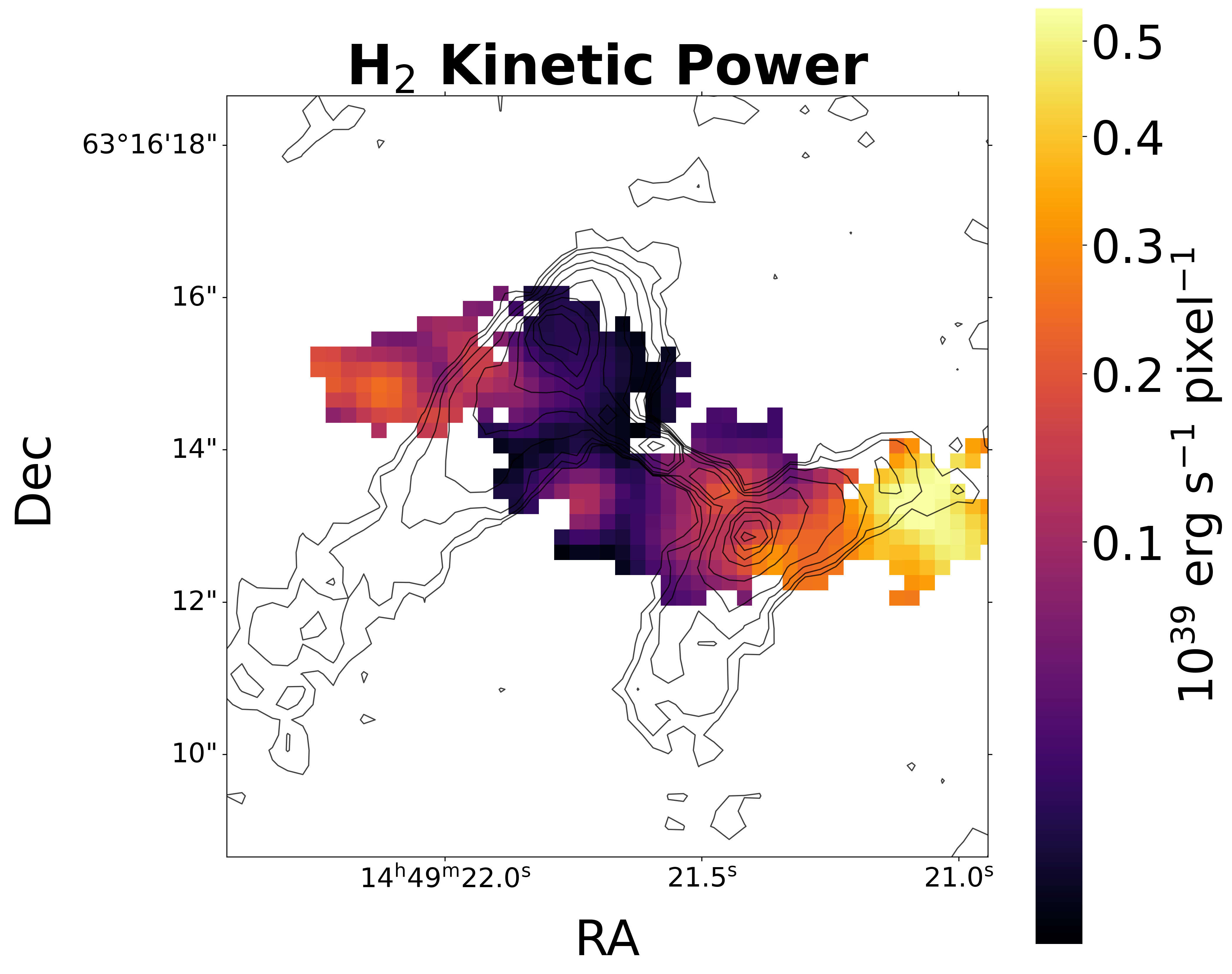}
    \caption{Left: Maps of the warm H$_2$ mass outflow rate, and Right: kinetic power in the outflow derived from the H$_2$ S(3) line assuming that the narrow component from the Gaussian decomposition is the outflowing component. A clear asymmetry is seen: the SW hotspot shows higher mass outflow rates due to larger outflowing gas mass in a denser medium, whereas the NE hotspot, though characterized by higher velocities, reveals lower mass and kinetic energy. Radio contours are overlaid.}
    \label{fig:MOR_H2}
\end{figure*}
\begin{figure*}
    \centering
    \includegraphics[width=0.52\linewidth]{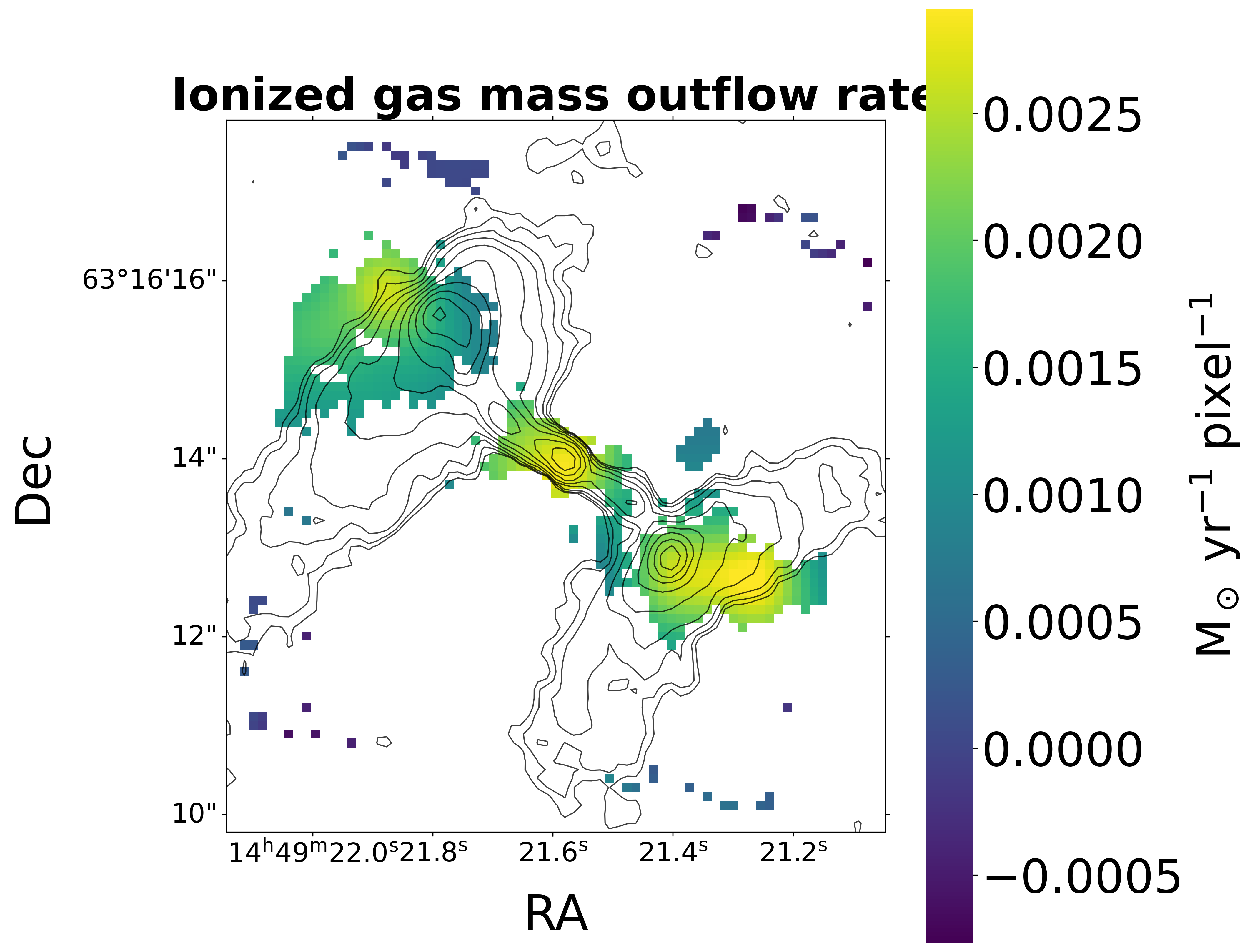}
    \includegraphics[width=0.46\linewidth]{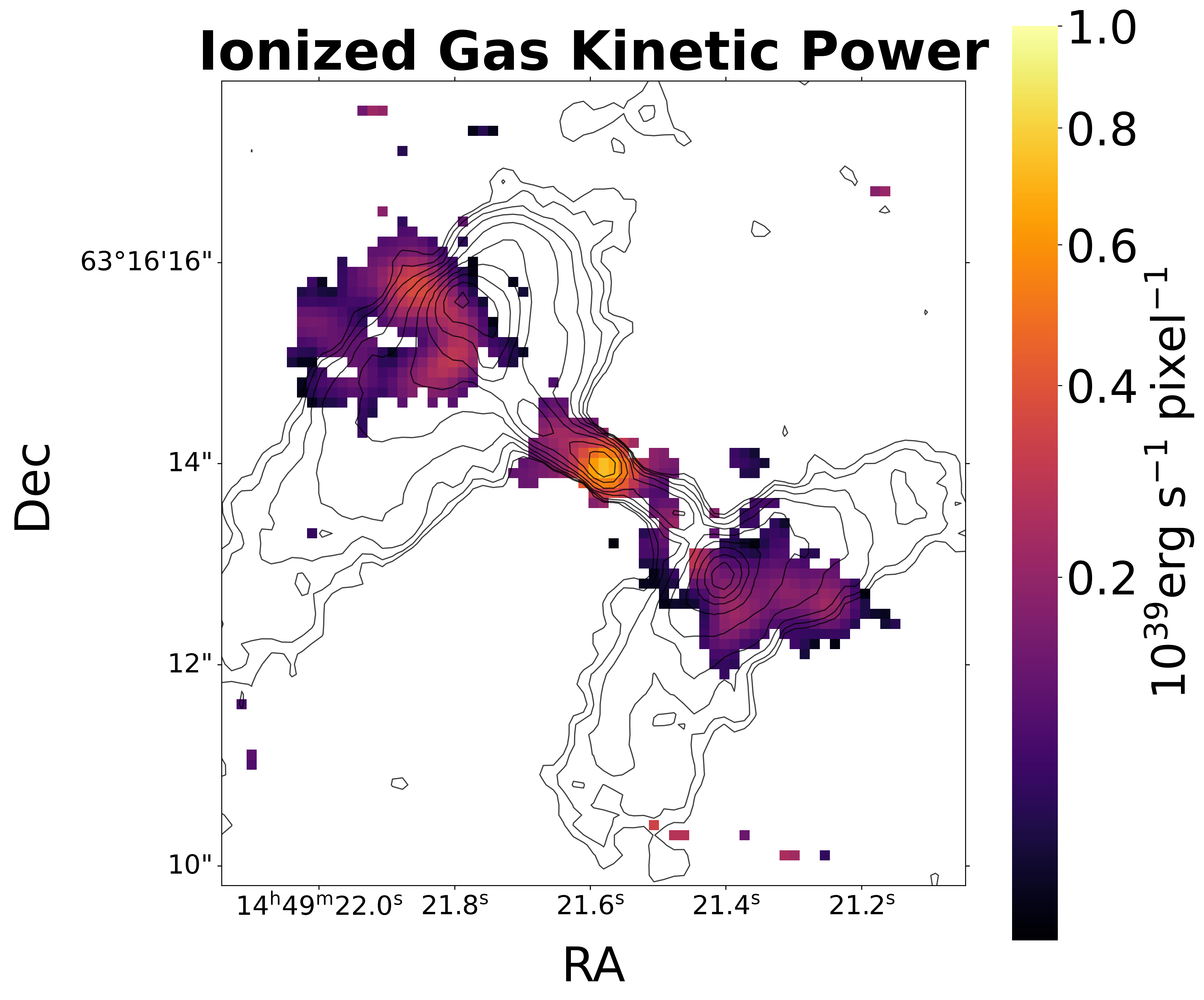}
    \caption{Left: Maps of the warm ionized gas mass outflow rate, and Right: kinetic power.  Radio contours are overlaid.}
    \label{fig:MOR_ionized}
\end{figure*}

In this section, we compare the estimated jet power to the kinetic power carried by various gas phases and the radiative cooling budget. We have summarized these values in Table~\ref{tab:energy_power_budget}. We present maps of spatially resolved mass outflow rates and kinetic power for both warm H$_2$ and ionized gas to examine their spatial variations. 

\subsubsection{Warm H$_2$ gas outflows}
For every spaxel, we derived the warm molecular gas mass above 300\,K (denoted as $M_{\text{H}_2, >300\,K}$) by fitting the H$_2$ excitation using a power-law distribution in temperature as described in Section~\ref{sec:excitation_model}. We derived the outflowing warm gas mass from the fraction of flux density in the outflow versus the total,  $f_{{\rm out}}=F_{{\rm out}}/(F_{{\rm out}}+F_{{\rm sys}})$ based on the Gaussian decomposition of the components we performed in Section~\ref{sec:kinematics}. We consider the narrow component as the outflowing component. The S(3) line was used for this calculation as it provides a good balance between S/N ratio and the sampling of the hot gas. Per-spaxel mass outflow  rate is derived as follows:
\begin{equation}
\dot{M}_i \;=\; \frac{f_{{\rm out},i}\,M_{\text{H}_2, >300\,K}\,|v_i|}{L},
\end{equation}
where $L$ is the physical depth of the outflow, which is assumed to be comparable to the galaxy disk scale height, with an adopted value of 280~pc. $v_i$ is the line-of-sight outflow speed in spaxel $i$, taken as the centroid offset of the outflow Gaussian component of H$_2$ S(3) relative to the systemic velocity.
The underlying assumptions include a plane-parallel escape of gas through the spaxel volume with no inclination correction, and that the S(3) velocity split represents all H$_2$ lines. The outflows are assumed to be locally produced within each spaxel volume and not to extend over a large radius spanning across multiple spaxels. 

Our outflow energetics depend on the characteristic depth of the emitting gas along the line of sight, which we assume to be comparable to the disk scale height (\(\sim280\) pc). Since \(\dot{M} \propto L^{-1}\), both the mass outflow rate and kinetic power scale inversely with the assumed depth \(L\). Hence, for an assumed range of \(L \sim 140\)--560 pc, the warm H$_2$ outflow rate could vary from 1.9 to 7.4 M$_{\odot}$~yr$^{-1}$. Variations in the assumed depth affect the absolute values of the outflow energetics, but not the spatial structure of the outflow properties. Therefore, the association of the strongest outflows with the radio hotspots is robust, although the absolute values of the mass outflow rate and kinetic power remain subject to geometric uncertainties.

Mass outflow rate and kinetic power maps are shown in Figure~\ref{fig:MOR_H2}. The smoothing in the figure is purely cosmetic, and we have used unsmoothed images for all integrals and sums.
 The figure shows that the molecular gas does not follow a simple radial outflow. 
A clear asymmetry is evident in the mass outflow rate maps: the southwest side shows higher mass outflow than the northeast side, even though the velocity is higher at the northeast.
As the mass outflow rate is proportional to \( \frac{M |v|}{L} \), these spatial variations track both the variation in the mass and velocity. The southwest hotspot, which has a larger outflowing gas mass in a denser medium, exhibits a higher mass outflow rate than the higher-velocity, lower-density regions in the northeast. Conversely, the northeast hotspot, which has higher velocity but lower mass loading, shows a smaller mass outflow rate but comparable kinetic energy to the southern hotspot. As discussed in Section 3.6, this matches the asymmetry found in the HI absorption map against the radio jet, where the HI column density and outflow rate are greatest at or near the SW hotspot.

The southwest hotspot terminates the shorter arm of the radio jet, where there is also more molecular gas and higher extinction, implying that it encounters greater resistance from a dense, clumpy medium through which the jet interacts. The frustrated jet deposits both momentum and energy locally. On the other hand, the longer northeast arm, where the jet propagates more freely, encounters less resistance, accelerating the gas to higher velocities but evacuating less H$_2$.

\subsubsection{Ionized gas outflows}
To estimate the ionized gas mass using the Pa\,$\alpha$ flux in every spaxel, we assumed the following formula,

\begin{equation}
\left( \frac{M_{\mathrm{HII}}}{M_\odot} \right)
= 2.3 \times 10^{18}
\left( \frac{F_{\mathrm{Pa}\alpha}}{\mathrm{erg\,s^{-1}\,cm^{-2}}} \right)
\left( \frac{D}{\mathrm{Mpc}} \right)^2
\left( \frac{n_e}{\mathrm{cm^{-3}}} \right)^{-1},
\label{eq:pa_alpha_mass}
\end{equation}
where $M_{\mathrm{HII}}$ is the ionized gas mass, $F_{\mathrm{Pa}\alpha}$ is the integrated Pa$\alpha$ flux,
$D$ is the luminosity distance,
and $n_e$ is the electron density \citep{deMellos2025}. Note that mass is inversely proportional to the electron density, which we do not have precise knowledge of, and that the observed emission likely comes from clouds with a large range in density.
We adopt an electron density of 1000 ${\mathrm{cm^{-3}}}$, which is suitable for dense regions undergoing jet-ISM interaction, typically found in AGN narrow-line region \citep{Holt2011,Davies2020,Santoro2020,Holden2023}. 
We assume all material with absolute velocities, $|v| > 200~\mathrm{km\,s^{-1}}$, is outflowing based on the convention that velocities in excess of the local rotational velocities indicate outflowing gas. For all channels in every spaxel following this condition, we derive a mass outflow rate, $\dot{M}_{\mathrm{HII}}(v) \;=\; \frac{M_{\mathrm{HII}}(v)\,|v|}{\Delta R},$
where $\Delta R$ is the galaxy disk scale height of 280~pc and $v$ is the line–of–sight velocity of the outflowing gas. The total mass outflow rates for each spaxel are then derived by summing $\dot{M}_{\mathrm{HII}}(v)$ over all outflowing velocity channels. We note our selection of only the material with $|v| > 200~\mathrm{km\,s^{-1}}$ might result in the exclusion of some low-velocity outflows from the sum. However, since kinetic energy and power scale as  $v^2$ and $v^3$ respectively, the impact of this on our result is limited. 
The estimated mass outflow rates and kinetic power maps are presented in Figure~\ref{fig:MOR_ionized}. 

While still asymmetric about the AGN, the ionized gas mass outflow is notably more symmetric than the H$_2$ mass outflow. The ionized outflows are also more collimated along the jet axis than the H$_2$ outflows. It may be that general turbulence adds to the H$_2$ kinetic power across the molecular disk, while jet-shock-ionized gas is localized along the jet.  The highest ionized gas mass outflow rates are found at the nucleus and exterior to the radio hotspots. We further discuss the tendency for the ionized outflow rates to peak outside of the active radio jet region in Section 4.3. The ionized gas kinetic power peaks in the core, but is distributed more broadly in the regions exterior to the radio hotspots, perhaps reflecting the capacity of the hot jet cocoon to launch outflows wherever it intersects the galaxy disk.


\subsubsection{Radiative cooling from MAPPINGS shock models}

We extracted the 3MdBs `shock+precursor' model that best matched our observations (see Section~\ref{sec:shock_ionization}) with the corresponding best fit parameters ({$v_s = 275$ km s$^{-1}$}, {$B = 100$ $\mu$G}, {$n = 1000$ cm$^{-3}$}, \cite{Allen2008} Solar abundances).  We note that the combination of magnetic field and electron densities is partially degenerate and that it is difficult to obtain independent constraints on $B$ for this high-density ionized gas phase. Moreover, different emission lines have their origin in regions with different densities. As a result, it is not possible to ascribe a single physical density to all emitting regions. In comparison, \cite{Hardcastle2012} find that $B> 16$ $\mu$G for the hot, low-density ($n_e \sim 0.2$ cm$^{-3}$) X-ray emitting plasma phase, based on radio depolarization. They also argue that the magnetic field must be $B< 100 \mu$G so that it does not exceed the thermal energy of the plasma. 

We retrieved the line fluxes of all emission lines in all wavebands (tables \texttt{emis\_IR}, \texttt{emis\_VI}, \texttt{emis\_UVA/UVB/UVC}) and summed them to obtain the total model emission line flux. To scale the model to our observations and to obtain the total radiative flux, we used the following set of bright MIR lines: 
$[\mathrm{Ar\,II}]\,6.99~\mu\mathrm{m},$ 
$[\mathrm{Ar\,III}]\,9.00~\mu\mathrm{m}, $
$[\mathrm{S\,IV}]\,10.51~\mu\mathrm{m},$
$[\mathrm{Ne\,II}]\,12.81~\mu\mathrm{m},$
$[\mathrm{Ne\,V}]\,14.32~\mu\mathrm{m},$
$[\mathrm{Ne\,III}]\,15.55~\mu\mathrm{m},$
$[\mathrm{S\,III}]\,18.71~\mu\mathrm{m},$
\text{and }
$[\mathrm{O\,IV}]\,25.89~\mu\mathrm{m}$. The total radiative flux from the full MAPPINGS model, including both UV and optical wavebands that are inaccessible to JWST, was estimated for various regions across the source. 
The factor to convert from the measured MIR lines to the total line flux ranges from 23 to 29 across the conditions in our models. For our best-fit model with n$_e$ = 1000 cm$^ {-3}$, we obtain a conversion factor of 29.

For the entire source, we estimate the model-derived total emission line luminosity from UV-IR, $L_{\rm line,tot} \approx {6.75\times10^{43}}~{\rm erg~s^{-1}}$. Importantly, this shows that the total emission line power from shock-ionized gas is $\sim 3$ times greater than the AGN total IR continuum luminosity of ${2\times10^{43}}~{\rm erg~s^{-1}}$, estimated from the {\it Spitzer} 24 $\mu$m flux \citep{Dicken2010, Hardcastle2012}. This rules out any substantial contribution from AGN photoionization to the integrated emission line power.

To estimate the radiation from warm molecular gas, we measured and totaled the mid-IR H$_2$ pure-rotational and near-IR rovibrational line luminosities for the entire source. While the total H$_2$ rotational line luminosity is a factor of 67.5 times smaller than the ionized gas line luminosity (Table~\ref{tab:energy_power_budget}), it is bigger than the kinetic power imparted to all gas phases except for the X-ray emitting hot phase. 

\subsubsection{Jet Power}

\cite{Hardcastle2012} and \cite{Lanz2016} estimate 3C 305 jet powers of $10^{43}~{\rm erg~s^{-1}}$ and $2.3\times10^{43}~{\rm erg~s^{-1}}$ respectively. \cite{Hardcastle2012} estimate the jet power by dividing the energy content of the X-ray emitting plasma by the jet expansion time, derived from their estimate of the deprojected shock velocity, divided by the jet half-length. This approach assumes that the radio hotspot has moved outward at an average speed set by the temperature of the X-ray emitting plasma. This is far from guaranteed, particularly if the jet is stalled by the dense ISM. \cite{Lanz2016} provide a jet power estimate by scaling the radio luminosity \citep{Punsly2005}. This latter approach has been derived for classical FR II radio sources, assuming that the radio lobes are far out of equipartition and that most of the energy is in particles rather than the magnetic field. It is unclear whether or not the same relation applies to a radio jet like 3C 305, trapped in the host galaxy ISM. Both of these values are within an order of magnitude of the emission line luminosity, but are not sufficient to power the observed emission. We conjecture that the jet power is $>L_{\rm line,tot}$, assuming the jet is primarily responsible for heating the ionized gas. If most of the energy is dumped into the X-ray emitting hot cocoon, then the emission line power may represent only a fraction of the jet power. This line of reasoning implies that the jet power is underestimated by a factor of $>6$ by \cite{Hardcastle2012} and a factor of $>2.6$ by \cite{Lanz2016}. 

\subsubsection{Total Outflow Kinetic Power}
The total kinetic power in ionized gas and warm H$_2$ outflows, as estimated above from our JWST observations, are 1.6$\times10^{41}~{\rm erg~s^{-1}}$
and 6$\times10^{40}~{\rm erg~s^{-1}}$ respectively. Hence, only $\lesssim$1\% of the jet power is used to drive these outflows in 3C\,305. We note that the ionized gas kinetic power is sensitive to the assumed density. An electron density of 100~cm$^{-3}$, would result in an increased kinetic power output and mass outflow rate by an order of magnitude. Even in this case, the kinetic power ($\sim8\%$) would still remain far below the radiative power output. We note that such low post-shock densities are extremely unlikely, given the measured densities in other jet-cloud interactions across a range of radio luminosities \citep[e.g., ][]{Santoro2020,Holden2023} and the likelihood of strong shock compression. In addition to the density, several other uncertainties remain, including temperature distribution uncertainties from power-law H$_2$ fits and projection effects. The turbulent motions add a term proportional to the square of the velocity dispersion, in addition to the bulk velocity. However, this term is expected to be small relative to the outflow kinetic energy term unless the velocity widths approach or exceed the bulk velocities. In our 3C\,305 dataset, we find that the velocity widths are smaller than the outflow velocities; hence, the turbulent power term only adds a minor correction.

\begin{table*}
\centering
\caption{Power Budget Across Gas Phases}
\label{tab:energy_power_budget}

\begin{tabular}{lccccc}
\hline\hline
Gas Phase &
Tracer(s) &
$\dot{E}_{\rm kin}$ &

$L_{\rm line}$ &
$\dot{M}$ &
$M$\\
 &
 &
(erg s$^{-1}$) &
(erg s$^{-1}$) &
$M_{\odot}~yr^{-1}$&
$M_{\odot}$\\
\hline
Warm Molecular (H$_2$) &
H$_2$ S(1)--S(7)$^*$ &
5.8$\times10^{40}$ &
{1$\times10^{42}$} &
3.7&
{4.8$\times10^{6}$}\\

Ionized Gas (UV-IR) &
Pa$\alpha$ &
{1.6$\times10^{41}$} &
{6.75$\times10^{43}$} &
1.6&
{1$\times10^{6}$}\\


Neutral Gas (H\,\textsc{i})$^{\rm a}$ &
21-cm absorption &
 {3.8$\times10^{41}$} &
-&
5--10&
{1.3$\times10^{7}$}\\


Hot X-ray Gas$^{\rm b}$ &
Thermal Bremsstrahlung &
{2.9$\times10^{42}$}  &
{4.8$\times10^{41}$}&
22&
{7$\times10^{7}$}\\
\hline
Total & --- &
3.5$\times10^{42}$ &
{6.85$\times10^{43}$} &
35&
8.9$\times10^{7}$\\
\hline
\end{tabular}

\vspace{0.2cm}
\begin{flushleft}
\footnotesize
$^{\rm a}$ Neutral H\,\textsc{i} outflow energetics \citep{Morganti2005}. \\
$^{\rm b}$ Hot X-ray gas energetics derived from \citep{Hardcastle2012}. \\
$^{*}$ Rovibrational lines are included while estimating the H$_2$ line luminosity. \\
\end{flushleft}

\end{table*}

\cite{Hardcastle2012} provide a detailed summary of the internal, kinetic, and total energies in various gas phases and find that the X-ray emitting hot gas dominates the energy budget. We extend their analysis by adding our JWST observations of the warm molecular and warm ionized gas phases (Table~\ref{tab:energy_power_budget}). The values for the neutral gas and the X-ray-emitting hot plasma are taken from the literature. We obtain the kinetic power of the hot plasma from kinetic energy using $\dot{E}_{kin}$=${E}_{kin}\times\frac{v}{R}$, where we assume a velocity of 650 km s$^{-1}$ and radius of 2.2~kpc following \cite{Hardcastle2012}. We note that the mechanical power of the cold, turbulent molecular gas disk is not provided in Table~\ref{tab:energy_power_budget}, and warm H$_2$ traces only a small fraction of the total molecular gas mass.  From the NOEMA CO moment-2 maps \citep{Morganti2023}, we derive an intensity-weighted line-of-sight velocity dispersion of $\sigma \simeq 70$~km~s$^{-1}$, which we treat as an upper limit on the turbulent velocity due to possible beam-smearing and unresolved velocity gradients. The turbulent kinetic power is $\sim 5\times10^{41}\ \mathrm{erg\,s^{-1}}$ for an assumed scale of turbulence, $\ell$ $\sim 2.2\ \mathrm{kpc}$. This value is of the same order of magnitude as the bulk kinetic power of both the ionized and warm molecular gas. 


Our analysis confirms that the warm H$_2$ and warm ionized gas carry only a minor fraction of the kinetic and internal energy. However, these dense phases are much more radiatively efficient than the low-density, hot, X-ray emitting gas, providing the vast majority of radiative dissipation. The hot phase is radiatively inefficient, such that most of its energy goes into inflating gas cavities and driving outflows rather than directly producing observable line emission.  The MIR-UV line luminosity exceeds the mechanical power output across all gas phases by an order of magnitude, indicating transfer of energy from the hot phase to the warm phase, most of which is dissipated via UV-IR emission lines.  Power must be transferred from the hot gas to the warm gas to explain the high emission line luminosity. The most likely mechanism, supported by our JWST observations, is shock-heating of the initially stationary, multiphase ISM by jet-driven hot outflows. Only a small fraction of that power goes into accelerating molecular, neutral, and warm ionized outflows (Table~\ref{tab:energy_power_budget}).


\subsection{AGN Jet vs. Radiative Feedback Efficiency}

Quantifying the efficiency of various modes of AGN feedback is important for understanding their relative impact on galaxy ISM and evolution. One useful measure of feedback efficiency is $f_e$, the fraction of jet (or AGN) power transferred to the thermal, turbulent, and kinetic energy of the ISM. The efficiency of transferring that energy to drive outflows $f_o$ is another important factor. Both of these considerations are crucial to understanding the impact on star formation, to the extent that hot, turbulent, or outflowing gas is unable to form stars. As discussed above, our JWST observations of line emission give a useful lower limit to the jet power transferred to the ISM. However, the existing estimates of total jet power are less than this number, giving a formal estimate of $f_e > 1.0$, an impossibility. 

Alternatively, we can estimate what fraction of jet power escapes via the relativistic particles and associated magnetic fields traced by the radio lobes. The bulk of the remaining jet power must go into the thermal and kinetic energy of the hot ISM, with a fraction of that going into shocking the warm ionized gas observed by JWST. According to \cite{Hardcastle2012}, the radio lobes carry a {\it minimum} energy of  $3 \times 10^{56}$ erg and the jet lifetime is 3.5 Myr. Alternatively, the 3.3 kpc half-length of the transverse X-ray lobes divided by their sound speed of 650 km s$^{-1}$  gives a comparable, slightly longer jet lifetime of 4.9 Myr. Combining these yields a lobe power of $\gtrsim 2 \times 10^ {42}$ erg s$^{-1}$. Remarkably, this is an order of magnitude smaller than the jet power estimates and a factor of $\sim 40$ times smaller than $L_\mathrm{rad}$, indicating that little energy escapes via the radio lobes and transfer of energy from the jet to the outflowing hot ISM is highly efficient ($O \sim 1$). Only a small fraction of the jet power goes into the kinetic energy of the warm H$_2$ and ionized outflows observed by JWST ($f_o \lesssim 0.01$) and the hot outflows observed by {\it Chandra} ($f_o \lesssim 0.06$). However, this is sufficient to drive massive outflows that deplete the ISM. The remainder of the jet power must go into heating and expanding the hot cocoon of the radio jet. 

Notably, AGN jet feedback can be much more efficient than AGN photoionization and dust heating, where most of the AGN power is simply absorbed and then re-emitted. Furthermore, AGN jet power can be transmitted to large distances, unlike AGN radiation power, which is diluted by $r^{-2}$ with radius. This is particularly true for jets that are trapped in the ISM, which appears to be the case for 3C 305, and for the early stages of more powerful radio jets, before they break out of the host galaxy. In simulations of low-power jets inclined along the galaxy disk and hence trapped, confinement enhances coupling efficiencies and strong local dissipation \citep{Meenakshi2022}. 



\subsection{Dynamics and Impact of Jet-Driven Multiphase Outflows}

The lack of complete spatial coincidence between the radio jet/lobes and the emission-line regions is expected for jet-driven outflows. In 3C305, we find that some outflows appear to be launched up to 2 kpc away from the radio jet hotspots. Although such outflows appear to have a jet-related origin, they are unlikely to have formed through direct momentum transfer between the jet and the ISM. Models instead show that the relativistic jet comes into contact with the dense disk material and is brought to a halt at the hotspot, leading to shock-heating of the gas in the immediate vicinity. The gas thus heated then expands adiabatically in a volume-filling cocoon at very high velocities. Simulations by \cite{Mukherjee2018} find that such a cocoon could possess temperatures of 10$^8$-10$^{10}$ K and densities in the range 10$^{-2}$-10$^{-5}$ cm$^{-3}$. It would not be possible to detect such a dilute, hot gas using the \textit{Chandra} telescope. However, wherever this volume-filling hot gas outflow impacts cold gas clouds, it drives shocks into the clouds and destroys them on a cloud-crushing timescale. 
The gas ablated from these clouds by their relative motion to the hot wind likely forms the ionized and molecular outflows observed with {\it Chandra} and JWST. In fact, recent simulations show that such a hot cocoon shreds the cold gas clouds, leading to the formation of turbulent mixing layers at the cloud-wind interface, which can in turn generate observable soft X-ray emission \citep{Ward2026}.

The ``z" shape of both the radio lobes and ionized outflows in 3C 305 is naturally explained by buoyancy. The outflows escape from the disk along the direction of the greatest pressure gradient, which is roughly perpendicular to the disk at their launch point. Similar jet-driven, z-shaped outflows are seen in NGC 4258 and 3C 293 \citep{Ogle2014,Lanz2015}. The radial offsets of the outflow launching points exterior to the jet hotspots indicate that the hot wind has a component of velocity along the disk. We conjecture that the fastest winds launched from these points may additionally impact other material before they escape the disk, leading to the observed cascading chains of outflows in 3C 305.

Our estimates of the molecular and ionized gas mass outflow rates enable us to determine the timescale for gas to be ejected from the 3C 305 host galaxy molecular disk. The gas-depletion timescale for the measured cold H$_2$ mass of $2\times 10^9 M_{\odot}$ and total mass outflow rate of 12 M$_{\odot}$~yr$^{-1}$ excluding X-ray gas is 170~Myr. In comparison, the star formation rate estimated from fitting the global SED is only $\sim 0.3 M_{\odot}$~yr$^{-1}$ and the corresponding gas depletion timescale by star formation is much longer at 7~Gyr \citep{Lanz2016}. The gas depletion times by outflows decrease further if smaller electron densities are assumed. 

Whether the outflowing molecular/ionized gas can escape the galaxy critically depends on the escape velocity at the given radius. If we assume that the molecular disk in 3C\,305 is moderately to highly inclined ($i\sim60^\circ$--$90^\circ$), the circular velocity derived from the observed projected rotation velocity is in the range of $v_{\rm circ}\sim260$--300~km~s$^{-1}$. This implies that the escape velocities will be $v_{\rm esc}\approx\sqrt{2}\,v_{\rm circ}\sim370$--425~km~s$^{-1}$ at 1.8~kpc (NE hotspot location) for an isothermal potential. Very fast outflows exceeding these velocity limits in 3C\,305 exist and will hence successfully escape, although a large fraction might still be recycled within the galaxy.

\cite{Hardcastle2012} estimate a much shorter jet dynamical timescale of 3.5~Myr in 3C\,305. We note that such an estimate of the jet dynamical timescale may be unreliable in a source such as 3C\,305, where the jet remains trapped within the galaxy due to strong interaction with the galactic disk. It is possible that jet expansion has stalled at the hotspots, and the jet ages are much longer. The jet will remain trapped within the dense gas disk until most of the gas has been effectively removed by the jet-driven outflows. Systems like 3C\,305 are not commonplace at low redshifts since elliptical galaxies spend little time in the gas-rich merger phase, along with the lower probability of the jets' alignment with the disk.

\section{Conclusion}
In this paper, we carry out a spatially resolved study of the impact of the compact jet in 3C\,305 on its multi-phase ISM with a focus on the warm molecular H$_2$ and warm ionized gas. Our main findings are summarized below. 
\begin{enumerate}

    \item H$_2$ outflows with velocities up to 400 km s$^{-1}$ are present near the locations of the radio hotspots. The spatial resolution of the JWST MIRI MRS instrument enabled the discovery of these previously unknown molecular outflows. Our results suggest that jet-driven outflows are capable of heating, entraining, and ejecting H$_2$ gas out of the disk.
    \item Despite the higher outflow velocity at the northeast hotspot, the mass outflow rate at the southwest hotspot seems to dominate. We conclude this is a result of differences in density. Hence, the nature of jet feedback to the ISM depends on the environment.
    \item Enhancement of [Fe\,II] 5.34$\mu$m emission near the radio hotspots supports the scenario where dust grains are destroyed by jet-driven shocks, releasing Fe atoms into the ISM.
    \item Our H$_2$ excitation modeling, assuming a power law distribution of temperature, reveals flatter power law indices ($n \sim 3.6$–4.4) at the hotspots, consistent with a higher fraction of hot molecular gas. We derive a warm H$_2$ mass ($T \gtrsim 200$ K) of $\sim 2.6 \times 10^{7} M_\odot$ corresponding to $\sim$1.3\% of the total molecular reservoir.
    \item The ionized gas shows even faster outflows near the radio hotspots than warm H$_2$, up to 1000 km s$^{-1}$.  Mid-IR line ratios are consistent with the `shock+precursor' models of the MAPPINGS line-ratio grids. The shock velocities are consistent with the observed ionized gas kinematics.
    \item Our energetics calculations show that the jet in 3C\,305 is highly dissipative, with most of the energy deposited locally in hot gas outflows, which drive shocks into the dense ISM. Most of this energy is radiated away via MIR-UV line emission. The jet also powers outflows, although it transfers $\lesssim$5–10\% of its power to outflow kinetic energy across all gas phases. 
    
\end{enumerate}
\label{sec:conclusion}

\begin{acknowledgments}
This work is based in part on observations made with the NASA/ESA/CSA James Webb Space Telescope. Some of the data presented in this paper were obtained from the Mikulski Archive for Space Telescopes (MAST) at the Space Telescope Science Institute. The specific JWST observations analyzed can be accessed via \dataset[https://doi.org/10.17909/m33s-hg28]{https://doi.org/10.17909/m33s-hg28}. This paper employs a list of Chandra datasets, obtained by the Chandra X-ray Observatory, contained in the Chandra Data Collection ~\dataset[DOI: 10.25574/cdc.595]{https://doi.org/10.25574/cdc.595}. STScI is operated by the Association of Universities for Research in Astronomy, Inc., under NASA contract NAS5–26555. Support to MAST for these data is provided by the NASA Office of Space Science via grant NAG5–7584 and by other grants and contracts. The Karl G. Jansky Very Large Array (VLA) is a facility of the National Science Foundation operated under cooperative agreement by Associated Universities, Inc.  This research has made use of data obtained from the \textit{Chandra} Data Archive provided by the Chandra X-ray Center (CXC). SGB acknowledges support from the Spanish grant PID2022-138560NB-I00, funded by
MCIN/AEI/10.13039/501100011033/FEDER, EU.
CO and SB acknowledge support from the Natural Sciences and Engineering Research Council (NSERC) of Canada. SB and PO gratefully acknowledge support from JWST grant JWST-GO-04237.

\end{acknowledgments}

%

\vspace{5mm}
\facilities{JWST(NIRSpec, MIRI MRS and NIRCam), CXO}


\software{astropy \citep{2013A&A...558A..33A,2018AJ....156..123A,2022ApJ...935..167A},  
          Cloudy \citep{2013RMxAA..49..137F},
          MAPPINGS\citep{Allen2008}
          }



\appendix

\bibliography{sample631}{}
\bibliographystyle{aasjournal}



\end{document}